\documentclass{JHEP3}
\usepackage{epsfig,amsmath,amssymb}

\def\lNN{l_{\rm n}}
\def\lFF{l_{\rm f}}
\def\lN{{l_{\rm n}}}
\def\lF{{l_{\rm f}}}

\def\NO{{\tilde\chi^0_1}} 
\def\NT{{\tilde\chi^0_2}} 
\def\NThree{{\tilde\chi^0_3}} 
\def\NQ{{\tilde\chi^0_4}} 
\def\lR{{\tilde l_R} }
\def\qL{{\tilde q_L}}
\def\gl{{\tilde g}}
\def\sq{{\tilde{q}}}
\def\lL{\tilde{l}_L}
\def\snu{\tilde{\nu}}
\def\stau{\tilde{\tau}}

\def\CO{{\tilde{\chi}_1^\pm}}
\def\CT{{\tilde{\chi}_2^\pm}}
\def\dR{{\tilde d_R}}
\def\uR{{\tilde u_R}}

\def\qR{{\tilde q_R}}
\def\bL{{\tilde b_L}}
\def\bR{{\tilde b_R}}
\def\sb{{\tilde b}}

\def\bT{{\tilde b_2}}

\def\sle{{\tilde l}}
\def\bO{{\tilde b_1}}
\def\tO{{\tilde t_1}}
\def\uL{{\tilde u_L}}
\def\dL{{\tilde d_L}}
\def\cL{{\tilde c_L}}
\def\sL{{\tilde s_L}}

\def\tauO{{\tilde\tau_1}}
\def\mNO{{m_\NO}}
\def\mlR{{m_\lR}}
\def\mNT{{m_\NT}}
\def\mqL{{m_\qL}}
\def\mgl{{m_\gl}}
\def\muL{{m_\uL}}
\def\mdL{{m_\dL}}
\def\mcL{{m_\cL}}
\def\msL{{m_\sL}}
\def\mbO{{m_\bO}}
\def\mbT{{m_\bT}}

\def\mlL{m_{\lL}} 
\def\msle{{m_\sle}}

\def\msq{{m_\sq}}
\def\sNO{{m_\NO^2}}
\def\slR{{m_\lR^2}}
\def\sNT{{m_\NT^2}}
\def\sqL{{m_\qL^2}}
\def\sgl{{m_\gl^2}}
\def\mlRfour{{m_\lR^4}}
\def\ssq{{m_\sq^2}}
\def\both{{\rm both}}
\def\low{{\rm low}}
\def\high{{\rm high}}
\def\equal{{\rm eq}}

\def\max{{\rm max}}
\def\min{{\rm min}}
\def\GeV{{\rm GeV}}
\def\mHalf{m_{1/2}}
\def\mZero{m_0}
\def\AZero{A_0}
\def\rqLNT{\frac{\mqL}{\mNT}}
\def\rNTlR{\frac{\mNT}{\mlR}}
\def\rlRNO{\frac{\mlR}{\mNO}}
\def\theory{{\rm th}}
\def\exp{{\rm exp}}
\def\mll{m_{ll}}
\def\mqll{m_{qll}}
\def\mqlN{m_{q\lN}}
\def\mqlF{m_{q\lF}}
\def\mql{m_{ql}}
\def\mqlLow{m_{ql(\low)}}
\def\mqlHigh{m_{ql(\high)}}
\def\mqllThres{m_{qll(\theta>\frac{\pi}{2})}}
\def\maxmll{m^\max_{ll}}
\def\maxmqll{m^\max_{qll}}
\def\maxmqlLow{m^\max_{ql(\low)}}
\def\maxmqlHigh{m^\max_{ql(\high)}}
\def\minmqllThres{m^\min_{qll(\theta>\frac{\pi}{2})}}
\def\mbll{m_{bll}}

\def\mblLow{m_{bl(\low)}}
\def\mblHigh{m_{bl(\high)}}
\def\mbllThres{m_{bll(\theta>\frac{\pi}{2})}}

\def\minmbllThres{m^\min_{bll(\theta>\frac{\pi}{2})}}
\def\SPSOaa{SPS~1a ${(\alpha)}$}
\def\SPSOab{SPS~1a ${(\beta)}$}
\def\s{\\ \vspace*{-3.5mm}}
\def\palpha{{($\alpha$)}}
\def\pbeta{{($\beta$)}}
\def\mqLowll{m_{q_{\low}ll}}
\def\mqHighll{m_{q_{\high}ll}}
\def\mqBothll{m_{q_{\both}ll}}
\def\mqLowlLow{m_{q_{\low}l(\low)}}
\def\mqLowlHigh{m_{q_{\low}l(\high)}}

\def\vvec#1{\mbox{\bf#1}}
\def\mat#1{\mbox{\bf{#1}}}
\def\av#1{\langle #1 \rangle}
\def\LSfun{\Sigma}
\def\scale{{\rm scale}}
\def\stat{{\rm stat}}
\def\nom{{\rm nom}}
\def\itB{\it}
\def\spcA{\mbox{\hspace{1.2ex}}}
\allowdisplaybreaks

\title{Measurement of SUSY Masses via Cascade Decays for SPS~1a}
\author{B.~K.~Gjelsten\\ 
Department of Physics, 
University of Oslo, 
P.O.B. 1048, 
Blindern, N-0316 Oslo, Norway\\ 
E-mail: \email{B.K.Gjelsten@fys.uio.no}}
\author{D.~J.~Miller\\ 
Department of Physics and Astronomy, 
University of Glasgow, 
Glasgow G12 8QQ, 
U.K. and \\ 
School of Physics,
University of Edinburgh,
Edinburgh EH9 3JZ,
U.K.\\
E-mail: \email{D.Miller@physics.gla.ac.uk}}
\author{P.~Osland\\ 
Department of Physics, 
University of Bergen, N-5007 Bergen, Norway\\ 
E-mail: \email{Per.Osland@ift.uib.no}}
\abstract{If R-parity conserving supersymmetry exists below the 
TeV-scale, new particles will be produced and decay in cascades 
at the LHC. The lightest supersymmetric particle will escape the 
detectors, thereby complicating the full reconstruction of the 
decay chains. 
In this paper we expand on existing methods for determining the 
masses of the particles in the cascade from endpoints of 
kinematical distributions. 
We perform scans in the mSUGRA parameter space to delimit the 
region where this method is applicable. 
From the examination of theoretical distributions for a wide 
selection of mass scenarios it is found that caution must be 
exerted when equating the theoretical endpoints with the 
experimentally obtainable ones. 
We provide analytic formulae for the masses in terms of the endpoints 
most readily available. 
Complications due to the composite nature of the endpoint 
expressions are discussed in relation to the detailed analysis 
of two points on the SPS~1a line. 
Finally we demonstrate how a Linear Collider measurement can 
improve dramatically on the precision of the masses obtained.}
\preprint{ATL-PHYS-2004-029\\
Edinburgh 2004/12}

\begin{document}

\section{Introduction}

The Standard Model (SM) of particle physics has been remarkably successful in
describing the physics probed by modern day particle accelerators. No
deviation from the SM has thus far been confirmed by experiment and only the
Higgs mechanism, the SM's instrument for the breaking of the electroweak
symmetry, remains to be discovered. Nevertheless, the SM suffers from
considerable theoretical difficulties, not least of which is the {\it
hierarchy problem}~\cite{Weinberg:1975gm}, the extreme sensitivity of the
electroweak scale to new physics. Such difficulties imply that the SM is only
an {\it effective low-energy theory} (albeit a highly successful one)
applicable only up to a few hundred GeV or so, and will need to be extended in
order to describe physics at higher scales.

One extension which has attracted a lot of attention is
supersymmetry~\cite{Fayet:1976cr,Dimopoulos:1981zb,Nilles:1983ge,Haber:1984rc}.
Supersymmetry not only solves the hierarchy problem but has many other
attractive features: it is the only non-trivial extension to the
Poincar\'e symmetry of space-time~\cite{Haag:1974qh}; it is essential
to the formulation of superstring theories~\cite{Green:1987sp}; it provides a
low-energy theory which is more amenable to the unification of the
fundamental forces into a Grand Unified Theory (GUT) at some high
energy scale~\cite{Sakai:1981pk}; it provides a {\it natural}
mechanism for generating the Higgs potential which breaks the
electroweak symmetry~\cite{Inoue:1982pi,Falck:1985aa,Ibanez:1982fr,
Ibanez:1982ee,Ellis:1982wr,Alvarez-Gaume:1983gj}; 
and it supplies a good candidate for
cold dark matter~\cite{Goldberg:1983nd}. 
Furthermore, if it is to be relevant in
solving the hierarchy problem it must exhibit experimental
consequences at the TeV-scale, and therefore can be tested by
experiment at the Large Hadron Collider (LHC). For an overview of
supersymmetry searches at LEP, the Tevatron and HERA, see
Ref.~\cite{Gianotti:vg}.

If supersymmetric particles are produced at the LHC, thus confirming
supersymmetry, it will become important to identify them and accurately
measure their masses. This will be essential for identifying the low
energy model and hopefully distinguishing the Minimal Supersymmetric
Standard Model (MSSM) from other non-minimal extensions.  Furthermore,
since no supersymmetric particles have so far been discovered,
supersymmetry must be broken by some as yet unknown mechanism. Only an
accurate determination of the supersymmetric particle masses and
couplings will allow us to determine the low energy soft supersymmetry
breaking parameters.  It is hoped that extrapolation of these masses
and couplings to high energies using the renormalisation group
equations will provide an insight into the mechanism of supersymmetry
breaking and, more generally, physics at the GUT
scale~\cite{Allanach:2004ud}. Since errors in the mass measurements
will be magnified by the renormalisation group running it is
absolutely essential that these masses be determined as accurately as
possible.

Here we will discuss supersymmetric mass measurements with reference
to one particular model of supersymmetry breaking, minimal
super-gravity
(mSUGRA)~\cite{Chamseddine:jx,Ibanez:1982ee,Inoue:1982pi,Ellis:1982wr,
Alvarez-Gaume:1983gj}. In this model, the supersymmetry is broken by
the interaction of new particles at high energy which are only linked
to the usual particles by gravitational interactions; this new sector
of physics is often referred to as the {\it hidden sector}. These
gravitational interactions transmit the supersymmetry breaking from
the hidden sector to our own sector, producing TeV scale effective
soft supersymmetry breaking terms in the GUT scale Lagrangian,
quantified by parameters which run logarithmically down to the probed
TeV scale. At the GUT scale, the scalar supersymmetric particles are
assumed to have a common mass, $\mZero$, while the gauginos have a
common mass $\mHalf$. The trilinear couplings are also taken to be
universal at the GUT scale and denoted $\AZero$.

Mass measurements in the MSSM are complicated by R-parity
conservation, which is introduced to prevent unphysical proton
decay. R-parity conservation requires that supersymmetric particles
are produced in pairs and causes the lightest supersymmetric
particle (LSP) to be stable. Consequently the LSP is inevitably the
end product of every supersymmetric decay and, if electrically
neutral, will escape the detector leaving no track or energy deposit. 
While this provides a very distinctive {\it missing energy} signature, 
it makes it very difficult to measure masses at the LHC since one 
cannot fully reconstruct decays.

Instead, mass measurements rely on continuous mass distributions
of decay products which attain extrema for certain configurations of
the particle momenta that are unambiguously determined by the masses
of initial, intermediate and final particles involved. 
These relations may often be inverted
to give the masses of unstable particles. This is analogous to the way
a bound on the neutrino mass can be obtained from the end-point of the
beta-decay spectrum of ${}^3\text{H}$ \cite{Lobashev:tp}, but is
usually more complex, since a long decay chain is often involved.

In this study we will consider supersymmetric mass measurements made
by examining the mass distribution endpoints or `edges' of the long decay 
chain\footnote{Throughout the text the notions lepton/slepton as 
well as $l$/$\sle$ will refer to the particles of the first 
and second generation. The third generation particles will be called 
tau/stau and denoted $\tau$/$\stau$.} 
$\tilde q \to \NT q \to \sle lq \to \NO llq$
in the ATLAS detector~\cite{AtlasTDRvol1}. 
In particular, we will focus on the
Snowmass mSUGRA benchmark line SPS~1a \cite{Allanach:2002nj}, but will
also include other mSUGRA parameters in a general discussion. In
addition to the usual SPS~1a point, which we will denote
SPS~1a~($\alpha$), we will also consider another point on the SPS~1a
line, denoted SPS~1a~($\beta$), which has a reduced branching ratio
for the decay. This will provide a counterpoint to the study of
SPS~1a~($\alpha$) where the branching ratio is rather high.  This
study differs from previous
reports~\cite{Baer:1995va,Hinchliffe:1996iu,Hinchliffe:1998zj,Bachacou:1999zb,
Polesello}
by (i) discussing theoretical distributions which arise for
different mass scenarios, (ii) providing inversion formulas, 
(iii) discussing ambiguities and complications related to the 
composite nature of the endpoint expressions. 
Furthermore, we provide an overview
of the mSUGRA parameter space, and consider a new point on
the SPS~1a line. Finally, we discuss the effects of including
Linear Collider data in the analysis.

In Sect.~\ref{sect:msugra} we will discuss general mSUGRA scenarios, paying
attention to supersymmetric decay branching ratios to gain some understanding
of how generally applicable these kinematic endpoint measurements are. We will
identify mass hierarchies for which these mass measurements are possible and
show that they occur over a large portion of the mSUGRA parameter space.  In
Sect.~\ref{sect:sps1a} we will outline the properties and mass spectra of the
Snowmass benchmark line SPS~1a and the associated points SPS~1a~($\alpha$) and
SPS~1a~($\beta$).  After defining the decay chain under investigation, we go
on in Sect.~\ref{sect:edge-meas} to discuss the theoretical framework of
cascade endpoint measurements, and present analytic expressions for masses in
terms of these endpoints. In Sect.~\ref{sect:data-gen-recon} the experimental
situation at the ATLAS detector will be studied, including event generation
and reconstruction, the removal of backgrounds, and results for the
measurement of the kinematic endpoints will be presented. The extraction of
masses from the kinematic endpoints will be described in
Sect.~\ref{sect:masses-from-edges}. Finally, the remarkable improvement of the
accuracy of the mass measurements obtained by using inputs from an $e^+e^-$
collider~\cite{lhc-lc,gjelsten-atlas} will be studied in
Sect.~\ref{sect:lhc-lc}, before drawing our conclusions in
Sect.~\ref{sect:conc}.

%%%%%%%%%%%%%%%%%%%%%%%%%%%%%%%%%%%%%%%%%%%%%%%%%%%%%%%%%%%%%%%%%%%%%%
\section{Cascade decays in mSUGRA scenarios} 
\label{sect:msugra}
%%%%%%%%%%%%%%%%%%%%%%%%%%%%%%%%%%%%%%%%%%%%%%%%%%%%%%%%%%%%%%%%%%%%%%

In this paper we will be examining the decay chain $\tilde q \to \NT q
\to \sle lq \to \NO llq$ in the Snowmass scenario SPS 1a for the
purpose of the extraction of the supersymmetric particle masses.
However, it is extremely unlikely that SPS 1a is exactly the parameter
choice of reality, and there would be little point to the study if our
methods were only applicable at SPS~1a, or a small region around it.
Therefore, in this section we will take a more general look at mSUGRA
scenarios to determine whether or not these methods may be used more
generally, over a wide parameter range.

At the LHC the main supersymmetric production will be sparticle pairs
$\gl\gl$, $\gl\sq$ and $\sq\sq$, as long as these are not too heavy.
Each sparticle immediately decays into a lighter sparticle, which in
turn decays further, until, at the end of the chain, an LSP will be
produced.  Since there are two parent supersymmetric particles, each
event will typically have two such chains, complicating their
reconstruction.  

What route is taken from the initial gluino or squark down to the LSP depends
on which decay channels are open, as well as their branching fractions.
Therefore there are two criteria which must be met in order to use the cascade
kinematic endpoint methods: firstly the sparticle mass hierarchy must be such
that the analysed decay chain is allowed; and secondly, the cross-section for
the entire decay chain must be large enough to allow analysis.

\subsection{The mSUGRA mass hierarchy}

In a general MSSM model there are few constraints on the sparticle
masses, so little can be assumed about their relative mass hierarchy.
However, if universal boundary conditions are imposed on the mass
parameters at the GUT scale, some mass orderings at the TeV scale are
natural, and some are even necessary. 

For example, in mSUGRA scenarios squarks and sleptons have a common
mass at the GUT scale, $\mZero$, but when the masses are evolved down
to the TeV scale, QCD interactions affect only the evolution of the
squarks and not the colourless sleptons. Consequently, squarks are
always heavier than sleptons at LHC energies.  For the same reasons,
although they may start off with different masses at the GUT scale,
running induced by QCD interactions is usually enough to result in the
squarks of the first two generations being heavier than the
neutralinos and charginos at the TeV scale. (Large mixing can however
bring $\tO$, and to a lesser extent $\bO$, quite low in mass.)

At the opposite end of the spectrum, there are a number of possible
candidates for the LSP: the lightest neutralino, $\NO$, the lightest
chargino $\CO$, or the lightest slepton (in mSUGRA models, the
gravitino is usually rather heavy). Which of these is the LSP depends
on the relative sizes of $\mZero$, $\mHalf$ and the derived low-energy
Higgs-higgsino mass parameter $\mu$. However, the assumption of gauge
unification at the GUT scale explicit in mSUGRA models leads to the
relation
\begin{equation}
M_1 \approx \frac{5}{3} \tan^2 \theta_W M_2
\label{eq:gauginomassrelation}
\end{equation}
between the $U(1)$ and $SU(2)$ gaugino masses, $M_1$ and $M_2$
respectively. As a result, $M_1$ tends to be rather low, significantly
lower than $\mHalf$.  Furthermore, the derived quantity $\mu$ is often
required to be much larger than $M_1$ in order to give the correct
electroweak symmetry breaking (at the SPS~1a~($\alpha$) reference
point $\mu = 357.4$~GeV). For the majority of parameter choices this
implies that the LSP will be $\NO$, with $\tauO$ being the LSP only if
$\mZero \ll \mHalf$, and $\CO$ only for a small region where $\mHalf \to
0$. The left-handed sneutrino, by virtue of its $SU(2)$ interactions,
is usually heavier than $\tauO$, and is anyway ruled out by direct
searches~\cite{Eidelman:2004wy}.  It is indeed fortunate that $\NO$ is the LSP for
most of the parameter space since it is clear that only an
electrically neutral LSP can play the role of the dark matter
constituent which is believed to fill the universe.  Finally, the
gaugino mass relation, Eq.~(\ref{eq:gauginomassrelation}), implies
that the LSP is usually bino-like. \s

%%%%%%%%%%%%%%%%%%%%%%%%%%%%%%%%%%%%%%%%%%%%%%%%%%%%%%%%%%%%%%%%%%%%%%
\FIGURE[ht]{
%\begin{figure}
%%Begin InstantTeX Picture
\let\picnaturalsize=N
\def\picsize{8.0cm}
%If you do not have the picture file add:
%\let\nopictures=Y
%to the beginning of the file.
\ifx\nopictures Y\else{
\let\epsfloaded=Y
\centerline{\hspace{9mm}{\ifx\picnaturalsize N\epsfxsize \picsize\fi
\epsfbox{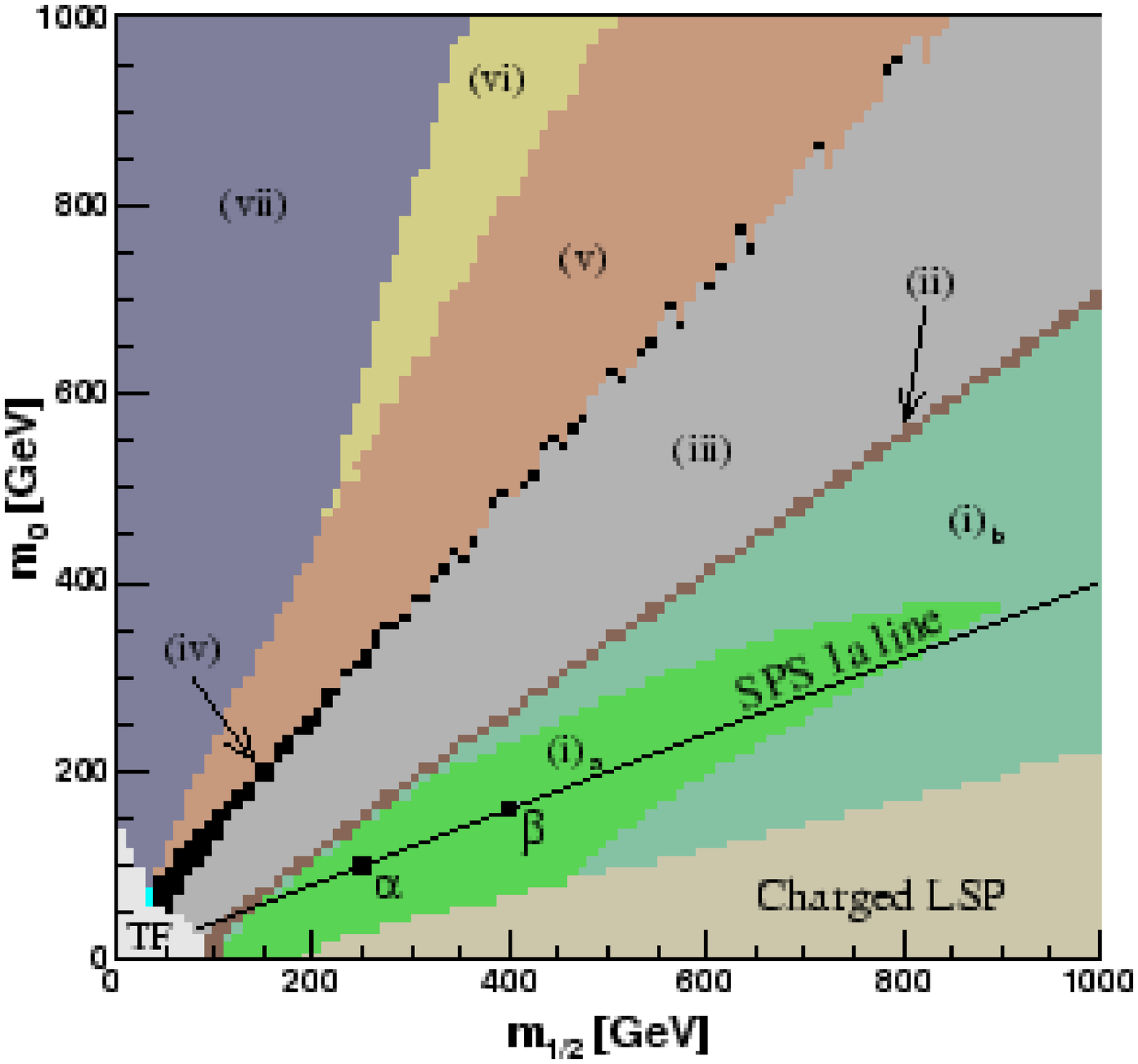}
            {\ifx\picnaturalsize N\epsfxsize \picsize\fi
\hspace{-5mm}
\epsfbox{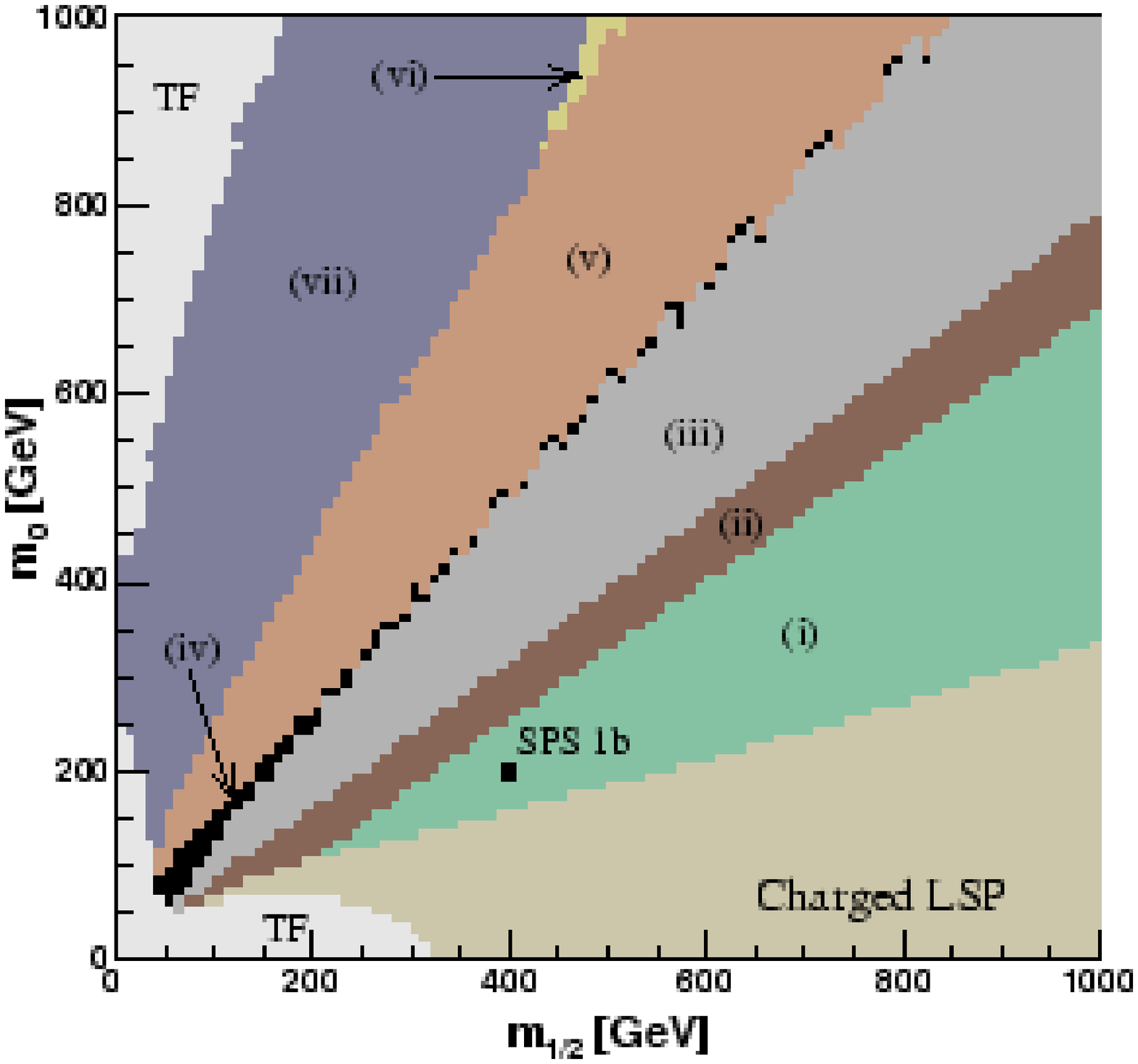}} } }
\centerline{\hspace{9mm}{\ifx\picnaturalsize N\epsfxsize \picsize\fi
\epsfbox{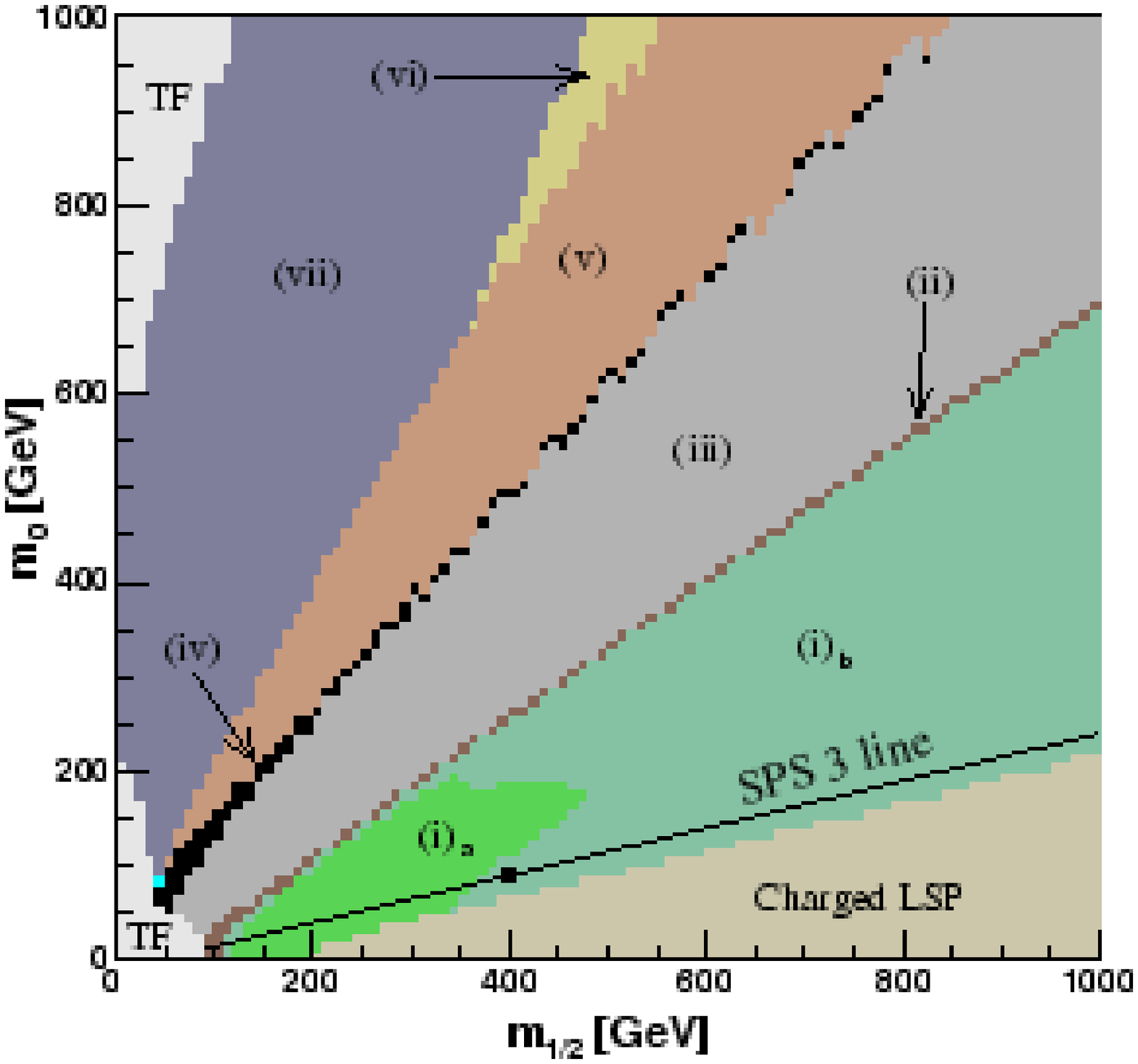}
            {\ifx\picnaturalsize N\epsfxsize \picsize\fi
\hspace{-5mm}
\epsfbox{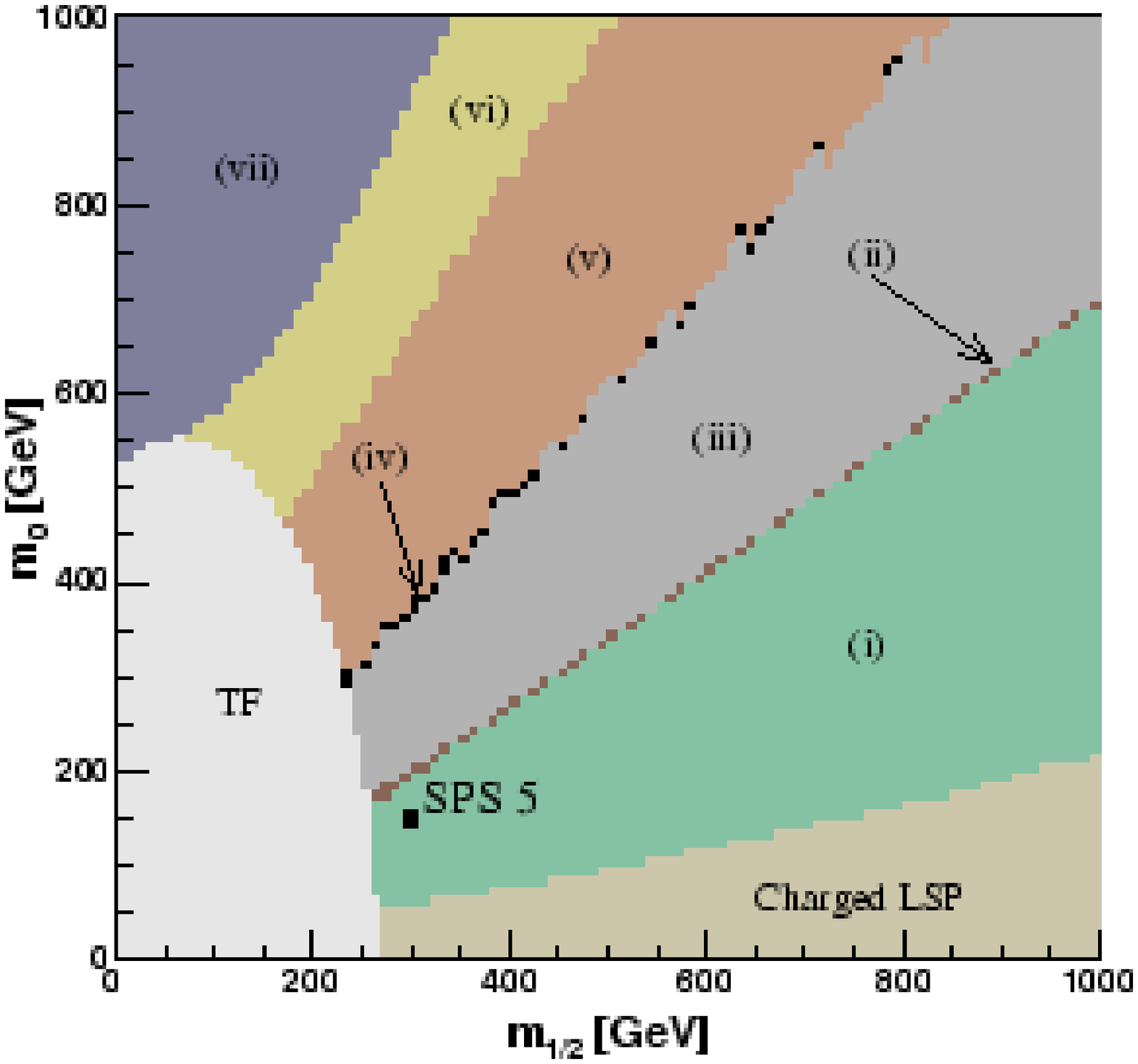}} } }
}\fi
%%End InstantTeX Picture
\vspace{-6mm}
\caption{Classification of different hierarchies, labeled (i)--(vii),
         for four combinations of $\tan\beta$ and $\AZero$, such that the
         four panels contain respectively the SPS~1a line, the SPS~1b
         point, the SPS~3 line, and the SPS~5 point. 
	 The regions marked `TF' are theoretically forbidden.
	 (See text for details.)
\label{Fig:m_half-m_0-scans}}}
%\end{figure}
%%%%%%%%%%%%%%%%%%%%%%%%%%%%%%%%%%%%%%%%%%%%%%%%%%%%%%%%%%%%%%%%%%%%%%

The first requirement for the decay chain $\tilde q \to \NT q \to \sle
lq \to \NO llq$ is that the gluino should be 
comparable to or heavier than the squark
initiating the decay chain.  If the gluino is sufficiently light,
then the squark will almost always choose to decay via its strong
interaction $\tilde q \to \tilde g q$ rather than by the electroweak
decay $\tilde q \to \NT q$. Of course, one does not need all of the
squarks to be lighter than the gluino; as long as one squark, for
example $\tilde b_1$, is lighter than the gluino, useful information
can potentially be obtained from its subsequent decay chain.  The
second important characteristic is that $\NT$ should be heavier than
$\sle$, thereby allowing the lower part of the chain to proceed, $\NT
\to \sle l \to \NO ll$. Otherwise $\NT$ will decay to $\NO Z$ or $\NO h$, 
or to $\NO f\bar f$ via a three-body decay, and the useful
kinematic endpoints are lost.

In order to understand where in the mSUGRA parameter space these
hierarchy requirements are realised, we have performed a scan over the
$\mHalf$--$\mZero$ plane for four different choices of $\AZero$ and
$\tan \beta$ (with $\mu>0$), and identified the different hierarchy
regions with different colours in Fig.~\ref{Fig:m_half-m_0-scans}. The
renormalisation group running of the parameters from the GUT scale to
the TeV scale has been done using version 7.58 of the program {\tt
ISAJET} \cite{Baer:1993ae}, which is inherent to the definition of
the `Snowmass Points and Slopes' 
(see Sect.~\ref{sect:sps1a}).

The upper left plot shows the $\mHalf$--$\mZero$ plane with
$\AZero=-\mZero$ and $\tan \beta=10$ and includes the SPS~1a line and
points (labeled ($\alpha$) and ($\beta$)). The upper right plot has
$\AZero=0$ and $\tan \beta=30$ and contains the benchmark point
SPS~1b. The lower left plot also has $\AZero=0$ but $\tan \beta=10$
and contains the SPS~3 benchmark line and point. Finally the lower
right plot has $\AZero=-1000$~GeV and $\tan \beta=5$ and contains SPS~5.

The different hierarchies themselves are combinations of the hierarchy
between the gluino and the squarks important to the upper part of the
decay chain, and that of $\NT$ and the sleptons relevant to the later
decays. 
Since $\mlR<\mlL$ for any set of mSUGRA parameters, 
we here use $\lR$. The seven numbered regions are defined by:
\begin{alignat}{23} \label{Eq:hierarchies}
&\text{(i)}&\quad &\gl>\max(\dL,\uL,\bO,\tO)&\quad &\text{and}\quad 
\NT>\max(\lR,\tauO) \nonumber \\
&\text{(ii)}&\quad &\gl>\max(\dL,\uL,\bO,\tO)&\quad &\text{and}\quad 
\lR>\NT>\tauO  \nonumber \\
&\text{(iii)}&\quad &\gl>\max(\dL,\uL,\bO,\tO)&\quad &\text{and}\quad 
\min(\lR,\tauO)>\NT \nonumber \\
&\text{(iv)}&\quad &\dL>\gl>\max(\uL,\bO)&\quad &\text{and}\quad 
\min(\lR,\tauO)>\NT \nonumber \\
&\text{(v)}&\quad &\min(\dL,\uL)>\gl>\bO&\quad &\text{and}\quad 
\min(\lR,\tauO)>\NT \nonumber \\
&\text{(vi)}&\quad &\min(\dL,\uL,\bO)>\gl>\tO &\quad &\text{and}\quad 
\min(\lR,\tauO)>\NT \nonumber \\
&\text{(vii)}&\quad &\min(\dL,\uL,\bO,\tO)>\gl &\quad &\text{and}\quad 
\min(\lR,\tauO)>\NT
\end{alignat}
where for fermions a particle's symbol represents its mass, while for
scalars a particle's symbol represents the sum of the masses of the
scalar and its SM partner.
Also shown (mauve) is a region where the LSP is charged and therefore
ruled out, as well as a theoretically forbidden (TF) region (gray)
for low $\mHalf$.

It is interesting to note that there are {\it no} regions where a squark is
heavier than the gluino {\it and} $\NT$ is heavier than one of the
sleptons. This is simply because the gluino and [the gaugino part of] the
neutralino have a common mass, $\mHalf$, at the high scale, so if the gluino
is light, the neutralinos will also be light.

However, for more general non-mSUGRA unification scenarios one could still
expect hierarchies of the type $\mqL>\mgl>\mNT>\msle>\mNO$ to be realised.  It
would then be important to be able to distinguish one hierarchy from the
other; this should be possible using the kinematic endpoints, number of
$b$-quarks in the final state, etc. Also it is possible
to distinguish $\NT\to\sle l \to\NO ll$ ($\mNT>\msle$) from $\NT\to\NO ll$
($\msle>\mNT$).  The first has the well-known triangular shape of $\mll$ while
the second has a typical 3-body shape.  All in all it should therefore be
possible to distinguish the various hierarchies (\ref{Eq:hierarchies}) before
continuing to determine the masses themselves.

Region (i) is the only one that has a `useful' squark decay together
with a decay of $\NT$ to a slepton, and is shown in 
light and dark green in Fig.~\ref{Fig:m_half-m_0-scans}. We see that
there is therefore a large region where the mass hierarchy is
compatible with the methods presented here, and even though we will
only perform the analysis for the points ($\alpha$) and ($\beta$) on
the SPS~1a line, one would expect these methods to be widely
applicable.

However, these plots of the mass hierarchies really only show the
regions in which the masses are such that the decay chain {\it may}
occur. If the full decay chain is to be useful, it must have a
sufficiently large branching ratio to be seen above the many
backgrounds. We will therefore go on to examine the branching ratios
of the pertinent decays over (a restricted range of) the mSUGRA
parameter space. As a first taste, we have highlighted in a brighter
green (and denoted (i)$_a$) the part of region (i) corresponding to
where the overall branching ratio for the decay chain $\tilde q \to
\NT q \to \sle lq \to \NO llq$ exceeds a tenth of that at the SPS~1a
$(\alpha)$ reference point. Although the decay chain is available over
a rather large region of the parameter space, using this decay for
large values of $\mHalf$ and $\mZero$ will be extremely challenging
due to the small branching ratio.

\subsection{Gluino and squark decays: the upper part of the chain}

%%%%%%%%%%%%%%%%%%%%%%%%%%%%%%%%%%%%%%%%%%%%%%%%%%%%%%%%%%%%%%%%%%%%%%
\FIGURE[ht]{
%\begin{figure}
%%Begin InstantTeX Picture
\let\picnaturalsize=N
\def\picsize{7.5cm}
%If you do not have the picture file add:
%\let\nopictures=Y
%to the beginning of the file.
\ifx\nopictures Y\else{
\let\epsfloaded=Y
\centerline{\hspace{4mm}{\ifx\picnaturalsize N\epsfxsize \picsize\fi
\epsfbox{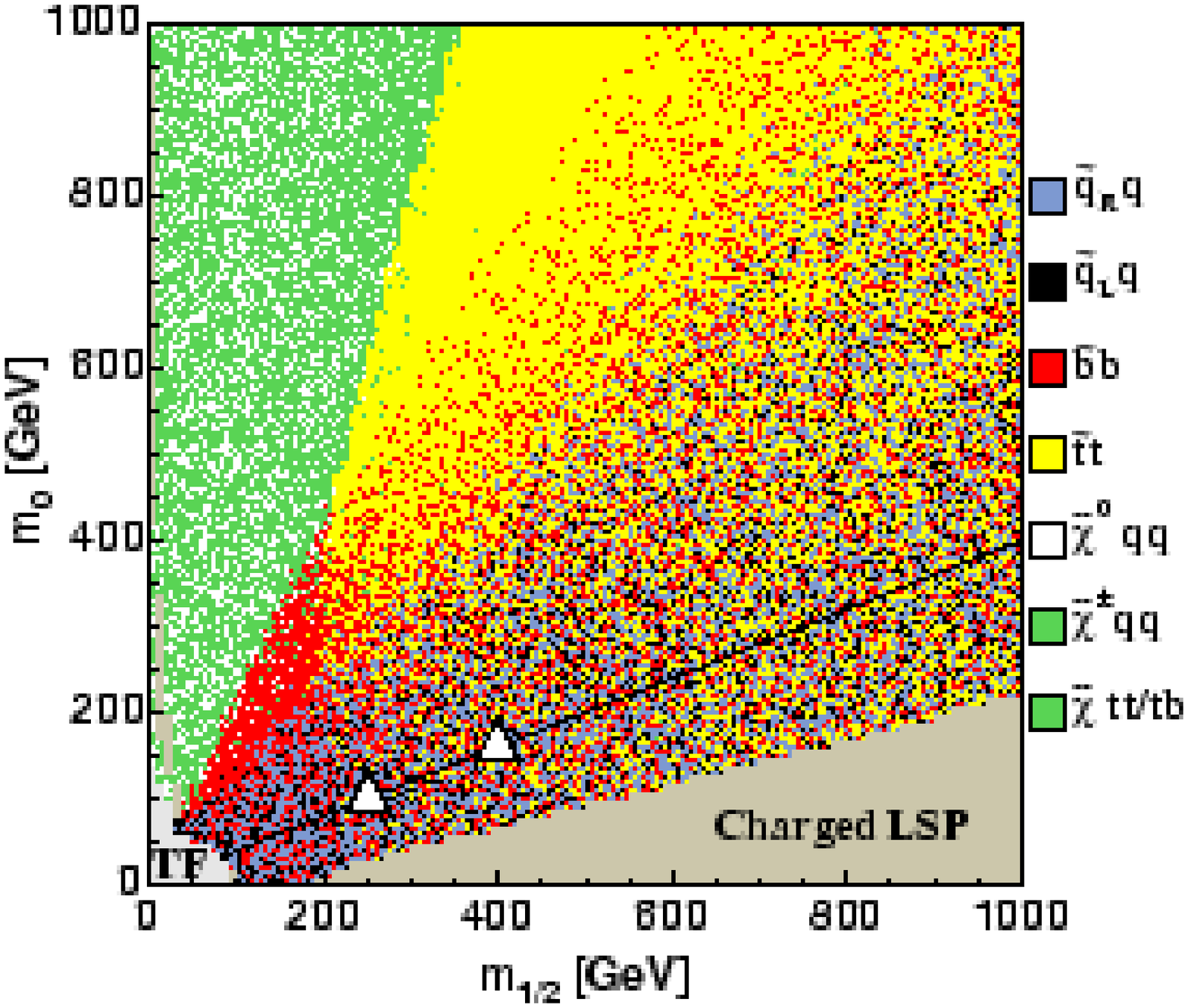}
            {\ifx\picnaturalsize N\epsfxsize \picsize\fi
\epsfbox{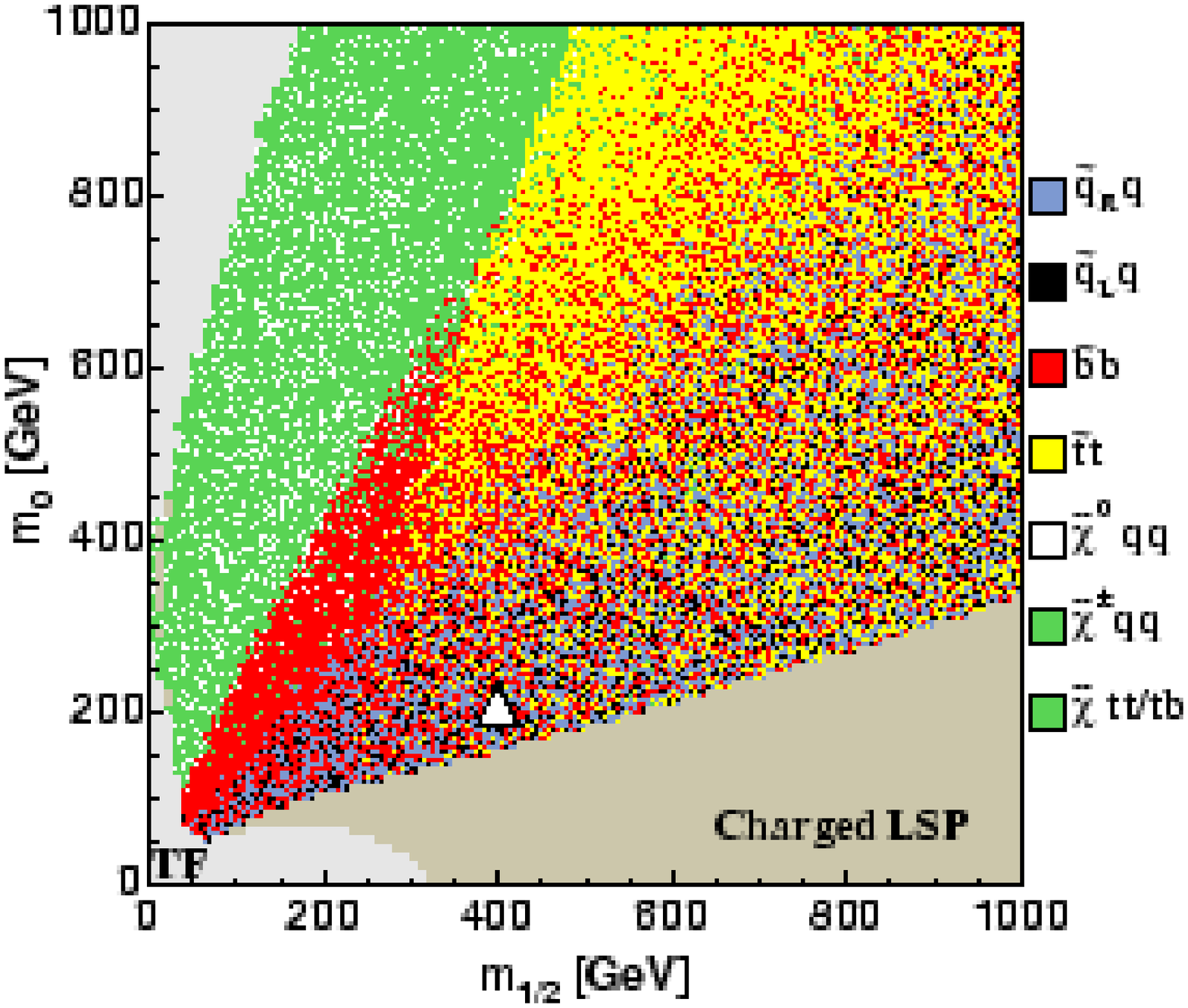}} } } 
}\fi
%%End InstantTeX Picture
\vspace{-6mm}
\caption{Decay channels of $\gl$ with $\AZero=-\mZero$, $\tan\beta=10$ (left)
and $A_0=0$, $\tan\beta=30$ (right). 
In the left panel the SPS~1a line is shown together with the two points 
\palpha\ and \pbeta\ marked with triangles. 
In the right panel the triangle marks the SPS~1b point. 
The branching ratio of a given decay channel in a small
neighbourhood of the $\mHalf$--$\mZero$ plane is equal to the fraction which
the corresponding colour occupies in that neighbourhood.  The region where
$\NO$ is not the LSP is denoted `Charged LSP' and is discarded.  Some
regions are also forbidden theoretically, in that e.g.\ it is not possible to
obtain electroweak symmetry breaking (labeled `TF').
\label{Fig:m_half-m_0-BRglscans}}}
%\end{figure}
%%%%%%%%%%%%%%%%%%%%%%%%%%%%%%%%%%%%%%%%%%%%%%%%%%%%%%%%%%%%%%%%%%%%%%

The decay branching ratios of the gluino are shown in
Fig.~\ref{Fig:m_half-m_0-BRglscans} over the $\mHalf$--$\mZero$ plane
for two different scenarios. The representation is such that the
branching ratio of a given decay channel in a small neighbourhood of
the $\mHalf$--$\mZero$ plane is equal to the fraction which the
corresponding colour occupies in that neighbourhood. Since the gluino
only feels the strong force, it has to decay into a quark and a
squark. If no squark is light enough, a three-body decay through an
off-shell squark will take place; this is what happens in the
green/white region of Fig.~\ref{Fig:m_half-m_0-BRglscans} at small
$\mHalf$.

For the rest of the $\mHalf$--$\mZero$ plane the gluino decays fairly
democratically into the accessible squarks. In considerable parts of
the plane only one two-body decay is open, $\bO b$ (red) or $\tO t$
(yellow), in which case the allowed decay takes close to the full
decay width. Although one can in principle obtain information about
the gluino mass by analysing its decay chain, we will only consider
here the decay chain starting from a parent squark, and leave the
gluino case for a separate publication~\cite{Gjelsten-2}. \s

%%%%%%%%%%%%%%%%%%%%%%%%%%%%%%%%%%%%%%%%%%%%%%%%%%%%%%%%%%%%%%%%%%%%%%
\FIGURE[ht]{
%\begin{figure}
%%Begin InstantTeX Picture
\let\picnaturalsize=N
\def\picsize{7.5cm}
%If you do not have the picture file add:
%\let\nopictures=Y
%to the beginning of the file.
\ifx\nopictures Y\else{
\let\epsfloaded=Y
\centerline{\hspace{6mm}{\ifx\picnaturalsize N\epsfxsize \picsize\fi
\epsfbox{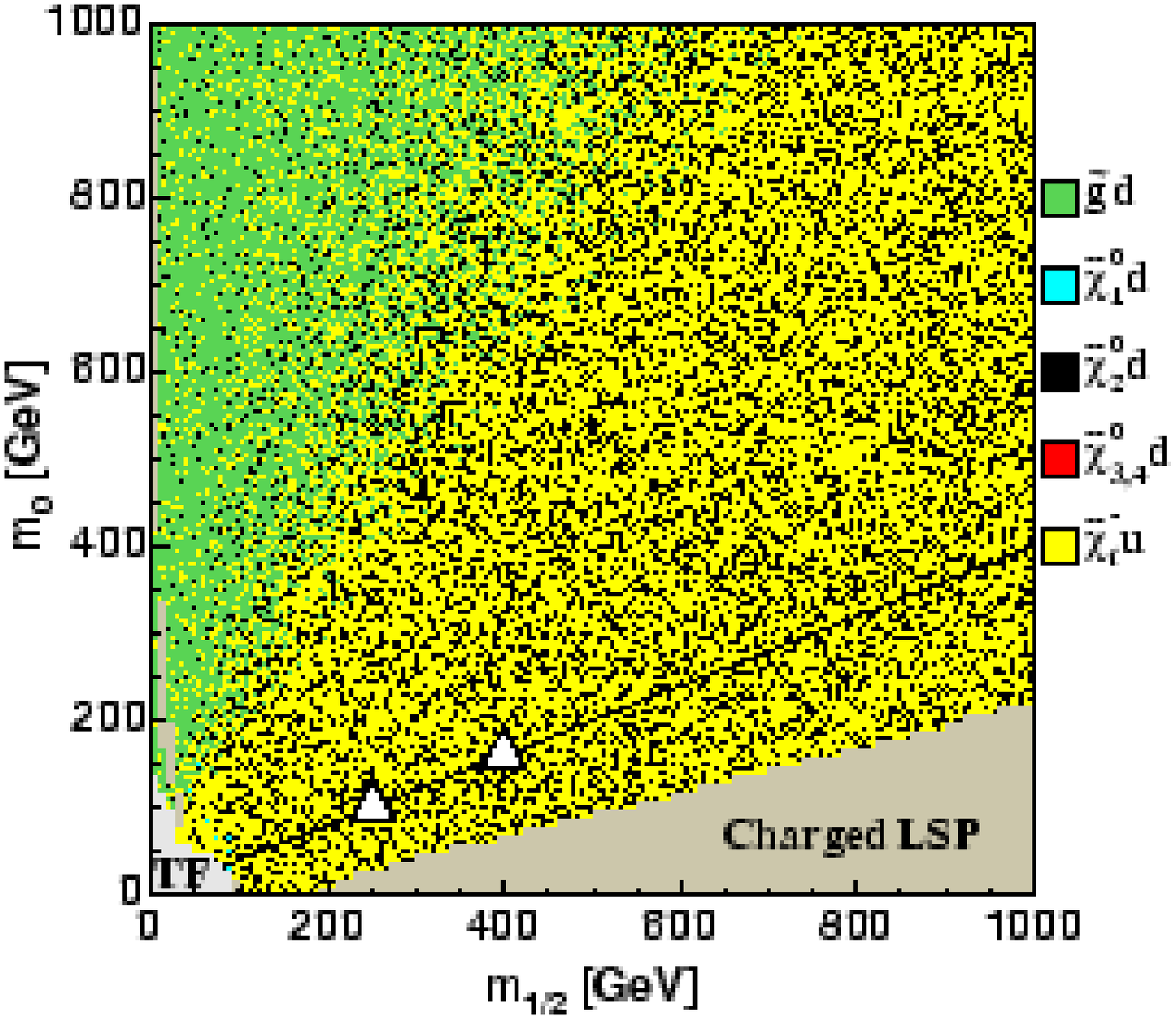}
            {\ifx\picnaturalsize N\epsfxsize \picsize\fi
\epsfbox{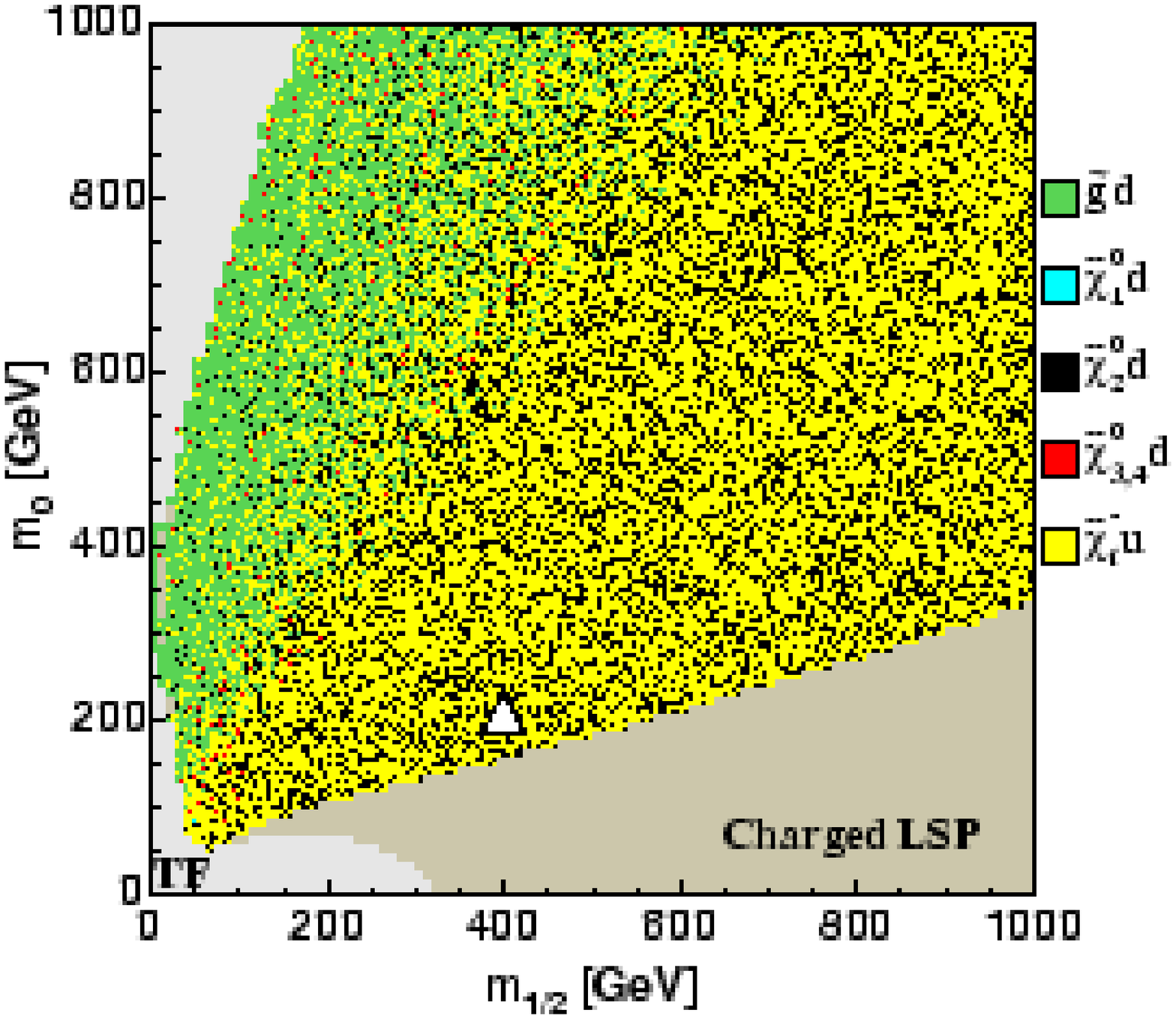}} } } 
\centerline{\hspace{6mm}{\ifx\picnaturalsize N\epsfxsize \picsize\fi
\epsfbox{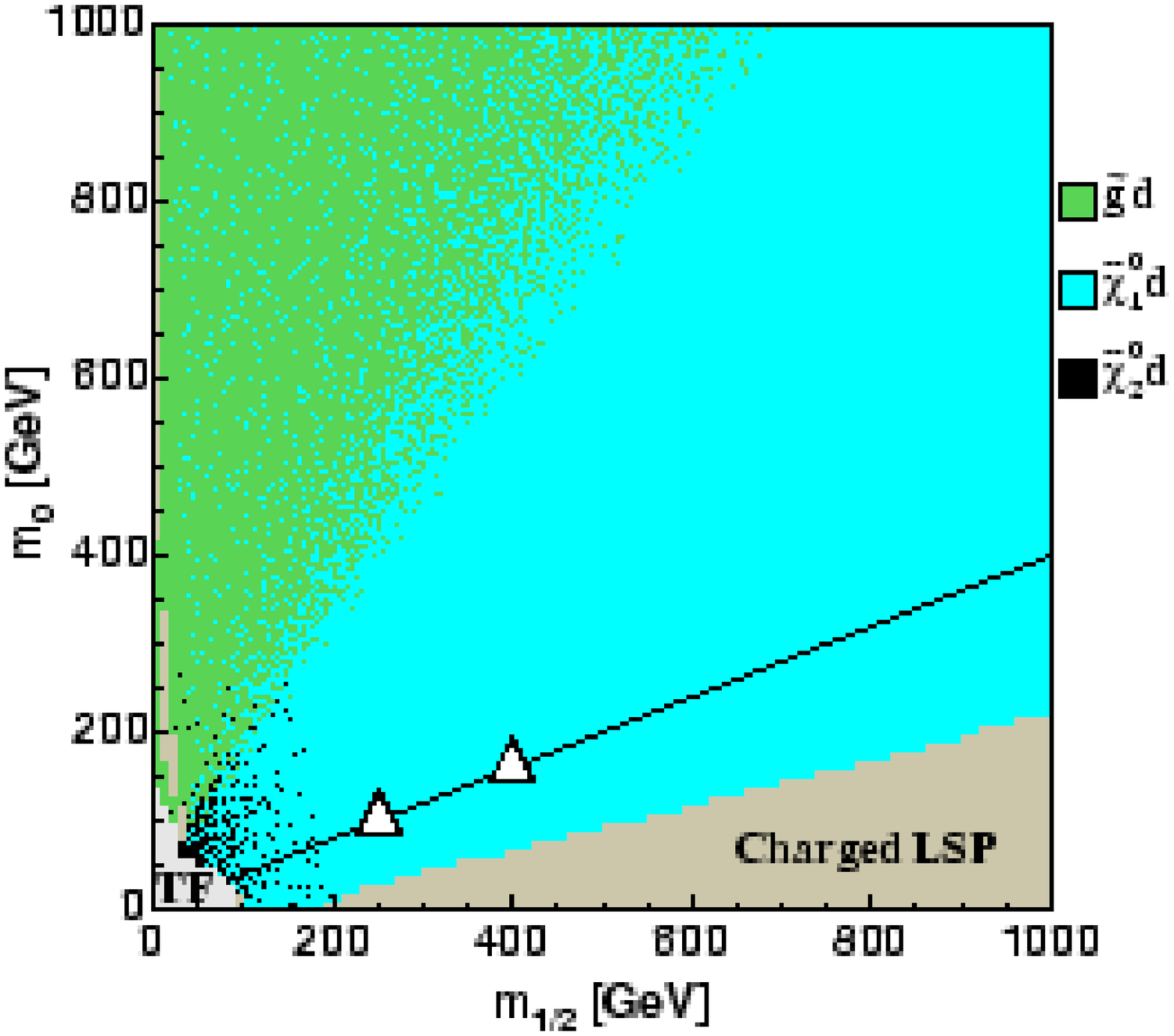}
            {\ifx\picnaturalsize N\epsfxsize \picsize\fi
\epsfbox{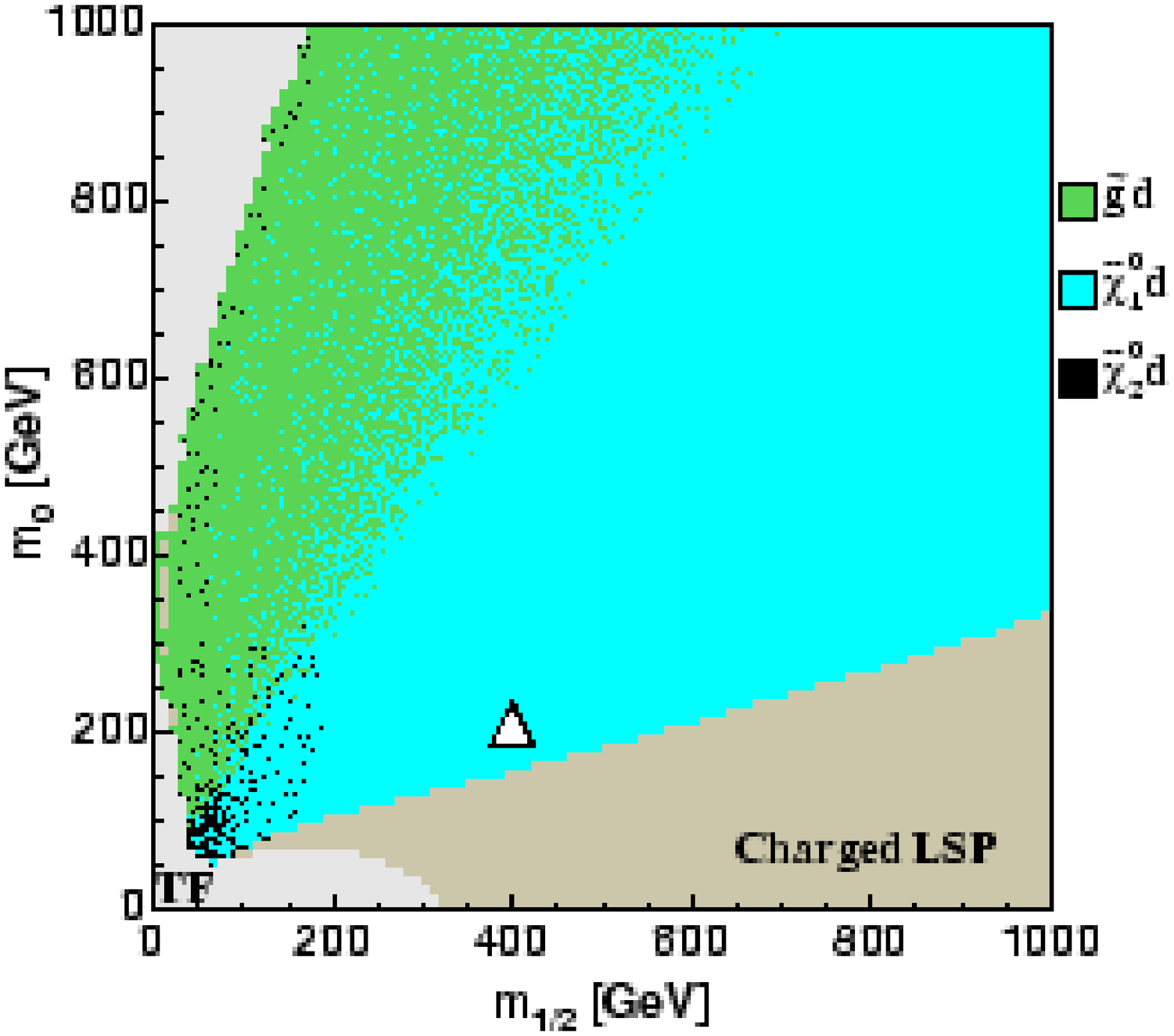}} } } 
}\fi
%%End InstantTeX Picture
\vspace{-6mm}
\caption{Decay channels of $\uL$ (top) and $\uR$ (bottom) with
$\AZero=-\mZero$, $\tan\beta=10$ (left) and $A_0=0$, $\tan\beta=30$ (right).
(See the caption of Fig.~\ref{Fig:m_half-m_0-BRglscans}.)  Note that the
branching ratios for all first and second-generation squarks are assumed to be
essentially the same.
\label{Fig:m_half-m_0-BRdLscans}}}
%\end{figure}
%%%%%%%%%%%%%%%%%%%%%%%%%%%%%%%%%%%%%%%%%%%%%%%%%%%%%%%%%%%%%%%%%%%%%%

As already intimated, squarks may decay by the strong force into a quark and a
gluino (if the gluino is lighter), or decay by weak interactions into a quark
and a chargino or neutralino, or via a loop into a gluon and a lighter squark.
If kinematically allowed, the strong interaction takes a large fraction of the
branching ratio, but since the (lighter) charginos and neutralinos are
typically much lighter than the gluino, there will always be some neutralino
production.

Within mSUGRA models, the squarks $\dL$ and $\uL$ are very close in
mass and behaviour.  Furthermore, the second generation squarks, $\sL$
and $\cL$, are almost identical copies of the former two.  It is
therefore useful to have the common notation, $\qL$, for these four
squarks.  In a similar manner $\qR$ is used for $\dR$, $\uR$ and their
second generation copies.  The right-handed squarks differ from the
left-handed ones in that they do not feel weak interactions, which
again makes their decay pattern different.  In
Fig.~\ref{Fig:m_half-m_0-BRdLscans} the decay branching ratios of
$\uL$ and $\uR$ are shown in the $\mHalf$--$\mZero$ plane for two
different scenarios.

For $\mHalf \ll \mZero$, when the gluino mass is smaller than the
squark mass, both $\qL$ and even more so $\qR$ have strong-interaction
decays.  In the rest of the $\mHalf$--$\mZero$ plane, when the strong
decay is forbidden or suppressed by phase space, their decay patterns
are very different: while $\qR$ decays directly into the LSP, $\qL$
prefers $\NT$ and $\CO$.

For both low and high $\tan\beta$, the $\NO$ is predominantly bino,
with only a tiny admixture of wino and higgsino, while $\NT$ and $\CO$
are mainly wino. For quite low mass parameters, $\mHalf\lesssim100$~GeV, 
they become more mixed.  
Since $\qL$ generally has a much larger SU(2) coupling than
U(1) coupling, decays to $\CO$ and $\NT$ will be preferred
unless the difference in phase space makes the decay to the lighter
$\NO$ competitive.  In contrast, since the $\qR$ has no SU(2)
interaction it will decay predominantly to the bino $\NO$, 
except at quite low mass parameters where the neutralinos change character. 

%%%%%%%%%%%%%%%%%%%%%%%%%%%%%%%%%%%%%%%%%%%%%%%%%%%%%%%%%%%%%%%%%%%%%%
\FIGURE[ht]{
%\begin{figure}
%%Begin InstantTeX Picture
\let\picnaturalsize=N
\def\picsize{7.5cm}
%If you do not have the picture file add:
%\let\nopictures=Y
%to the beginning of the file.
\ifx\nopictures Y\else{
\let\epsfloaded=Y
\centerline{\hspace{5mm}{\ifx\picnaturalsize N\epsfxsize \picsize\fi
\epsfbox{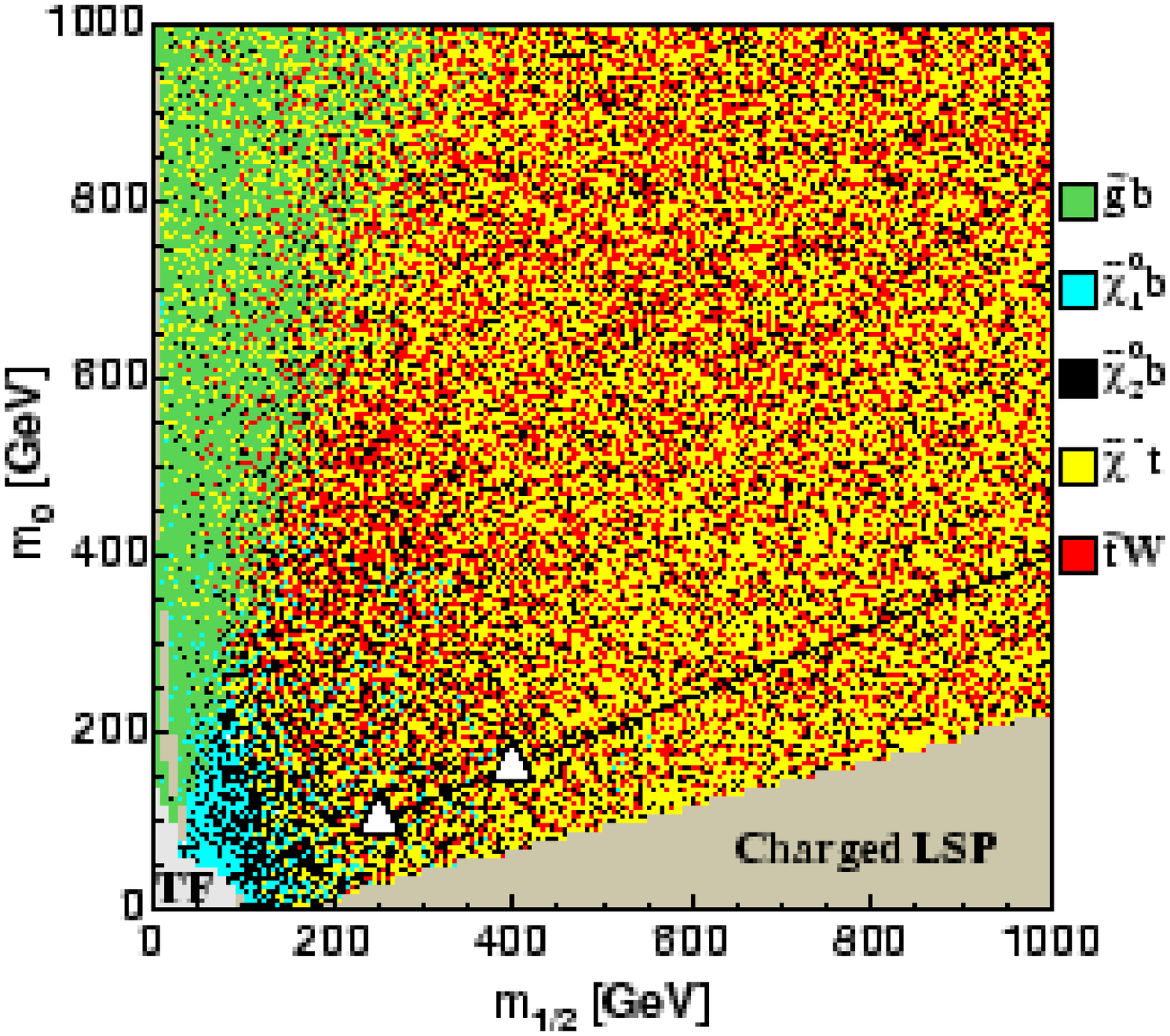}
            {\ifx\picnaturalsize N\epsfxsize \picsize\fi
\epsfbox{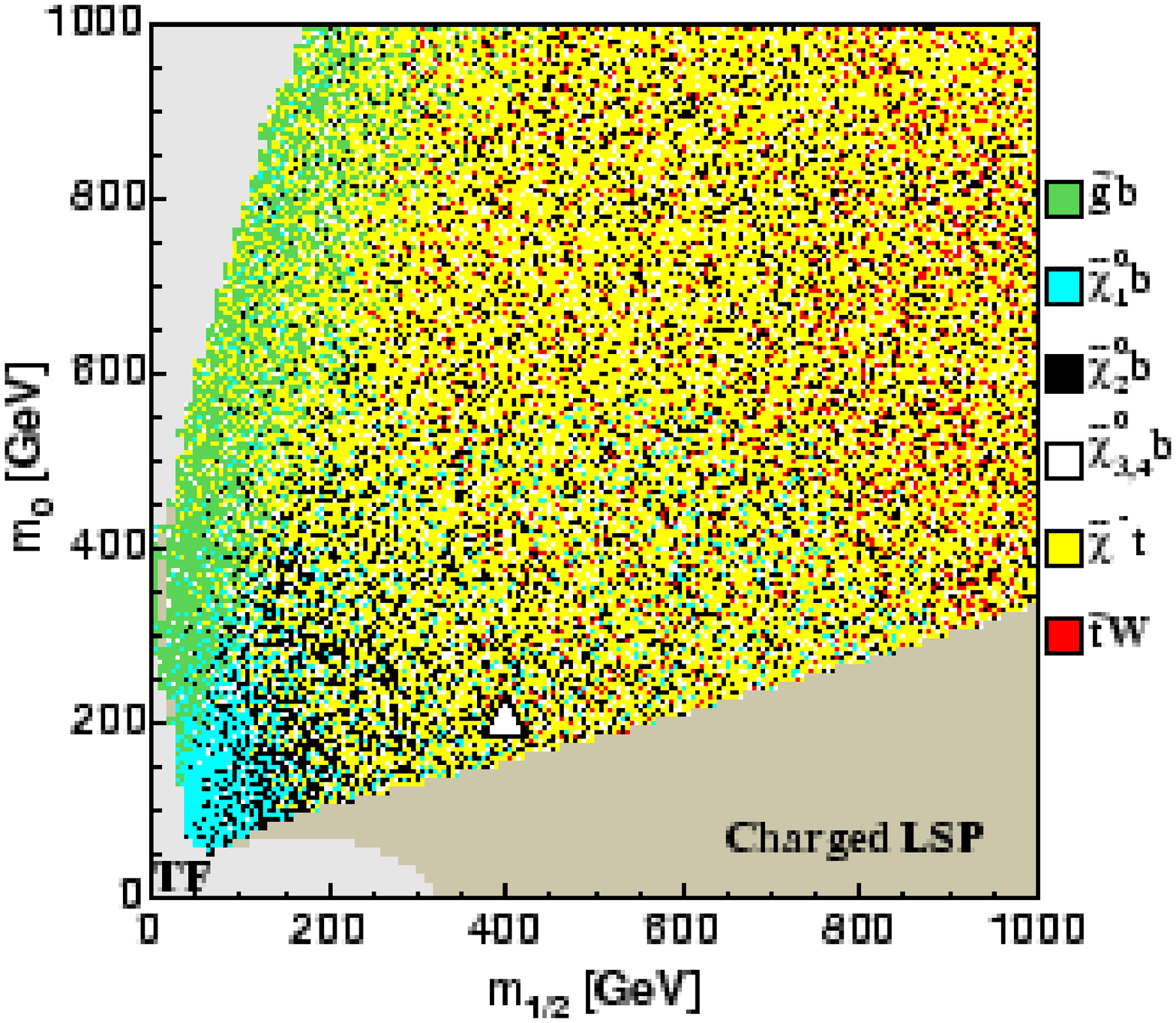}} } } 
\centerline{\hspace{5mm}{\ifx\picnaturalsize N\epsfxsize \picsize\fi
\epsfbox{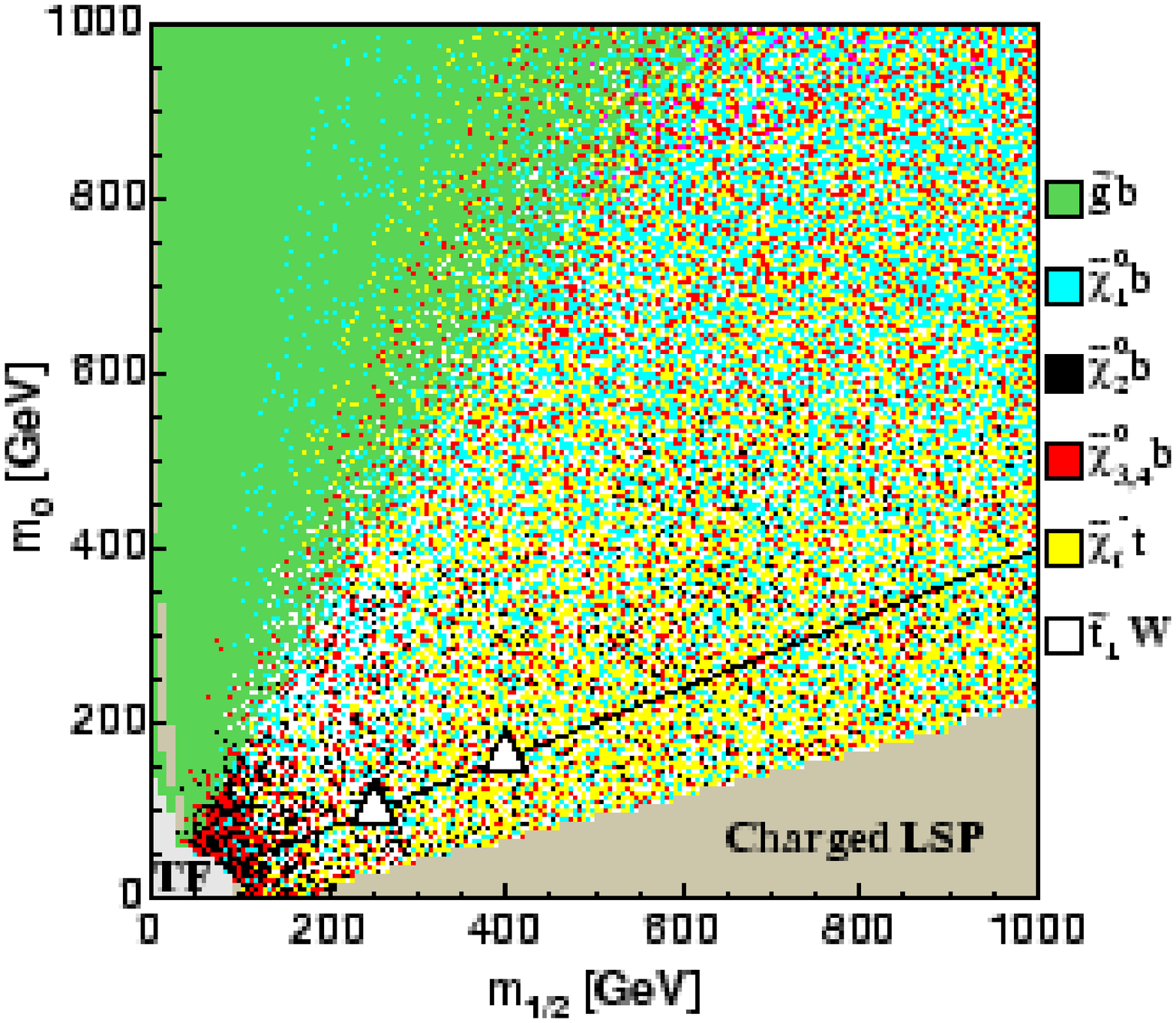}
            {\ifx\picnaturalsize N\epsfxsize \picsize\fi
\epsfbox{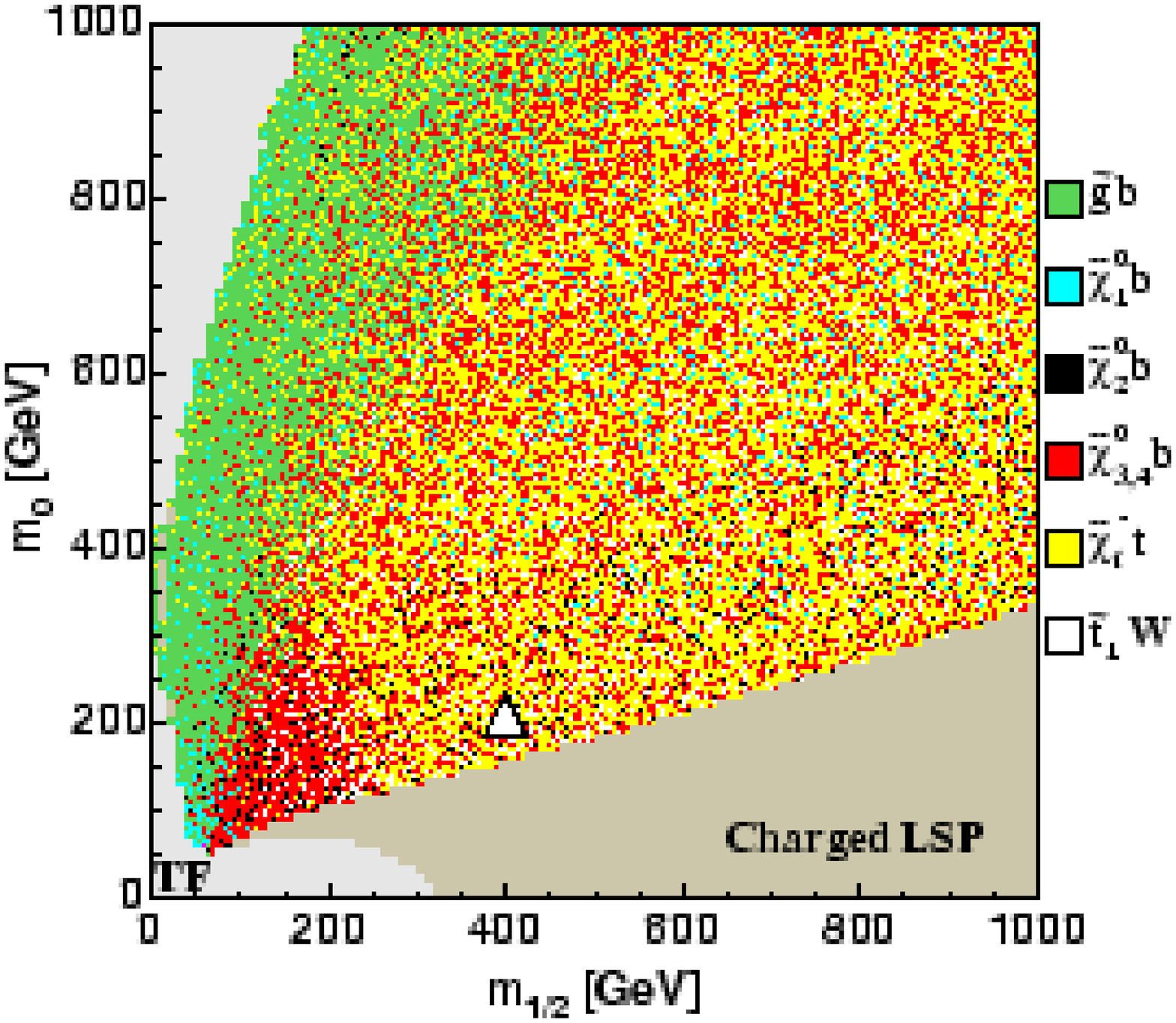}} } }
}\fi
%%End InstantTeX Picture
\vspace{-6mm}
\caption{Decay channels of $\bO$ (upper panels)
and $\tilde b_2$ (lower panels) with $\AZero=-\mZero$,
$\tan\beta=10$ (left) and $A_0=0$, $\tan\beta=30$ (right).
(See the caption of Fig.~\ref{Fig:m_half-m_0-BRglscans}.)
\label{Fig:m_half-m_0-BRb1t1scans}}}
%\end{figure}
%%%%%%%%%%%%%%%%%%%%%%%%%%%%%%%%%%%%%%%%%%%%%%%%%%%%%%%%%%%%%%%%%%%%%%

The third generation squarks differ from the others in two aspects. 
First, the mass eigenstates can have more even admixtures of both 
handedness components. 
For $\sb$ this is the case for low mass parameters, 
$\mZero,\mHalf\lesssim 200$~GeV, where the branching ratios into $\NO$ 
and $\NT$ are of comparable size. 
At higher masses $\bO\approx\bL$ and $\bT\approx\bR$, giving a $\bO$ 
which prefers to go to $\NT$ rather than $\NO$. 
Second, due to large splitting, the third generation squarks can decay 
into other third generation squarks together with a weak gauge boson. 
The drastic change observed in figure~\ref{Fig:m_half-m_0-BRb1t1scans} 
as $\mZero,\mHalf$ 
become less than $\sim200$~GeV, is due both to the more 
mixed mass eigenstates for lower masses, and to the closing 
of certain channels involving $t$ or a heavy gauge boson. 

While $\bO$ has a large branching ratio into $\NT$ throughout the 
entire plane, $\bT$ produces $\NT$ at a much smaller rate, except 
for small mass parameters.

To summarize, the squark decays that are `useful' for kinematic endpoint
analyses, are those of left-handed first and second-generation squarks, as
well as those of $\bO$ and to a lesser extent $\bT$.  These occur in the
entire $\mHalf$--$\mZero$ plane, except for extreme values $\mHalf\ll\mZero$,
and for both low and high $\tan\beta$ values.  For quite low mass parameters
also $\qR$ contributes.

\subsection{Neutralino and slepton decays: the lower part of the chain}

The reasons why $\NT$ often plays an important role 
in the reconstruction of SUSY events are many. 
Kinematically situated midway between the initially produced 
gluinos/squarks and the LSP, it is abundantly decayed into, 
as we have seen. 
What makes it so useful, usually more so than the $\CO$, which is produced in
similar ways and amounts, is the fact that its decay products, in addition to
easily setting off the trigger, also reconstruct well.

%%%%%%%%%%%%%%%%%%%%%%%%%%%%%%%%%%%%%%%%%%%%%%%%%%%%%%%%%%%%%%%%%%%%%%
\FIGURE[ht]{
%\begin{figure}
%%Begin InstantTeX Picture
\let\picnaturalsize=N
\def\picsize{7.5cm}
%If you do not have the picture file add:
%\let\nopictures=Y
%to the beginning of the file.
\ifx\nopictures Y\else{
\let\epsfloaded=Y
\centerline{\hspace{3mm}{\ifx\picnaturalsize N\epsfxsize \picsize\fi
\epsfbox{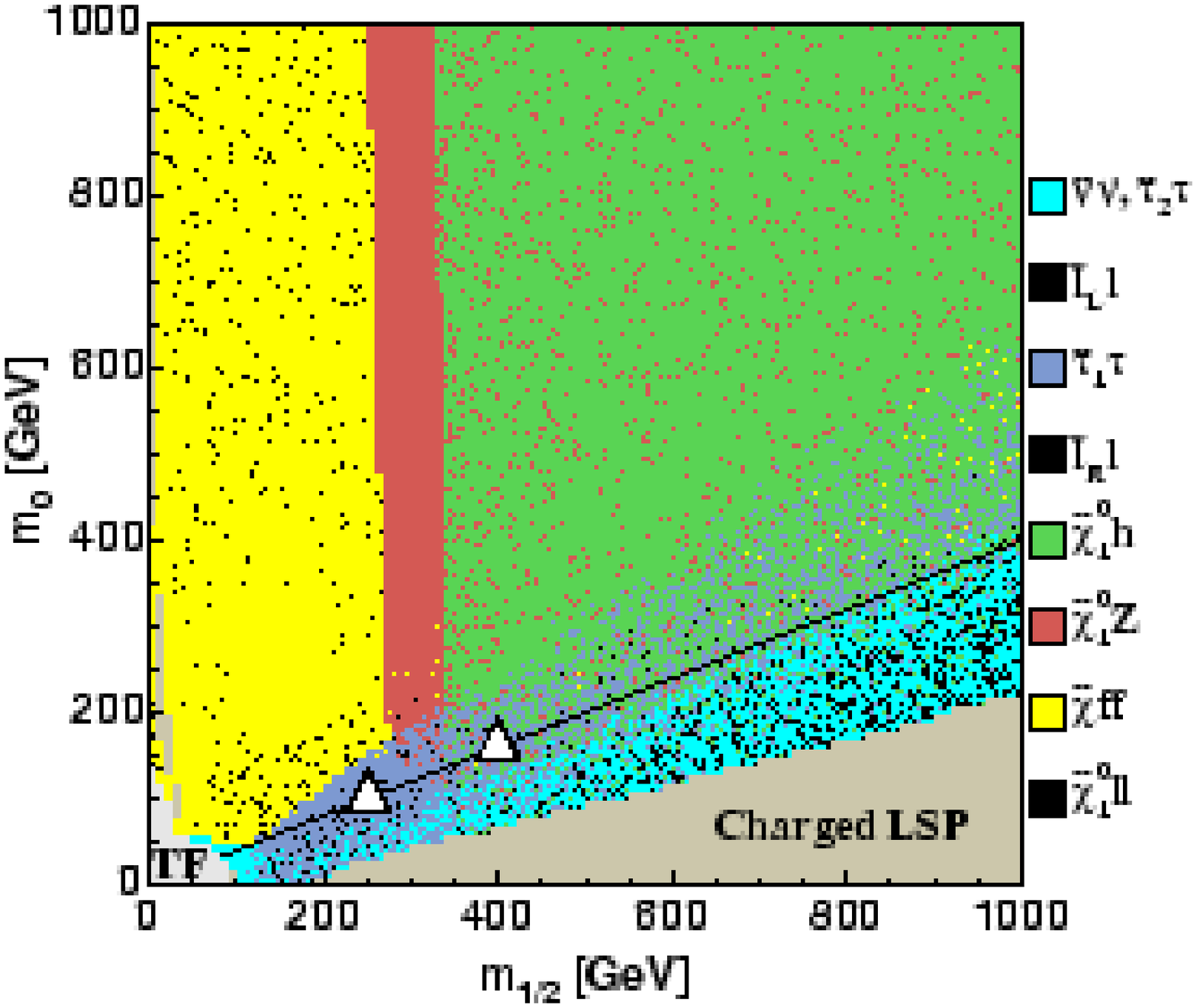}
            {\ifx\picnaturalsize N\epsfxsize \picsize\fi
\epsfbox{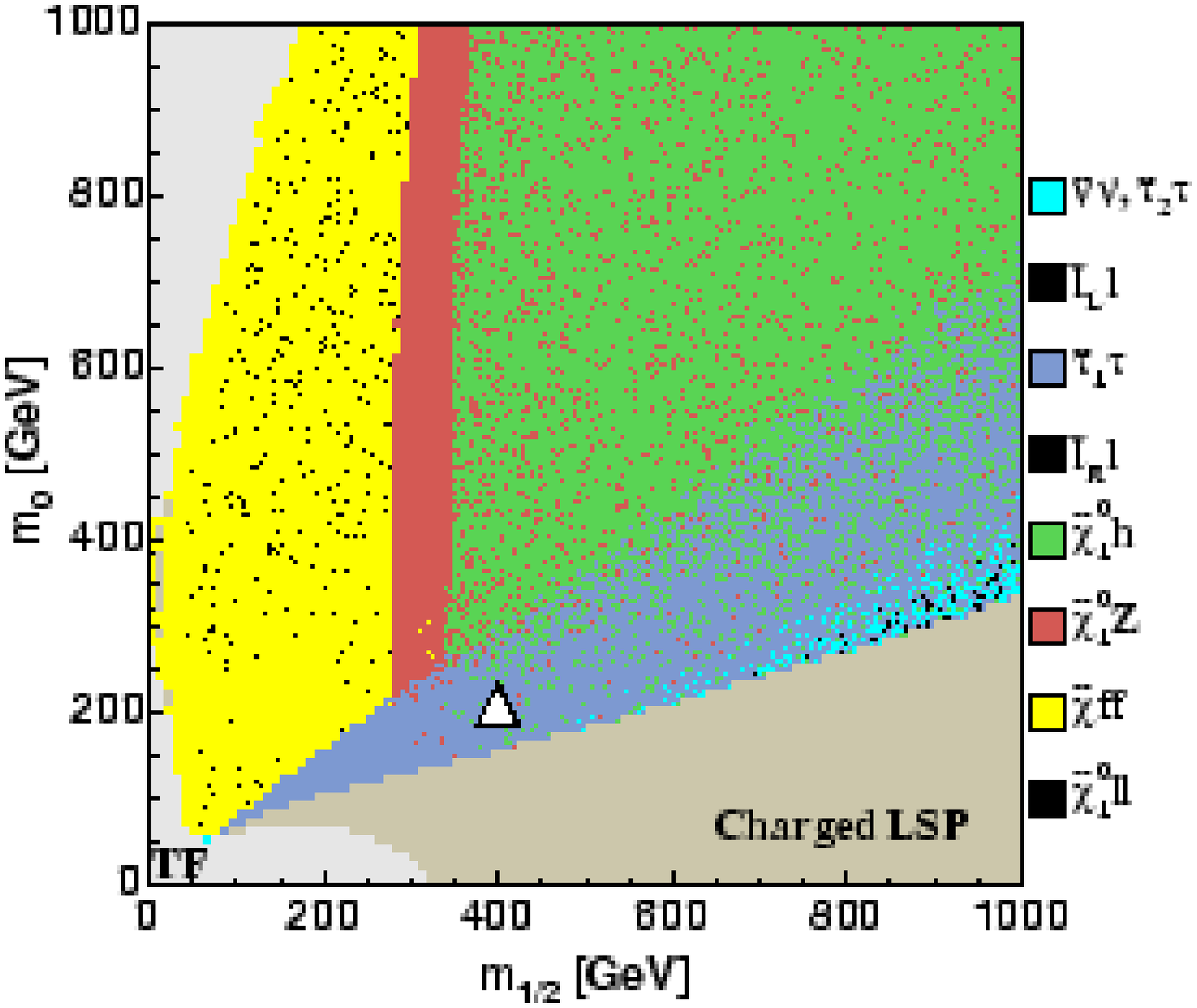}} } } 
}\fi
%%End InstantTeX Picture
\vspace{-6mm}
\caption{Decay channels of $\NT$ with $\AZero=-\mZero$, $\tan\beta=10$ (left) 
and $A_0=0$, $\tan\beta=30$ (right).
(See the caption of Fig.~\ref{Fig:m_half-m_0-BRglscans}.)
\label{Fig:m_half-m_0-BRN2scans}}}
%\end{figure}
%%%%%%%%%%%%%%%%%%%%%%%%%%%%%%%%%%%%%%%%%%%%%%%%%%%%%%%%%%%%%%%%%%%%%%

In Fig.~\ref{Fig:m_half-m_0-BRN2scans} the main decay 
channels of $\NT$ are shown for two values of $\tan\beta$. 
A two-body decay is preferred over a three-body decay, and 
the coupling to $\sle$, $\stau$ or $\snu$ is usually stronger 
than the coupling to $\NO$. 

For $\mZero\gtrsim0.5\mHalf$ all of $\sle$, $\stau$ and $\snu$ 
are heavier than $\NT$, so only the decay into the LSP is possible. 
In yellow, to the very left, no two-body channel is open, 
and $\NT$ undergoes a three-body decay, proceeding through 
an off-shell squark or stau/slepton, or involving an off-shell 
$Z$, $W$ or $h$. 
As $\mHalf$ increases, also $\mNT-\mNO$ increases and 
more decay channels become available. 
First the $Z$ channel opens and takes the full decay width, 
then the $h$ channel opens to dominate. 
The mass difference between $\NO$ and $\NT$ is mostly independent 
of $\mZero$, which is why, to a good approximation, 
the yellow, red and green regions are stacked horizontally. 

In the blue regions decays into $\sle/\stau/\snu$ are kinematically 
allowed. 
Following a clockwise movement, the scalar masses are reduced 
relative to $\NT$. The right-handed scalars are lighter and become 
available first (dark blue region). 
In the light blue region the left-handed scalars have become 
available and, despite less phase space, take most of the width 
due to their SU(2) coupling. 
The black part of the blue regions shows the decay into $\lR$ 
(dark blue region) and $\lL$ (light blue region). 
These are the decays of interest to us. 
At low $\tan\beta$ (left panel) the slepton channels can be used 
in most of the blue regions. 
At high $\tan\beta$ the situation is less optimistic. The $\tauO$ 
channel totally dominates the $\lR$ channel, and only in a small 
region is the $\lL$ channel open. 

Decay products which involve tau particles are more difficult to use
since their reconstruction is always incomplete due to undetected
neutrinos. However, in some parts of the parameter space, especially
at high $\tan\beta$, these channels take the full decay width, so one
must be prepared to use them. In any case, whatever the MSSM
parameters should turn out to be, in order to measure the stau mass itself,
tau particles must be reconstructed.

In the non-blue regions of the plane, information on the sparticle 
masses can still be retrieved, although to a lesser extent. 
From an experimental point of view final states involving two 
leptons are preferable. In the yellow region this fraction is 
marked in black. In the red region $Z$ decays leptonically 
in 7\% of the cases. 
In the green region, no leptonic decay is available. 
Here the $b\bar b$ final state of the Higgs can be used. 
This channel may even serve as a discovery channel for the Higgs boson.

Returning once more to our chosen decay chain, $\sq \to \NT q \to \sle
lq \to \NO llq$, it is clear from
Figs.~\ref{Fig:m_half-m_0-BRglscans}--\ref{Fig:m_half-m_0-BRb1t1scans}
that the initiating squark must be $\qL$ or $\sb$. 
Furthermore Fig.~\ref{Fig:m_half-m_0-BRN2scans} shows that for low 
$\tan\beta$ the sleptonic decay of $\NT$ is open in a large fraction 
of the $\mHalf$--$\mZero$ plane.

\subsection{Other constraints}

Much of these parameter planes considered in
Figs.~\ref{Fig:m_half-m_0-scans}--\ref{Fig:m_half-m_0-BRN2scans} are
actually excluded or disfavoured by observations.  The left part
(small $\mHalf$) is typically excluded by the lower bound on the Higgs
mass, and the region of `large' $\mZero+\mHalf$ is excluded by the
WMAP data \cite{Bennett:2003bz,Spergel:2003cb}, since a too high
contribution to the Cold Dark Matter (relic LSP) density is produced. Such
bounds have been explored in considerable detail for the so-called
`Post-LEP' SUSY benchmark points \cite{Ellis:2002rp,Ellis:2003cw}.
For the SPS points, see Ref.~\cite{Raklev}.
Also, there are constraints from the non-emergence of unphysical
vacua during the renormalisation-group running from the high scale 
\cite{Casas:1995pd}.
These typically rule out a sector at low $\mZero$ and high $\mHalf$
that may extend beyond that excluded by charged LSP.
We shall here ignore such additional constraints, since they
depend somewhat on the assumptions which are adopted.
In particular, tunneling into them may take longer than 
the age of the Universe.

%%%%%%%%%%%%%%%%%%%%%%%%%%%%%%%%%%%%%%%%%%%%%%%%%%%%%%%%%%%%%%%%%%%%%
\section{Summary of SPS~1a} \label{sect:sps1a}
%%%%%%%%%%%%%%%%%%%%%%%%%%%%%%%%%%%%%%%%%%%%%%%%%%%%%%%%%%%%%%%%%%%%%

We have shown that the squark and gluino initiated cascade decays seen
at the benchmark point and slope SPS~1a are not atypical of a large
portion of the parameter space. In this section we will go on to
explore SPS~1a in more detail, and introduce our second SPS~1a point
on the line.

\subsection{The SPS~1a line and points}

The SPS~1a benchmark {\it line} is defined as the masses and couplings
of supersymmetric particles at the TeV scale as evolved from the GUT
scale mSUGRA inputs
\begin{eqnarray}  
m_0=-A_0=0.4\, m_{1/2}, \nonumber \\
\tan\beta=10, \qquad \mu>0,
\label{eq:slopedefinition}
\end{eqnarray}  
by version 7.58 of the program {\tt ISAJET}~\cite{Baer:1993ae}. Elsewhere 
in this report when the mSUGRA GUT scale parameters are referred to it
is to be understood that the low energy parameters are obtained in
this way.

In addition, we define two points ($\alpha$) and ($\beta$) on the
SPS~1a line according to:
\begin{alignat}{3}  \label{eq:sps1a-al-be}
&(\alpha):\quad
m_0&=100~\text{GeV}, \qquad
m_{1/2}&=250~\text{GeV}, \nonumber \\
&(\beta):\quad
m_0&=160~\text{GeV}, \qquad
m_{1/2}&=400~\text{GeV}.
\end{alignat}
The first point ($\alpha$) is the `basic' SPS~1a point of
Ref.~\cite{Allanach:2002nj} and studied in Ref.~\cite{lhc-lc,gjelsten-atlas},
while the second ($\beta$) is a new, less optimistic scenario with a reduced
cross-section for the decay chain.

The masses of particles relevant for our analysis are shown in
Fig.~\ref{fig:masses-line} (left),
\FIGURE[ht]{
\centerline{
\epsfxsize 7.5cm
\epsfbox{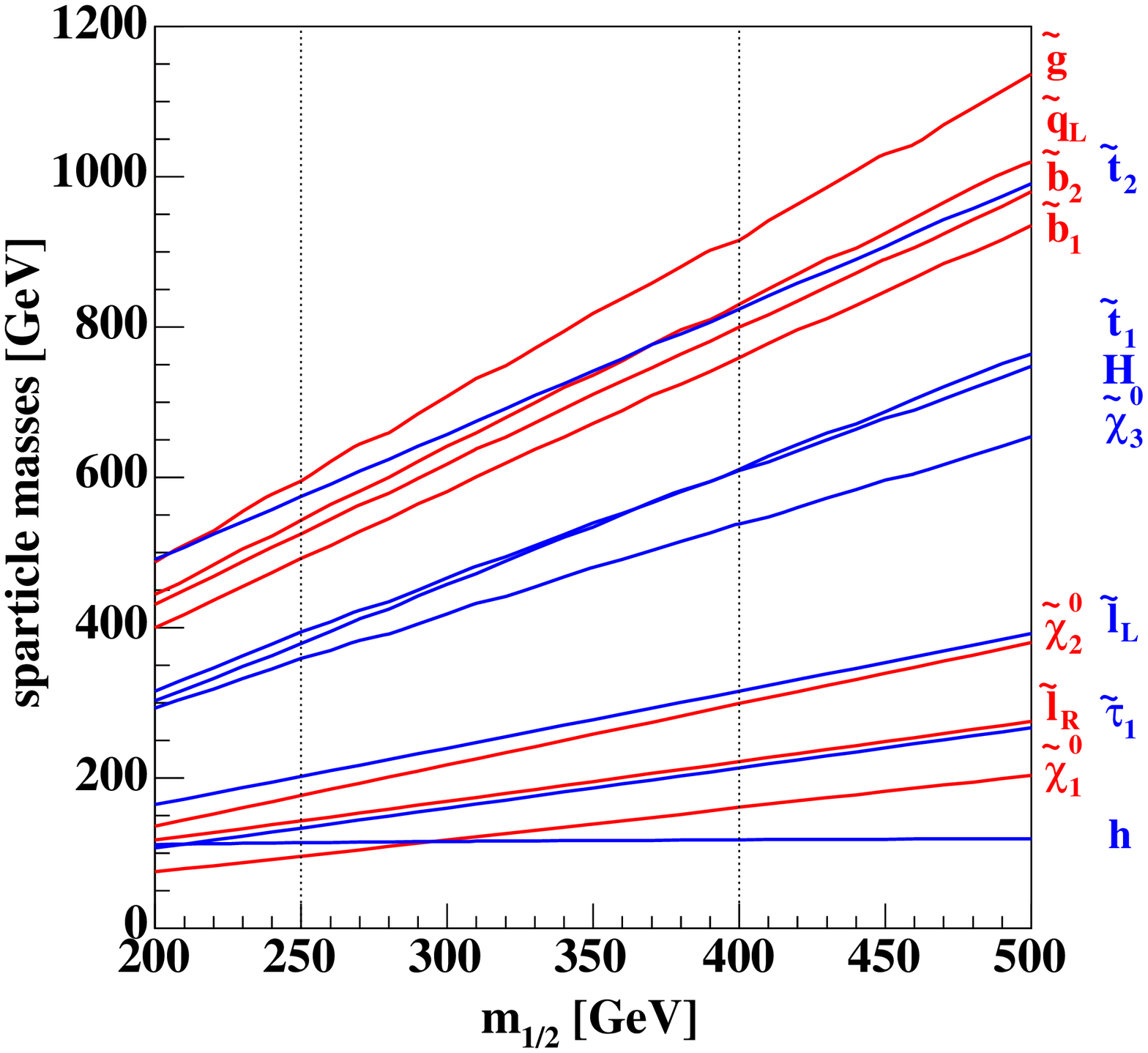}
\epsfxsize 7.5cm
\epsfbox{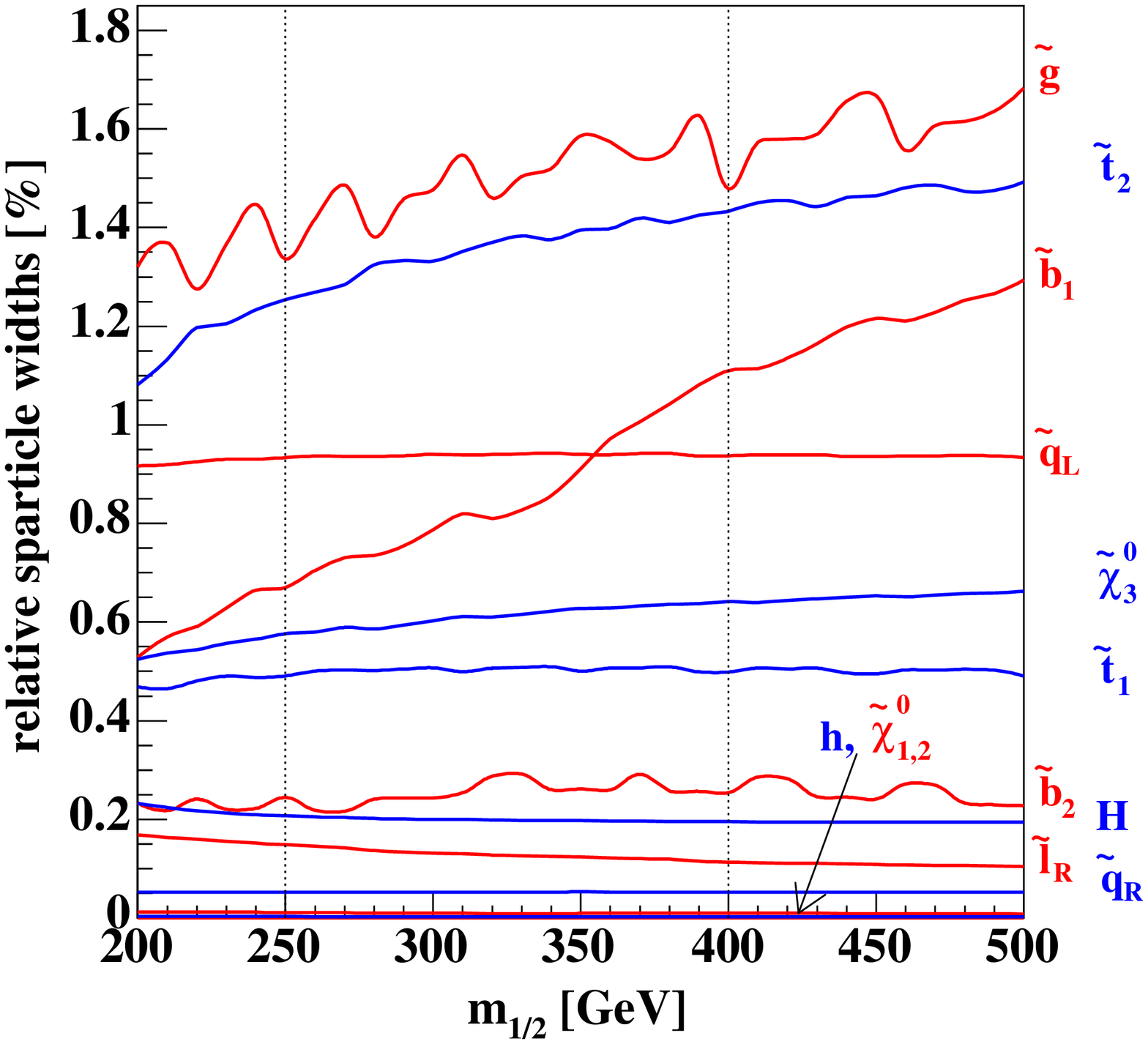} }
\vspace{-6mm}
\caption{Masses (left) and relative widths (right) of relevant
sparticles as $\mHalf$, $\mZero$ and $\AZero$ are varied along
the SPS~1a slope, defined by
Eq.~(\ref{eq:slopedefinition}). The vertical dotted lines
represent SPS~1a points ($\alpha$) and ($\beta$).
\label{fig:masses-line}}}
moving along the SPS~1a line by varying $\mHalf$. The values of the
masses at points ($\alpha$) and ($\beta$) can be seen from the
vertical dotted lines.  As expected, all the masses except the
lightest Higgs boson mass increase linearly with $\mHalf$. 
Neither the heaviest neutralino mass nor the chargino masses are shown;
to a good approximation, 
$m_{\tilde\chi_2^\pm} \simeq m_{\tilde\chi_3^0} \simeq 
m_{\tilde\chi_4^0}$ and $m_{\tilde\chi_1^\pm} \simeq m_{\tilde\chi_2^0}$.
Similarly, the masses of $H^\pm$ and $A$ are not shown, but 
$m_{H^\pm} \simeq m_A \simeq m_H$. 

For the points ($\alpha$) and ($\beta$), these masses are further
detailed in Table~\ref{table:spectra}, with the masses of the
particles in our chosen decay chain displayed in bold.
\begin{TABLE}{
\label{table:spectra}
\begin{tabular}{|l|c|c|c|c|c|c|c|c|c|}
\hline
Point &$\tilde g$ &$\tilde d_L$ &$\tilde d_R$ &$\tilde u_L$&$\tilde u_R$
&$\tilde b_2$&$\tilde b_1$&$\tilde t_2$&$\tilde t_1$\\
\hline
($\alpha$) & \bf 595.2& \bf 543.0 &520.1& \bf 537.2 & 520.5&\bf 524.6
&\bf 491.9& 574.6& 379.1 \\
($\beta$)  & \bf 915.5& \bf 830.1 &799.5& \bf 826.3 & 797.3&\bf 800.2 
&\bf 759.4& 823.8& 610.4\\
\hline\hline
 &$\tilde e_L$ &$\tilde e_R$ &$\tilde \tau_2$&$\tilde \tau_1$ 
&$\tilde \nu_{e_L}$
&$\tilde \nu_{\tau_L}$&  & $H^\pm$&$A$\\
\hline
($\alpha$) & 202.1& \bf 143.0 &206.0& 133.4 & 185.1& 185.1 & &401.8 &393.6\\
($\beta$)  & 315.6& \bf 221.9 &317.3& 213.4 & 304.1& 304.1 & &613.9 &608.3\\
\hline\hline
 &$\tilde \chi_4^0$ &$\tilde \chi_3^0$ &$\tilde \chi_2^0$
&$\tilde\chi_1^0$ &$\tilde \chi_2^\pm$
&$\tilde \chi_1^\pm$&  &$H$& $h$ \\
\hline
($\alpha$)& 377.8& 358.8 & \bf 176.8 &\bf  96.1& 378.2& 176.4 && 394.2 & 114.0\\
($\beta$) & 553.3& 538.4 & \bf 299.1 &\bf 161.0& 553.3& 299.0 && 608.9 & 117.9\\
\hline
\end{tabular}
\caption{Masses [GeV] for the considered SPS~1a points ($\alpha$) and
($\beta$) of Eq.~(\ref{eq:sps1a-al-be}).}  }
\end{TABLE}

The relative widths (width divided by the mass) of the decaying
sparticles are shown in Fig.~\ref{fig:masses-line} (right), and are
everywhere less than 2\% of the mass.  The wiggles in some of these
curves (as well as in some of the branching ratio curves below) are
due to limited precision in {\tt ISAJET}~\cite{Allanach:2001hm}. 
As will be discussed, these widths
contribute to a blurring of the kinematical endpoints, and will thus be
reflected in the mass determination.

\subsection{Sparticle production}

The cross-sections for producing supersymmetric particles at the LHC 
are for moderate values of $\mHalf$ rather high. This can be seen in
Fig.~\ref{fig:XSECAlongSlope} which shows 
the dominating sparticle pair production cross-sections, 
as $\mHalf$ is
varied along the SPS~1a line. Notice that these cross-sections fall
very rapidly as $\mHalf$ is increased, which will cause repercussions
in the analysis of SPS~1a ($\beta$).
\FIGURE[ht]{
\epsfxsize 9cm
\epsfbox{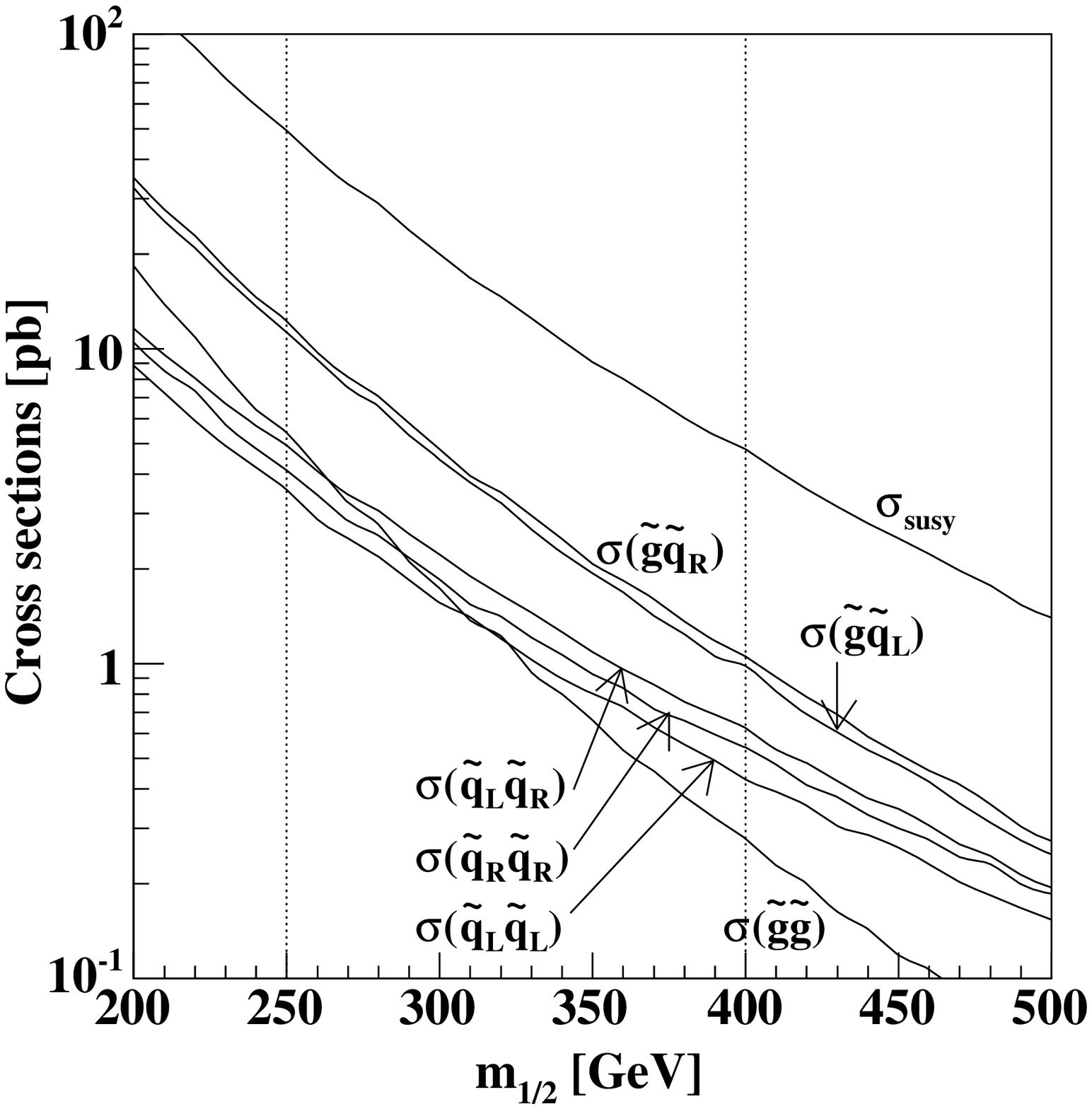}
\vspace{-2mm}
\caption{Cross-sections as $\mHalf$, $\mZero$ and $\AZero$ are varied
         along the SPS~1a slope, defined by
         Eq.~(\ref{eq:slopedefinition}). The vertical dotted lines
         represent SPS~1a points ($\alpha$) and ($\beta$).
\label{fig:XSECAlongSlope}}}

The cross-sections for gluino--gluino, gluino--squark and
squark--squark pair productions are detailed in
Table~\ref{table:sigma} for the two chosen analysis points, together
with the SUSY total rate.  Of course since other supersymmetric
particle pairs may contribute to the total SUSY rate it is {\it not}
simply a sum of the other numbers in the table. 

These supersymmetric particle pairs are predominantly produced by QCD
interactions of quarks and gluons in the colliding protons. For gluino
pairs this is mainly due to $gg \to \tilde g \tilde g$ via t-channel
gluino exchange and s-channel gluons, and at a much smaller rate $q
\bar q \to \tilde g \tilde g$ via s-channel gluons.  Squark pairs with
the same handedness have the dominant production process $qq \to
\tilde q_{L/R} \, \tilde q_{L/R}$ with a t-channel gluino exchange,
but can also be produced via $gg \to \tilde q_{L/R} \, \tilde q_{L/R}$
with an s-channel gluon, or via $q \bar q \to \tilde q_{L/R} \,
\tilde q_{L/R}$ with both t-channel gluino and s-channel gluons. For
opposite handedness, $\tilde q_L \tilde q_R$ production, the s-channel
gluon exchange processes [and thus $gg \to \tilde q_L \, \tilde q_R$
at tree-level] are disallowed. Lastly, the $\tilde g \tilde q$ final
states are produced by the process $gq \to \tilde g \tilde q$ mediated
by t-channel squark or gluino exchanges and s-channel quarks.
\begin{TABLE}{
\label{table:sigma}
\begin{tabular}{|l|r|r|r|r|r|r|r|}
\hline
&$\sigma$(SUSY) &$\sigma(\tilde g\tilde g)$ &$\sigma(\tilde g\tilde q_L)$
&$\sigma(\tilde g\tilde q_R)$
&$\sigma(\tilde q_L\tilde q_L)$&$\sigma(\tilde q_L\tilde q_R)$
&$\sigma(\tilde q_R\tilde q_R)$ \\
\hline
($\alpha$) & 49.3  & 5.3  & 11.4  & 12.3  & 3.5  & 4.8  & 4.1  \\
($\beta$)  &  4.76 & 0.29 &  0.97 &  1.06 & 0.44 & 0.61 & 0.53 \\
\hline
\end{tabular}
\caption{Selected supersymmetry cross-sections in pb.}  }
\end{TABLE}

However, these particle pair cross-sections are not the production
cross-sections that are relevant to our analysis of the decay chain,
since it does not matter from where the parent squark originates. Therefore
we should also be counting, for example, gluinos which decay into
squarks as possible sources of the decay chain. In
Table~\ref{table:sparticlerate} we divide the sparticle productions
rates into `direct' and `indirect' contributions, reflecting
production rates from sparticles in the `initial' supersymmetric state
as opposed to their generation from the decay of a parent
sparticle. Furthermore, if the `initial' supersymmetric state contains
two possible parents then the chance of generating the desired decay
chain is doubled\footnote{The total branching ratio for the decay is
sufficiently small that the chance of generating {\it two} of the
desired decay chains is tiny.}.  Therefore the pair-production rates
with two possible parent particles are counted twice. Again, one
cannot simply add the various contributions from
Table~\ref{table:sigma} to obtain the `direct' rates of
Table~\ref{table:sparticlerate}, since they include contributions from
squarks or gluinos produced in association with other supersymmetric
particles. \s
\begin{TABLE}{
\label{table:sparticlerate}
\begin{tabular}{|ll|r|r|r|r|r|r|r|}
\hline
& &$\Sigma(\gl)$ &$\Sigma(\qL)$ &$\Sigma(\qR)$
&$\Sigma(\bO)$&$\Sigma(\bT)$&$\Sigma(\tO)$
&$\Sigma(\NT)$\\
\hline
($\alpha$) & Direct    & 35.4  & 24.6  & 25.8  & 1.4  & 0.9  & 3.4  &  1.8  \\
           & Indirect  &    -  &  8.2  & 14.6  & 6.3  & 3.5  & 5.6  & 16.0  \\
           & Total     & 35.4  & 32.8  & 40.4  & 7.7  & 4.3  & 9.0  & 17.8  \\
\hline
($\beta$)  & Direct    &  2.71 &  2.64 &  2.80 & 0.10 & 0.06 & 0.23 &  0.23 \\
           & Indirect  &    -  &  0.58 &  1.00 & 0.40 & 0.25 & 0.64 &  1.44 \\
           & Total     &  2.71 &  3.21 &  3.79 & 0.50 & 0.31 & 0.87 &  1.67 \\
\hline
\end{tabular}
\caption{Selected sparticle production rates in pb.}
}
\end{TABLE}

\FIGURE[ht]{
%\begin{figure}
%%Begin InstantTeX Picture
\let\picnaturalsize=N
\def\picsize{8.5cm}
%If you do not have the picture file add:
%\let\nopictures=Y
%to the beginning of the file.
\ifx\nopictures Y\else{
\let\epsfloaded=Y
            {\ifx\picnaturalsize N\epsfxsize \picsize\fi
\epsfbox{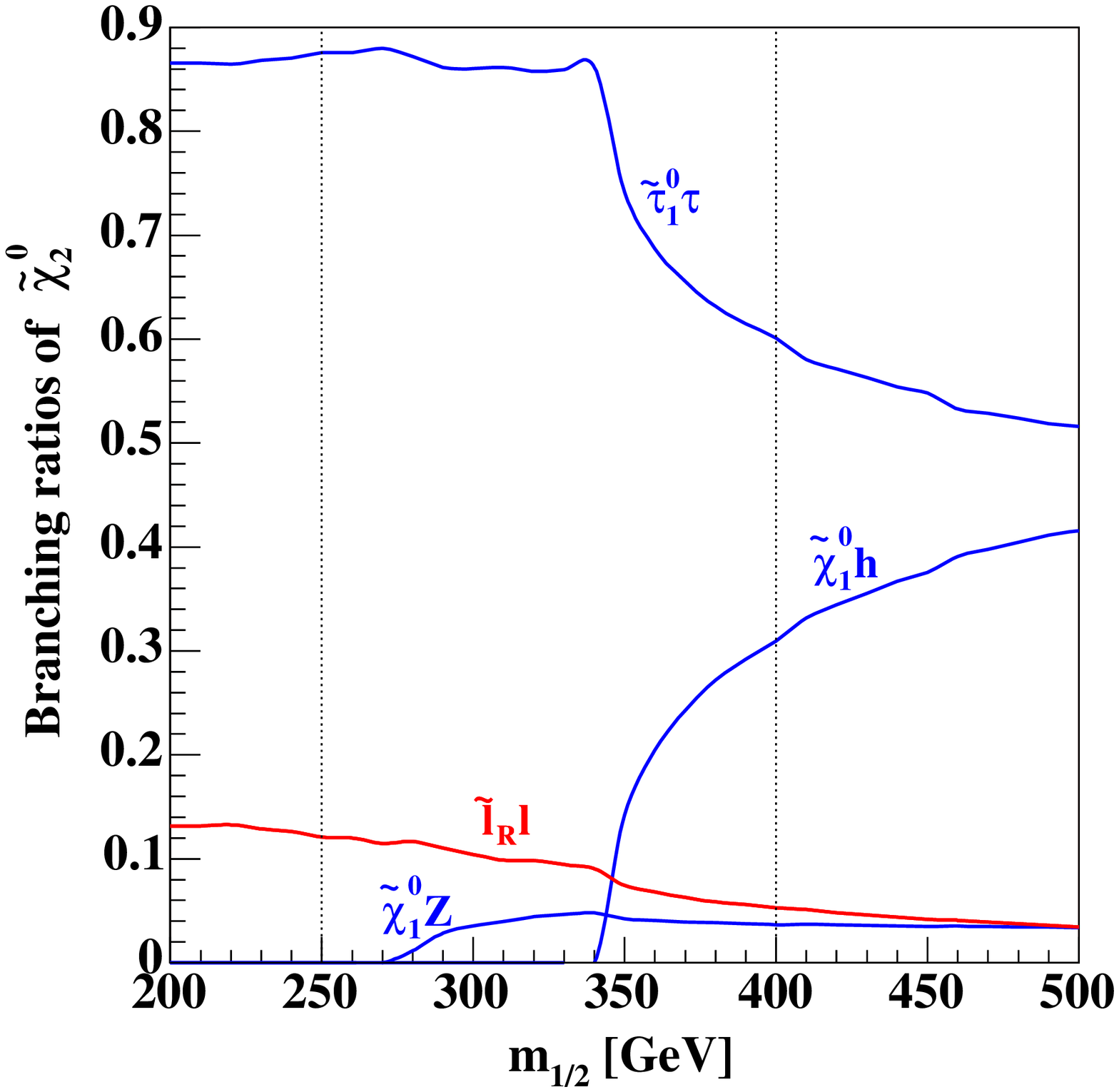}} 
%} }
}\fi
%%End InstantTeX Picture
\caption{Branching ratios of 
         $\tilde\chi_2^0$ as $\mHalf$, $\mZero$ and $\AZero$
         are varied along the SPS~1a slope. The vertical dotted lines
         represent SPS~1a points ($\alpha$) and ($\beta$).
\label{Fig:BRN2AlongSlope}}}
%\end{figure}

For low $\mHalf$ values $\NT$ will decay dominantly to a stau 
and a tau, see Fig.~\ref{Fig:BRN2AlongSlope}. 
Here, the decay mode of interest is to a right-handed slepton, 
$\tilde l_R$, and a lepton, and is at the level of 10\%.  
For higher values of $m_{1/2}$, the mass difference between $\NO$ and 
$\NT$ grows sufficiently to allow the decay $\NT\to \NO h$. 
At \pbeta, even though the Higgs channel takes a significant 30\% of the 
decay width, it was not used in this study. 
Due to the accuracy of lepton reconstruction compared to jet reconstruction, 
only in the case of very low lepton channel statistics can   
the Higgs channel improve on the results obtained with the slepton channel.

\subsection{The cascade}

%%%%%%%%%%%%%%%%%%%%%%%%%%%%%%%%%%%%%%%%%%%%%%%%%%%%%%%%%%%%%%%%%%%%%%
\FIGURE[ht]{
%\begin{figure}
%%Begin InstantTeX Picture
\let\picnaturalsize=N
\def\picsize{15.0cm}
\def\picfilename{squarkchain.eps}
%If you do not have the picture file add:
%\let\nopictures=Y
%to the beginning of the file.
\ifx\nopictures Y\else{
\let\epsfloaded=Y
\centerline{\ifx\picnaturalsize N\epsfxsize \picsize\fi
\epsfbox{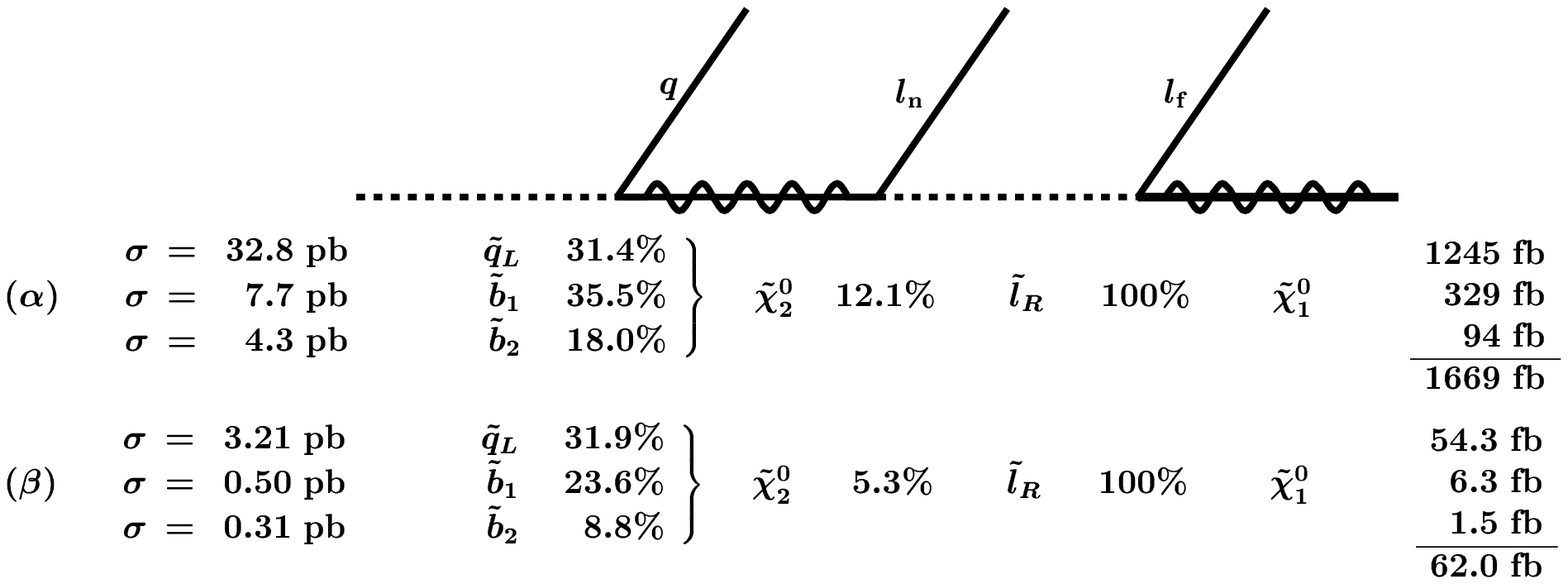}}}\fi
%%End InstantTeX Picture
\caption{The SPS~1a cascade decay chain.
\label{fig:squarkchain}}}
%\end{figure}
%%%%%%%%%%%%%%%%%%%%%%%%%%%%%%%%%%%%%%%%%%%%%%%%%%%%%%%%%%%%%%%%%%%%%%

The cross-sections and branching ratios of our chosen decay chains
\begin{eqnarray}
\tilde q_L &\to& \NT q \to \lR^\mp \lNN^{\pm} q \to \NO \lFF^{\mp} \lNN^{\pm} q 
\label{eq:squarkchainq} \\
\tilde b_1 &\to& \NT b \to \lR^\mp \lNN^{\pm} b \to \NO \lFF^{\mp} \lNN^{\pm} b 
\label{eq:squarkchainb1} \\ 
\tilde b_2 &\to& \NT b \to \lR^\mp \lNN^{\pm} b \to \NO \lFF^{\mp} \lNN^{\pm} b 
\label{eq:squarkchainb2} 
\end{eqnarray}
are summarized in Fig.~\ref{fig:squarkchain} for the two SPS~1a
points.  Since the left-handed up and down squarks, $\tilde u_L$ and
$\tilde d_L$, have very similar masses (at $537.2~\text{GeV}$ and
$543.0~\text{GeV}$ respectively), for \palpha, they are in the above jointly
referred to as $\tilde q_L$, and for this analysis will be grouped
together.  For the fraction of decay chains which commence
from a sbottom, $\tilde{b}_1$
is responsible for 78\% or so, leaving us rather insensitive to the
contribution from $\tilde{b}_2$.

%%%%%%%%%%%%%%%%%%%%%%%%%%%%%%%%%%%%%%%%%%%%%%%%%%%%%%%%%%%%%%%%%%%%%%
\section{Mass distributions} \label{sect:edge-meas}
%%%%%%%%%%%%%%%%%%%%%%%%%%%%%%%%%%%%%%%%%%%%%%%%%%%%%%%%%%%%%%%%%%%%%%

The longer a decay chain is, the more information it contains. To extract the
masses of the supersymmetric particles in the decay we require at least as
many kinematic endpoint measurements as unknown masses.  
In the lower part of the decay
chain, where the second-lightest neutralino decays via $\NT\to\lR l\to\NO ll$,
there are three unknown masses: $\mNT$, $\mlR$ and $\mNO$. 
However, only two particle momenta are
measured, those of the two leptons, from which only one mass distribution can
be constructed, $\mll$.  The system is highly underdetermined; one cannot
extract the three masses, only a relation between them.

When a squark is added to the head of the decay chain, $\tilde q \to 
\NT q \to \lR\lNN q \to \NO\lFF\lNN q$, three particles can be collected,
and one can construct four mass distributions, $\mll$, $m_{q\lNN}$,
$m_{q\lFF}$ and $\mqll$, where following the notation of
Refs.~\cite{Allanach:2000kt,Lester}, we denote the first emitted lepton 
$\lNN$ (`n' for `near')  and the second $\lFF$ (`f' for `far').  
In principle this is just sufficient for
extracting the four unknown masses:
$\msq$, $\mNT$, $\mlR$ and $\mNO$.  However, in order to use the
distributions $m_{q\lNN}$ and $m_{q\lFF}$, we need to be able to
distinguish $\lNN$ from $\lFF$. Since this is usually not possible, two
alternative distributions are defined, $\mqlHigh$ and $\mqlLow$
\cite{Allanach:2000kt}, constructed by selecting for each event the
largest and smallest values of $\mql$ respectively. 

As will be detailed later in this section, the
expressions for these kinematic endpoints are not always linearly
independent, so these four endpoints are not always sufficient to
determine the masses in the decay chain.  In this circumstance one
must look for other endpoint measurements.

Correlations between different mass distributions can provide further
measurements.  For example, one may define the mass distribution
$\mqllThres$ identically to the $\mqll$ distribution but with the
additional constraint
\begin{equation}
\label{eq:thres-constraint}
\maxmll/\sqrt{2}<\mll<\maxmll.
\end{equation}
This cut on $\mll$ translates directly into a cut on the angle
$\theta$ between the two leptons in the rest frame of $\lR$
\cite{Nojiri:2000wq}. In terms of this angle, $\mll$ is given by
\begin{equation}
\label{eq:mll(theta)}
\mll = \maxmll\sqrt{(1-\cos\theta)/2}
\end{equation}
so a constraint of the form (\ref{eq:thres-constraint}) directly
corresponds to $\theta > \frac{\pi}{2}$. 
The simplicity of this constraint
allows one to find an analytic expression for the minimum of the
$\mqllThres$ distribution. 

In principle, other correlations between mass distributions could be
used, but they are limited by the lack of analytic expressions for the
associated extrema. It is no doubt possible to construct simple
constraints for which analytic expressions for minima or maxima of
mass distributions are possible, but this will not be investigated
further in this study. 

If we were to also include a parent gluino at the head of the decay
chain, $\gl \to \sq q \to \NT qq \to \lR lqq \to \NO llqq$, 
we would have an extra quark momentum at our disposal and could
construct with it seven more (`primary') mass distributions.

\subsection{Theory curves of invariant mass distributions
\label{subsect:theorycurves}}

\FIGURE[ht]{
%\begin{figure}
%%Begin InstantTeX Picture
\let\picnaturalsize=N
\def\picsize{15.0cm}
\def\picfilename{TheoryCurves.eps}
%If you do not have the picture file add:
%\let\nopictures=Y
%to the beginning of the file.
\ifx\nopictures Y\else{
\let\epsfloaded=Y
\centerline{\ifx\picnaturalsize N\epsfxsize \picsize\fi
\epsfbox{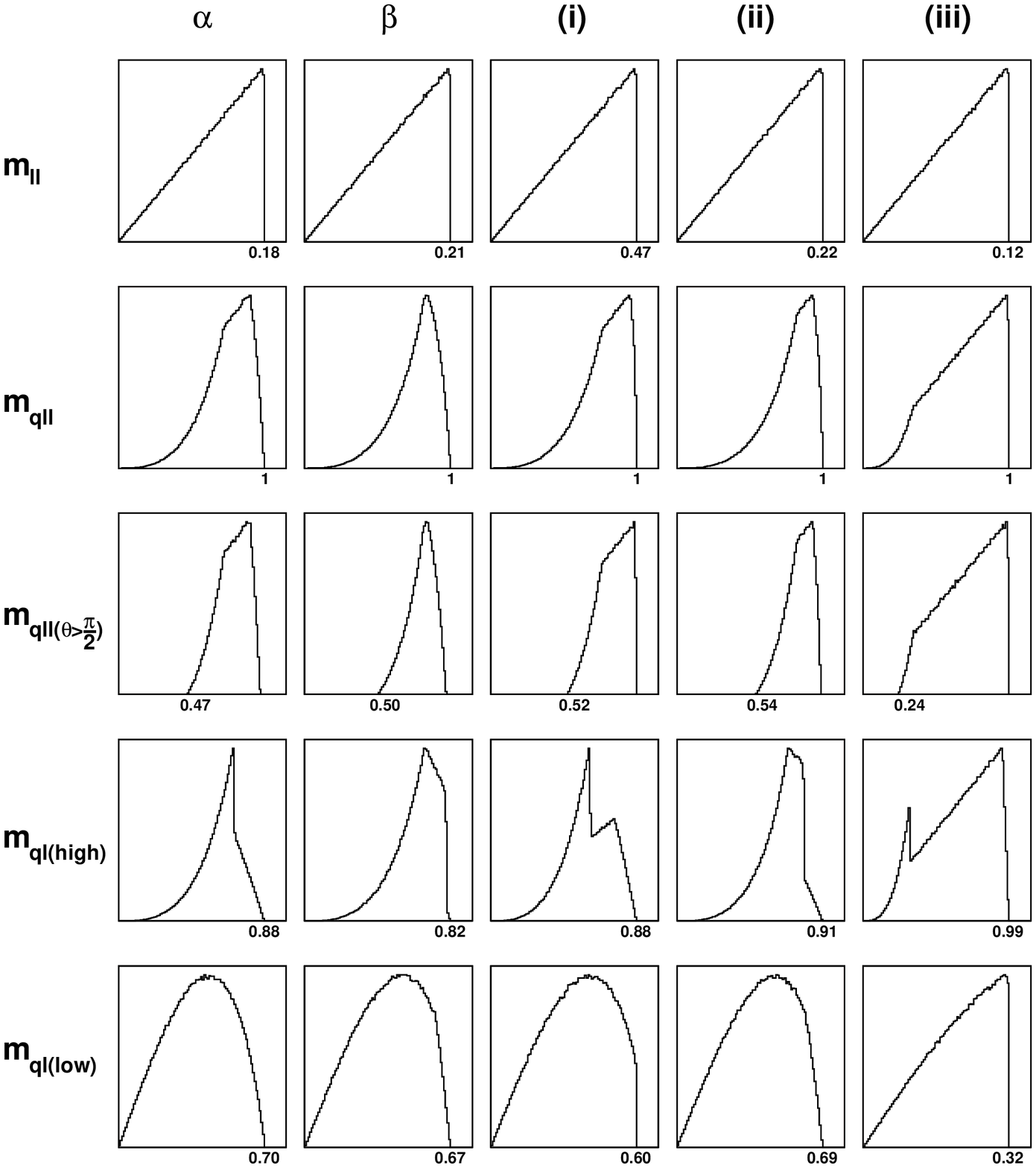}}}\fi
%%End InstantTeX Picture
\caption{Theoretical mass distributions for SPS~1a \palpha\ 
and \pbeta, as well as for three other mass scenarios, 
denoted (i), (ii) and (iii). 
Kinematic endpoints are given in units of $m^{\rm max}_{qll}$.
(More details will be given in \cite{gjelsten-thesis}.)
\label{Fig:th-curves}}}
% Masses used:
%        mN1,mlR,mN2,mqL
% alpha:  96,143,177,543
% beta : 161,222,299,830
% (i)  : 100,158,186,260
% (ii) : 100,188,279,906
% (iii): 100,225,235,605
%\end{figure}

In Fig.~\ref{Fig:th-curves} we show `theory' versions of the five mass
distributions discussed above for SPS~1a \palpha\ and \pbeta, and three other
mass scenarios.  These distributions reflect the parton level only, where the
quark and leptons are assumed to be perfectly reconstructed, and particle
widths have been neglected, suppressing a mild smearing of the distributions.
Leptons and quarks are assumed massless; at worst, i.e.\ for an endpoint
involving a 4.8~GeV $b$-quark, this approximation gives a value for $m^{\rm
max}_{qll}$ which is wrong by 5.7~MeV at \SPSOaa, which is negligible.
Furthermore a common squark mass is used; in reality, the experimental
distributions would be a sum of several similar distributions shifted
typically by a few percent, depending on the differences between the squark
masses.

These theory curves have been generated from the phase space only,
with no matrix elements inserted. For the two-body decay of scalar
particles this is not a problem; the daughter particles are always
emitted back-to-back in the rest frame of the parent and since a
scalar particle provides no intrinsic direction the decay will be
isotropic. The $m_{ll}$ distribution, for example, is very well
described by phase-space considerations alone since the scalar $\tilde
l_R$ removes any spin-correlations between the two leptons.  For the
decays of fermions this is not so clear; the parent's spin picks out a
particular direction so the decay need not be isotropic. One would
expect that spin-correlations must be fully taken into account by
calculating the full matrix elements for the decay chain.  In
practice, while these spin-correlations are indeed very significant in
individual processes, when one sums up final states containing 
leptons with positive and negative charge (which are produced in
equal proportion) these spin correlations cancel 
out~\cite{Richardson:2001df}. We have confirmed this assertion
analytically.

Most of the distributions of Fig.~\ref{Fig:th-curves} show a strong
dependence on the scenario. Only the $\mll$ distribution is
independent of the masses involved, which is easily understood from
(\ref{eq:mll(theta)}) together with the isotropic decay of $\lR$.  The
shape is easy to fit, and since only leptons are involved, the
experimental resolution is high.  The presence of the quark reduces
the precision with which the other distributions may be measured.  For
$\mqll$ there is a clear dependence on the masses involved, the main
feature being the length of the straight section, ranging from
infinitesimal to the entire distribution.  However, the maximum edge
itself is quite well described by a straight line, at least for the
latter part.  Next, the constrained $\mqll$ distribution differs from
$\mqll$ for smaller invariant masses.  Its rise from threshold is not
very well described by a straight line making the measurement of the
minimum value rather imprecise.  Both $m_{ql}$ distributions are
`composite' distributions, based on entries from $\mqlN$ and
$\mqlF$.  This double nature is readily apparent for $\mqlHigh$, which
has a wide variety of shapes.  Particularly dangerous are mass
scenarios like \pbeta\ or (ii), where the `foot' which forms the 
last part of the edge can
be hidden by backgrounds, giving a false maximum.  In Sections 
\ref{sect:data-gen-recon}--\ref{sect:masses-from-edges} these 
considerations will come up in relation to \pbeta. For $\mqlLow$ a
similar danger arises in cases like (i) where what appears to be an
approximately linear descent to zero suddenly turns into a vertical
drop towards the end.

These theory curves must serve as guidance for our determination of 
the endpoints. 
As a first stage they should be the basis for the choice of functions 
with which the edges of the different distributions are fitted.
As a second stage the more ambitious goal may be to fit not only the
endpoint, but the entire distribution.

\subsection{Formulae for kinematic endpoints}

The invariant masses of various subsets of particles can be determined
from kinematical endpoints and thresholds, as discussed in
\cite{Allanach:2000kt}. We confirm the results relevant for our
analysis:
\begin{eqnarray}
\left(\maxmll\right)^2 
&=& \big(\sNT-\slR\big)\big(\slR-\sNO\big)/\slR 
\label{Eq:edge-ll} \\[4mm]
\left(\maxmqll\right)^2 
&=& \left\{ 
\begin{array}{llcc}
\frac{\big(\sqL-\sNT\big)\big(\sNT-\sNO\big)}{\sNT} 
& \ {\rm for } \quad 
& \rqLNT > \rNTlR\rlRNO & \quad {\itB (1)} \\[4mm]
\frac{\big(\sqL\slR-\sNT\sNO\big)\big(\sNT-\slR\big)}{\sNT\slR} 
& \ {\rm for } \quad
& \rNTlR > \rlRNO\rqLNT & \quad {\itB (2)}\\[4mm]
\frac{\big(\sqL-\slR\big)\big(\slR-\sNO\big)}{\slR} 
& \ {\rm for } \quad
& \rlRNO > \rqLNT\rNTlR & \quad {\itB (3)} \\[4mm]
\big(\mqL-\mNO\big)^2 
& \ \lefteqn{\rm otherwise} && \quad {\itB (4)} 
\end{array} 
\right\}  \label{Eq:edge-qll-max} \\[4mm]
\big(\maxmqlLow,\maxmqlHigh\big) &=& \left\{
\begin{array}{llcc}
\big(m^\max_{q\lN},m^\max_{q\lF}\big) & 
\quad {\rm for } \quad
& 2\slR > \sNO+\sNT > 2\mNO\mNT & \quad {\itB (1)} \\[4mm]

\big(m^\max_{ql(\equal)},m^\max_{q\lF}\big) & 
\quad {\rm for } \quad
& \sNO+\sNT > 2\slR > 2\mNO\mNT & \quad {\itB (2)}\\[4mm]
\big(m^\max_{ql(\equal)},m^\max_{q\lN}\big) & 
\quad {\rm for } \quad
& \sNO+\sNT > 2\mNO\mNT > 2\slR & \quad {\itB (3)}
\end{array} 
\right\} \nonumber \\
&& \label{Eq:edge-ql-lowhigh} \\[4mm]
\big(m^\max_{q \lN}\big)^2 
&=& \big(\sqL-\sNT\big)\big(\sNT-\slR\big)/\sNT
\label{Eq:edge-qlN} \\[4mm]
\left(m^\max_{q \lF}\right)^2 
&=& \big(\sqL-\sNT\big)\big(\slR-\sNO\big)/\slR
\label{Eq:edge-qlF} \\[4mm]
\big(m^\max_{ql(\equal)}\big)^2 &=& 
\big(\sqL-\sNT\big)\big(\slR-\sNO\big)/\big(2\slR-\sNO\big)
\label{Eq:edge-ql-equal} \\[4mm]
\big(\minmqllThres\big)^2 &=& 
\Big[\big(\sqL+\sNT\big)\big(\sNT-\slR\big)\big(\slR-\sNO\big) 
\nonumber \\[2mm]
&& -\big(\sqL-\sNT\big)\sqrt{\big(\sNT+\slR\big)^2\big(\slR+\sNO\big)^2
-16\sNT\mlRfour\sNO} 
\nonumber \\[2mm]
&& + 2\slR\big(\sqL-\sNT\big)\big(\sNT-\sNO\big)\Big]/\big(4\slR\sNT\big) 
\label{Eq:edge-qll-min} 
\end{eqnarray}
where `low' and `high' on the left-hand side in 
Eq.~(\ref{Eq:edge-ql-lowhigh})
refer to minimising and maximising with respect to the
choice of lepton.  Furthermore `min' in Eq.~(\ref{Eq:edge-qll-min}) 
refers to the threshold in the subset
of the $\mqll$ distribution for which the angle between the two lepton
momenta (in the slepton rest frame) exceeds $\pi/2$, corresponding to
the mass range (\ref{eq:thres-constraint}).

Notice that the different cases listed in Eq.~(\ref{Eq:edge-qll-max}) 
are distinguished by {\it mass ratios} of neighbouring particles 
in the hierarchy, 
$\mqL/\mNT$, $\mNT/\mlR$ and $\mlR/\mNO$. 
Since each decay in the chain involves two massive particles
and one massless one, the boosts from one rest frame to another are
conveniently expressed in terms of such mass ratios.

%%%%%%%%%%%%%%%%%%%%%%%%%%%%%%%%%%%%%%%%%%%%%%%%%%%%%%%%%%%%%%%%%%%%%%
\subsection{Inversion formulae}
%%%%%%%%%%%%%%%%%%%%%%%%%%%%%%%%%%%%%%%%%%%%%%%%%%%%%%%%%%%%%%%%%%%%%%
Once the endpoints of the distributions have been measured, the masses may be
extracted. In principle, this can be done in two ways.  
If the number of endpoint measurements coincides with the number of unknown
masses, one may analytically invert the expressions for the endpoints to give
explicit formulae for the masses in terms of the endpoints. 
If more endpoints are known, the system becomes overconstrained and 
the measurements must be weighted according to their uncertainties. 
The analytic method cannot easily handle such a situation. 
Instead the more flexible approach of numerical fit must be used. 

However, the analytic inversion is more transparent than the numerical
treatment, and reveals some interesting features of how the masses are
related to the endpoints.  It is also practical to use the analytical
method in combination with the numerical method to provide initial
values for the fits. Consequently we will discuss the inversion of the
endpoint formulae in some detail in this subsection.

The four principal endpoints $\maxmll$, $\maxmqll$, $\maxmqlLow$ and
$\maxmqlHigh$ are given by
Eqs.~(\ref{Eq:edge-ll})--(\ref{Eq:edge-ql-lowhigh}).  While $\maxmll$ is given
by one unique expression, the others have different representations for 
different
mass spectra. To perform the inversion, each combination of the
endpoint expressions must be considered separately. There are four
representations of $\maxmqll$ and three representations of
$(\maxmqlLow,\maxmqlHigh)$.  Each overall combination corresponds to a unique
region in mass space $(\mqL,\mNT,\mlR,\mNO)$ and can be labeled by 
{\itB (i,j)}, where {\it i} and {\it j} denote the region of applicability of 
the $\mqll$ and $\mql$ endpoints, respectively, as given in
Eqs.~(\ref{Eq:edge-qll-max})--(\ref{Eq:edge-ql-lowhigh}).  

However, not all of the $4\times 3$ combinations are physical.  In
particular the regions {\itB (2,1)}, {\itB (2,2)} and {\itB (3,3)} are
not possible. For {\itB (2,1)} this can be seen by simultaneously 
trying to satisfy 
the mass constraints {\itB (2)} of Eq.~(\ref{Eq:edge-qll-max}) and 
{\itB (1)} of Eq.~(\ref{Eq:edge-ql-lowhigh}). 

Additionally, for the regions {\itB (2,3), (3,1)} and {\itB (3,2)} the
endpoint expressions are not linearly independent. In each of these
three regions, one of the mass ratios is dominant, and a lepton
(rather than a quark) is emitted from the vertex involving the
dominant mass ratio. The maximum of $\mqll$ occurs when this lepton
travels in one direction and the other lepton together with the
quark travels in the opposite direction, all observed from the rest
frame of any one of the sparticles in the decay.  For massless quarks
and leptons one can always write
\begin{equation}
m^2_{qll} 
= 2p_qp_\lN+2p_qp_\lF+2p_\lN p_\lF 
= m^2_{ql(\high)} + m^2_{ql(\low)} + m^2_{ll}
\label{Eq:sqll=sql+sql+ll}
\end{equation}
In the three regions mentioned above, for maximum values this equation
reduces to
\begin{equation}
(m^\max_{qll})^2 = (m^\max_{ql(\high)})^2 + (m^\max_{ll})^2
\label{Eq:sqll=sql+sql}
\end{equation}
Since one of the endpoints can be expressed in terms of two others,
the four endpoint measurements only transform into three independent
conditions.  Hence, in order to determine the masses an additional
endpoint would be required for these three regions. The endpoint of
$\mqllThres$ is therefore particularly important in such cases.
However, the complicated nature of Eq.~(\ref{Eq:edge-qll-min}) leads
to exceptionally cumbersome inversion formulae and will not be
studied here.

For the remaining six regions the inversion of the four endpoint
expressions Eqs.~(\ref{Eq:edge-ll})--(\ref{Eq:edge-ql-lowhigh}) is
possible.  Coincidentally these are actually the regions most likely
to be realised in SUSY scenarios with sparticle mass unifications at
the GUT scale. Inversion formulae for these regions are detailed
below.  For simplicity of notation we will write:
\def\rA{a} %\maxmll
\def\rB{b} %\maxmqll
\def\rC{c} %\maxmqlLow
\def\rD{d} %\maxmqlHigh
\def\A{\rA^2} 
\def\B{\rB^2} 
\def\C{\rC^2} 
\def\D{\rD^2} 
\def\sA{\rA^4} 
\def\sB{\rB^4} 
\def\sC{\rC^4} 
\def\sD{\rD^4} 
\begin{equation}
\rA=\maxmll,\ \rB=\maxmqll,\ \rC=\maxmqlLow,\ \rD=\maxmqlHigh. 
\end{equation}

\noindent {\itB Region (1,1):}
\begin{eqnarray}
\label{eq:inv(11)mN1}
\sNO & = & \frac{(\B-\D)(\B-\C)}{(\C+\D-\B)^2} \A \\
\slR & = & \frac{\C(\B-\C)}{(\C+\D-\B)^2} \A \\
\sNT & = & \frac{\C\D}{(\C+\D-\B)^2} \A \\
\sqL & = & \frac{\C\D}{(\C+\D-\B)^2} (\C+\D-\B+\A) 
\end{eqnarray}

\noindent {\itB Region (1,2):} 
\begin{eqnarray}
\label{eq:inv(12)mN1}
\sNO & = & \frac{(\B\C-\B\D+\C\D)(2\C-\D)}{(\D-\C)^2(\B-\D)} \A \\
\slR & = & \frac{(\B\C-\B\D+\C\D)\C}{(\D-\C)^2(\B-\D)} \A \\
\sNT & = & \frac{(2\C-\D)\C\D}{(\D-\C)^2(\B-\D)} \A \\
\label{eq:inv(12)mqL}
\sqL & = & \frac{\C\D}{(\D-\C)^2(\B-\D)} [\A(2\C-\D)+(\D-\C)(\B-\D)] 
\end{eqnarray}

\noindent {\itB Region (1,3):}
\begin{eqnarray}
\sNO & = & \frac{(\D-\C)(\B-\D)(\B-2\C)}{(\C+\D-\B)^2\D} \A \\
\slR & = & \frac{(\D-\C)^2(\B-\D)}{(\C+\D-\B)^2\D} \A \\
\sNT & = & \frac{(2\D-\B)(\D-\C)\C}{(\C+\D-\B)^2\D} \A \\
\sqL & = & \frac{(2\D-\B)\C}{(\C+\D-\B)^2\D} [\A(\D-\C)+\D(\C+\D-\B)]
\end{eqnarray}

\noindent {\itB Region (4,1):}
\def\SIGN{+} %plus sign is chosen
\begin{eqnarray}
\hspace{-1.0cm}
\sNO & = & \sNT - ({\rD\rA/\rC}+{\rA\rC/\rD})\mNT + \A \\
\hspace{-1.0cm}
\slR & = & \sNT - {(\rA\rC/\rD)}\mNT \\
\hspace{-1.0cm}
\label{eq:inv(41)mNT} 
\mNT & = & \frac{\big[(\A+\B)\C\D-(\B-\A)\A(\C+\D)  \SIGN  2\rA\rB\rC\rD\sqrt{(\A+\C-\B)(\A+\D-\B)}\big]}{[(\A\C+\A\D+\C\D)^2-4\A\B\C\D]/(\rA\rC\rD)} \\
\hspace{-1.0cm}
\mqL & = & \mNO + {\rB} 
\end{eqnarray}

\noindent {\itB Region (4,2):}
\def\SIGN{-} %minus sign is chosen
\begin{eqnarray}
\label{eq:inv(42)mNO}
\mNO & = & \frac{2\C-\D}{\D-\C}
\left({\rB}  \SIGN  \rC\sqrt{\frac{\B-\A-\D} {2\C-\D} }\right)
\\
\slR & = & {\frac{\C}{2\C-\D}} \sNO
\\
\sNT & = & \frac{\C}{2\C-\D} \sNO + \frac{\A\C}{\D-\C}
\\
\mqL & = & \mNO + \rB
\end{eqnarray}

\noindent {\itB Region (4,3):}
\begin{eqnarray}
\sNO & = & \slR + \rA\big[{\rA(\D-\C) - \sqrt{\A(\D-\C)^2 +
    4\C\D\slR}}\big]/({2\D})\\
\nonumber
\slR & = & 
\A \big\{
\hspace{-0.15cm}
-
\hspace{-0.05cm}
(\rD^4-\rC^4)^2\rA^8 
+ (\D-\C) [\rC^6(2\D-3\B)+7\rC^4\B\D+\rD^4(\C(3\B-2\D)+\B\D)]\rA^6 
\\&&\nonumber
+ \C\D[(5\rC^4-6\C\D-3\rD^4)\rB^4+(7\C\rD^4+5\rC^6-5\rC^4\D+\rD^6)\B+4\rD^4\rC^4] \rA^4
\\&&\nonumber
+ \rD^4\rC^4[4\D\C\B-10\C\rB^4+2\C\rD^4+2\rC^4\D+\rC^4\B+4\rB^6-2\D\rB^4-\rD^4\B] \A 
\\&&\nonumber
+ \rC^6\rD^6(\B-\D)(\B-\C)
\pm 2\rA\rB\rC\sqrt{(\B-\A-\D)(\B\D-2\A\D-2\C\D+2\A\C)}
\\&&\nonumber
\times[{(\D+\C)(\D-\C)^2\sA + 2\C\D(\C\B-\B\D+\sC+\sD)\A + \sC\sD(\D+\C-2\B)}]
\big\}
\\&&
/[\sA\sD + (\A+\D)^2\rC^4 + 2\A\C\D(\A+\D-2\B)]^2
\\
\sNT & = & \slR (\A+\slR-\sNO)/(\slR-\sNO) \\
\mqL & = & \mNO + {\rB}
\label{eq:inv(43)mqL}
\end{eqnarray}

For a given set of endpoint measurements the inversion formulae of 
most regions will return unacceptable masses. In some regions the masses 
returned will be in contradiction with the presupposed hierarchy, 
$\mqL>\mNT>\mlR>\mNO$, in some there will 
even be negative or imaginary masses. Such solutions must be discarded. 
If the masses returned by the inversion formulae of a region {\itB(i,j)}
do obey the hierarchy, they constitute a physical solution if they 
themselves belong to {\itB(i,j)}.
Solutions which do not satisfy this last constraint will be referred 
to as `unphysical'. 
In principle it should be possible to construct for each region  
a number of conditions on the endpoint values which, beforehand, would 
tell if the inversion formulae would return a physical solution or not. 
Such conditions
will however become quite complicated, and have not been sought in 
this study. 
Instead the more straightforward approach is taken: try the inversion 
formulae of each region, then discard any unacceptable solutions. 

Originating from realisation {\itB(4)} of $\maxmqll$, see 
Eq.~(\ref{Eq:edge-qll-max}), where masses appear unsquared, the 
inversion formulae of region {\itB(4,j)} in principle come in two 
versions, the difference being a sign within one of the mass expressions. 
In regions {\itB(4,1)} and {\itB(4,2)} it turns out that only one 
of the signs returns masses in the appropriate region. 
In Eqs.~(\ref{eq:inv(41)mNT}) and (\ref{eq:inv(42)mNO}) 
the physical sign is therefore chosen. 
For {\itB(4,3)} both versions must be considered. 

Once the required endpoints are measured and the inversion formulae
used to determine the masses, we encounter a delicate feature of
the entire method of obtaining masses from endpoints. While the
endpoints are given by single-valued functions of the masses, albeit
with different expressions for different mass regions, the inverse is
not true. A given set of endpoint values can in principle correspond
to several sets of mass values.  This is equally true for the
numerical fit method, and has not received much attention previously
(see, however, Ref.~\cite{Lester}).
This complication will be faced in 
Section~\ref{sect:masses-from-edges}.

%%%%%%%%%%%%%%%%%%%%%%%%%%%%%%%%%%%%%%%%%%%%%%%%%%%%%%%%%%%%%%%%%
\section{`Data' generation and reconstruction}\label{sect:data-gen-recon}
%%%%%%%%%%%%%%%%%%%%%%%%%%%%%%%%%%%%%%%%%%%%%%%%%%%%%%%%%%%%%%%%%

\subsection{Event generation\label{subsect:eventgen}}

The SPS points are defined by the low-energy MSSM parameters 
produced by {ISAJET 7.58} \cite{Baer:1993ae}, given a set of high-energy 
input parameters. 
In our analysis {PYTHIA 6.2} \cite{PYTHIA} with CTEQ~5L \cite{Lai:1999wy}
is used to generate the 
Monte Carlo sample.\footnote{The main parts of the analysis have been 
confirmed with {HERWIG} \cite{HERWIG}, see \cite{gjelsten-atlas}.}
To allow for this the low-energy parameters from {ISAJET} are fed into 
{PYTHIA} via the standard interface. {PYTHIA} in turn calculates
the decay widths and cross-sections. 
Each event produced is passed through 
{ATLFAST 2.60} \cite{ATLFAST}, a fast simulation of the ATLAS detector. 
In {ATLFAST} the output particles of {PYTHIA} are mapped onto a grid 
of `calorimetric cells' of a given spacing in pseudorapidity $\eta$ and 
azimuth angle $\phi$. Next, the cells are combined into clusters, 
after which particle identification takes place, including smearing of 
the four-momenta according to particle type. 
Jets are built by a cone algorithm with $\Delta R=0.4$, 
where $\Delta R=\sqrt{(\Delta\eta)^2+(\Delta\phi)^2}$.
Acceptance requirements are imposed: $|\eta| < 2.5$ for $e/\mu$ 
and $|\eta| < 5$ for jets as well as $p_T>5/6/10$~GeV for $e/\mu/$jets. 
Leptons are marked as isolated if there is no other cluster within a distance 
$\Delta R=0.4$ of the lepton, and if additional energy inside a cone $\Delta R=0.2$ 
is below 10~GeV.
While {ATLFAST} captures quite well the main features of the full 
simulation, some important effects are left out. 
Lepton identification efficiencies are not parametrized. 
A conventional 90\% efficiency per lepton is therefore included by hand
in the analysis. 
Also, the possibility of misidentifying a jet as a lepton is absent in the 
fast simulation, and has not been included in our analysis. 
The effect of pile-up on the jet energy resolution 
is accounted for in {ATLFAST} when run at high luminosity, 
as in this analysis, 
but pile-up events have not been simulated, 
and the underlying event is probably too `slim'. 
However, as the selection criteria on jets and leptons are quite hard, 
we do not expect a more realistic detector simulation to change 
the results very much.
Nevertheless, the numbers quoted at the end of this section should be 
validated with these effects included. 

The signature of a signal event is two opposite-sign same-flavour (SF)
leptons, considerable missing $p_T$ from the escaping LSPs, and 
at least two hard jets, one from the signal chain, the other from the
decay of the squark nearly always present in the other decay chain. 
The most important Standard Model process to have the same features as the
signal, is $t\bar t$ production. 
Also $W/Z$ together with jets, one of which is a $b$-jet, 
can mimic the signal, 
and in combination with the underlying event, pile-up and detector effects, 
other processes will also now and then result in the given signature. 
Together with
$t\bar t$ we therefore include QCD, $Z/W$+jet as well as $ZZ$/$ZW$/$WW$
production. 
No $K$-factors have been used. 

The precuts (not tuned) used to isolate the chain are the following, 
\begin{itemize}
\item At least three jets, satisfying: $p_T^\textrm{jet}>150, 100, 50$~GeV
\item $E_{T,\rm miss} > {\rm max}(100\ \GeV,0.2 M_{\rm eff})$ with 
$M_{\rm eff}\equiv E_{T,\rm miss} + \sum_{i=1}^{3}p_{T,i}^{\textrm{jet}}$  
\item Two isolated opposite-sign same-flavour leptons ($e$ or $\mu$), 
satisfying $p_T^\textrm{lep} > 20,10\ \GeV$  
\end{itemize}

The QCD background is cut away by the requirement of two leptons and of 
considerable missing $p_T$. For the processes involving $Z$ and $W$ 
the requirement of high hadronic activity together with the missing $p_T$ 
removes nearly all events. After the rather hard cuts listed above, 
the Standard Model background consists of approximately 95\% $t\bar t$.

\subsection{Different Flavour (DF) subtraction\label{subsect:DFsubtraction}}

At this stage the main background events, in addition to $t\bar t$, will come
from other SUSY processes which have two opposite-sign same-flavour leptons.
It is useful to distinguish between two types.  The first, which will also be
referred to as `lepton-correlated', produces correlated leptons, e.g.\
leptonic decay of $Z$.  In these processes the leptons always have the same
flavour.  The other type produces uncorrelated leptons which need not be of
the same flavour.  Typically the uncorrelated leptons are produced in
different decay branches within the same event.  Lepton universality implies
that electrons and muons are produced in equal amounts (apart from negligible
mass effects).  This means that for events which produce uncorrelated leptons,
there should be as many opposite-sign different-flavour (DF) lepton pairs
produced as there are opposite-sign same-flavour lepton pairs, and their event
characteristics should be the same.  The same-flavour leptons are already part
of the selected events.  If one then subtracts the different-flavour events
from the total same-flavour sample, this corresponds statistically to removing
the non-signal same-flavour events which come from uncorrelated leptons.  The
residual of the subtraction is a larger statistical uncertainty in the new
signal distribution.  Clearly the different-flavour subtraction is a very
effective tool which takes care of both $t\bar t$ and most of the SUSY
background.  Only processes with correlated leptons remain.

In a more detailed study one would need to correct for the different 
acceptances of electrons and muons as functions of $(p_T,\eta,\phi)$ 
in the different-flavour subtraction.
Here we have assumed equal acceptance for electrons and muons, which
even for a fast simulation is only approximately true. 
Also we have assumed that the efficiencies for reconstructing 
two same-flavour and two different-flavour leptons are the same. 
For close leptons this is not the case in the {ATLFAST} reconstruction. 
This will have some visible effects in the different-flavour-subtracted 
distributions, see below.

%%%%%%%%%%%%%%%%%%%%%%%%%%%%%%%%%%%%%%%%%%%%%%%%%%%%%%%%%%%%%%%%%%%%%%
\FIGURE[ht]{
%\begin{figure}
%%Begin InstantTeX Picture
\let\picnaturalsize=N
\def\picsize{7.1cm}
%If you do not have the picture file add:
%\let\nopictures=Y
%to the beginning of the file.
\ifx\nopictures Y\else{
\let\epsfloaded=Y
\centerline{{\ifx\picnaturalsize N\epsfxsize \picsize\fi
\epsfbox{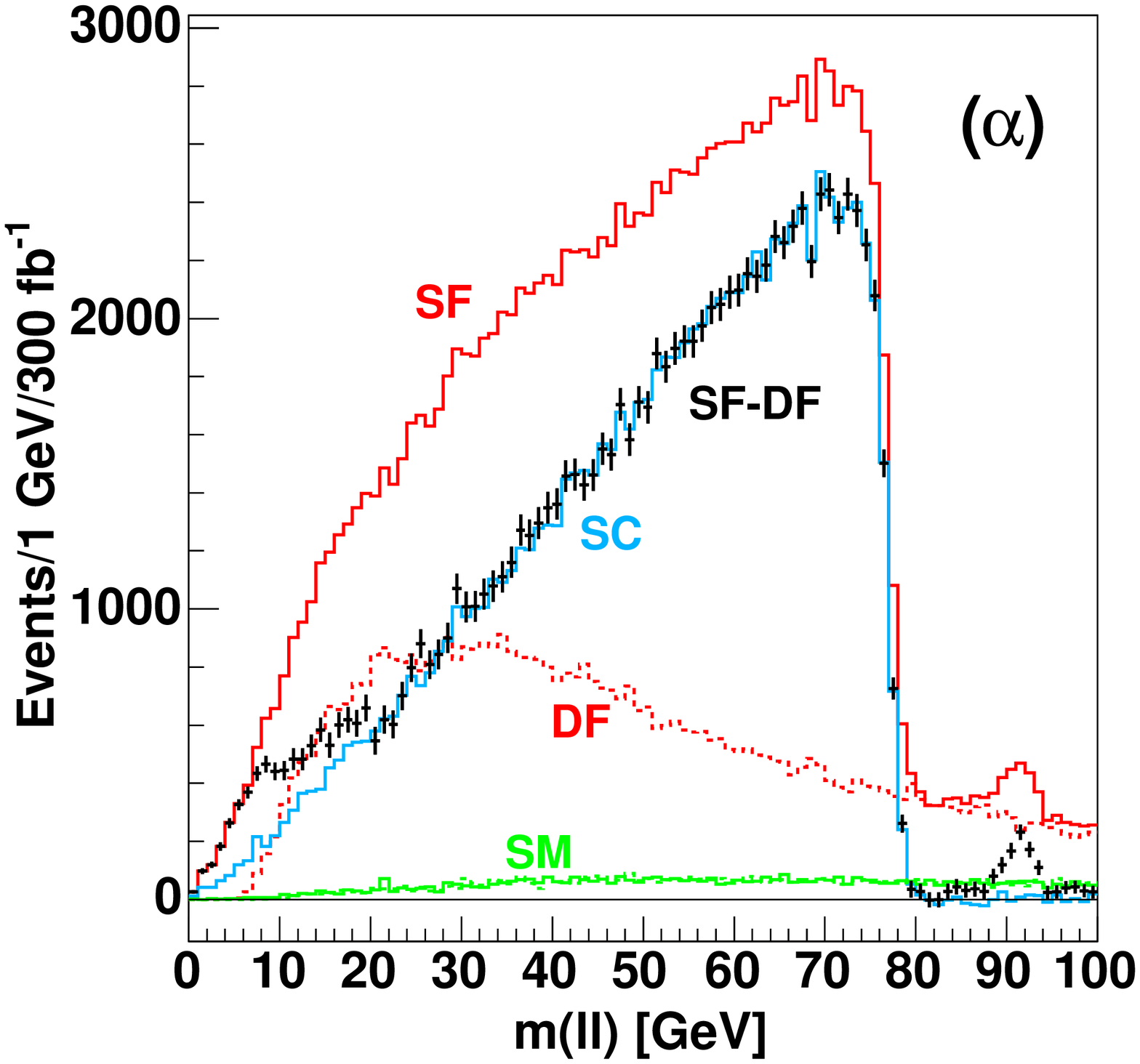}
            {\ifx\picnaturalsize N\epsfxsize \picsize\fi
\epsfbox{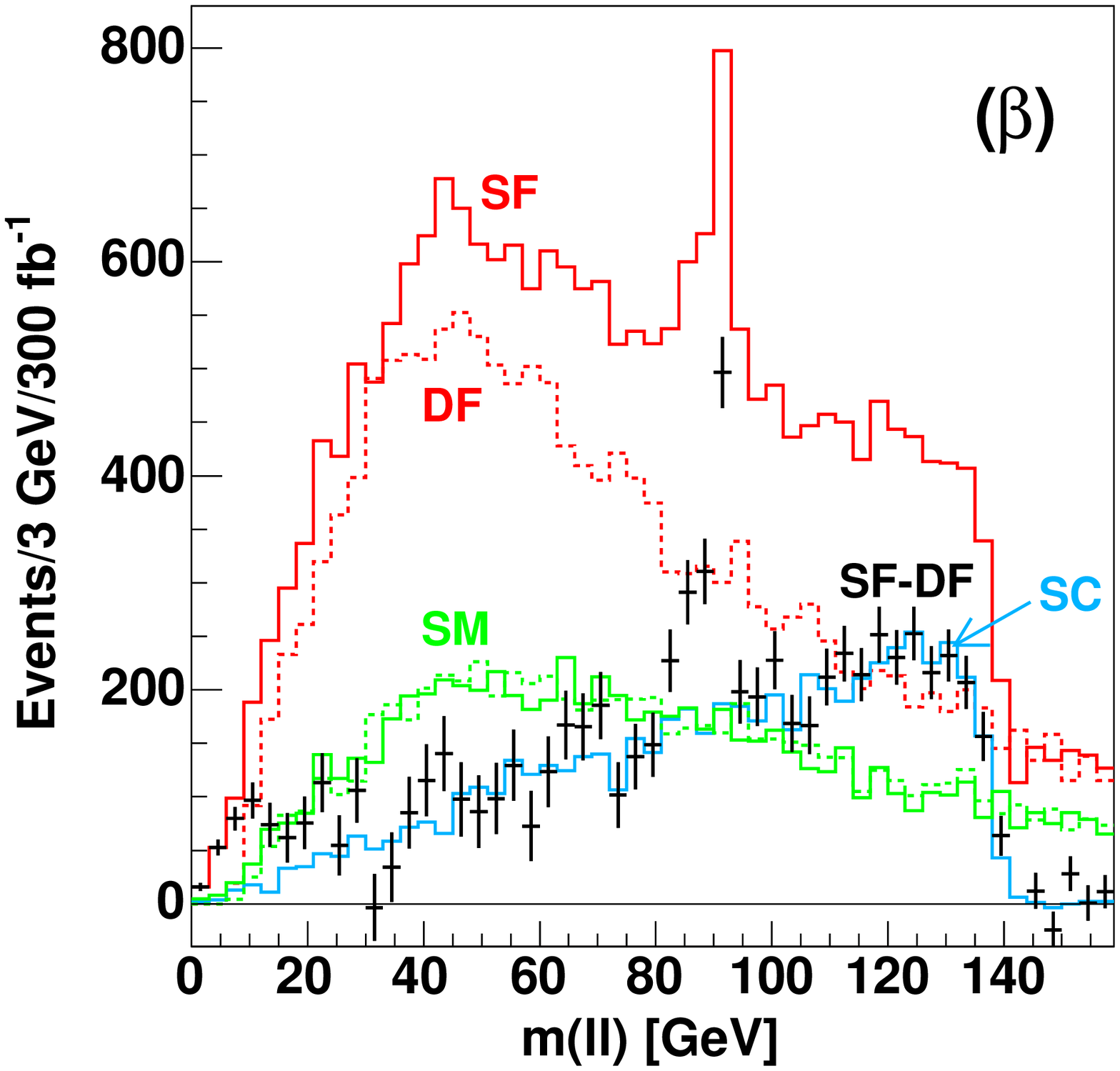}
} } }
\centerline{{\ifx\picnaturalsize N\epsfxsize \picsize\fi
\epsfbox{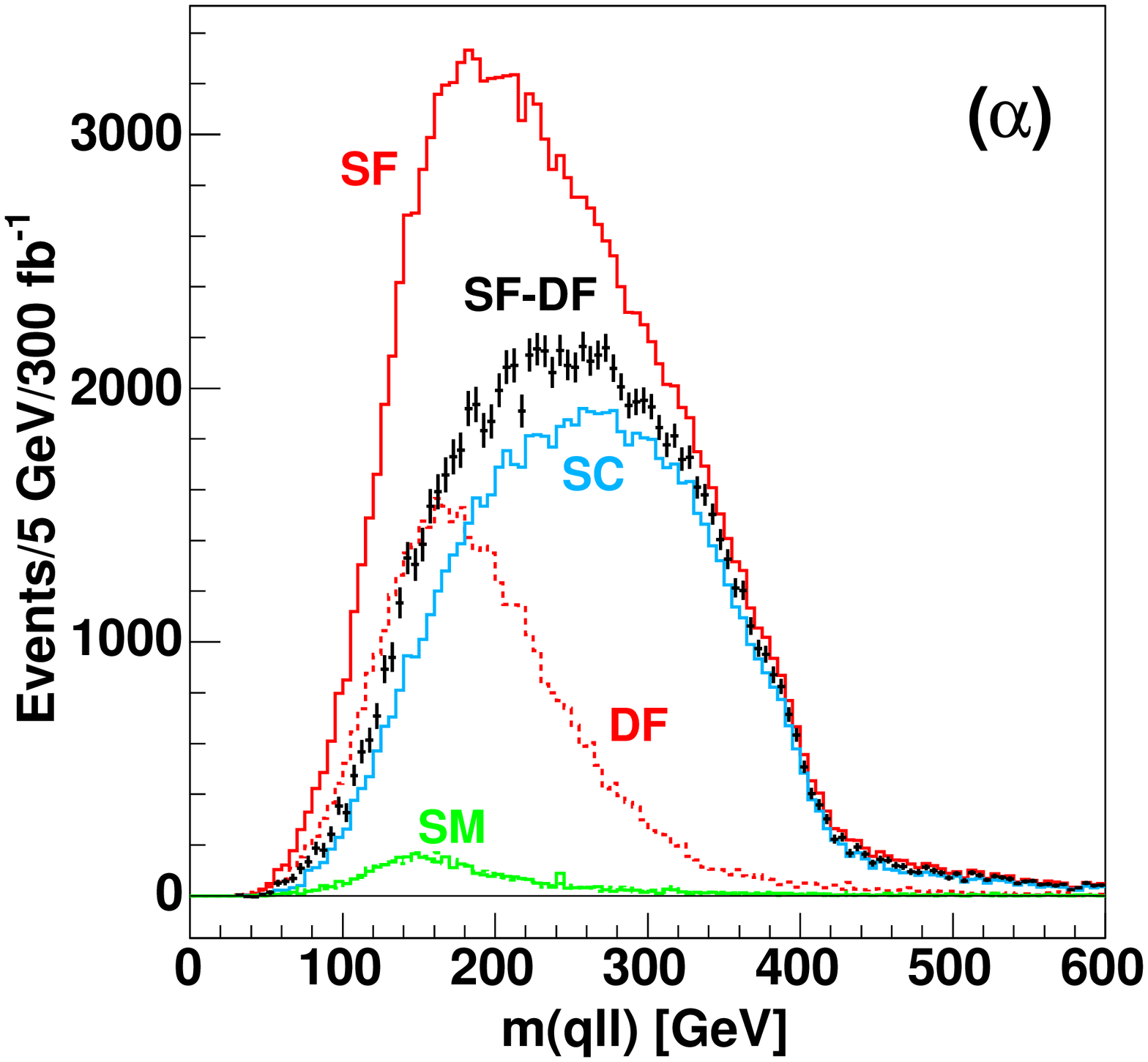}
            {\ifx\picnaturalsize N\epsfxsize \picsize\fi
\epsfbox{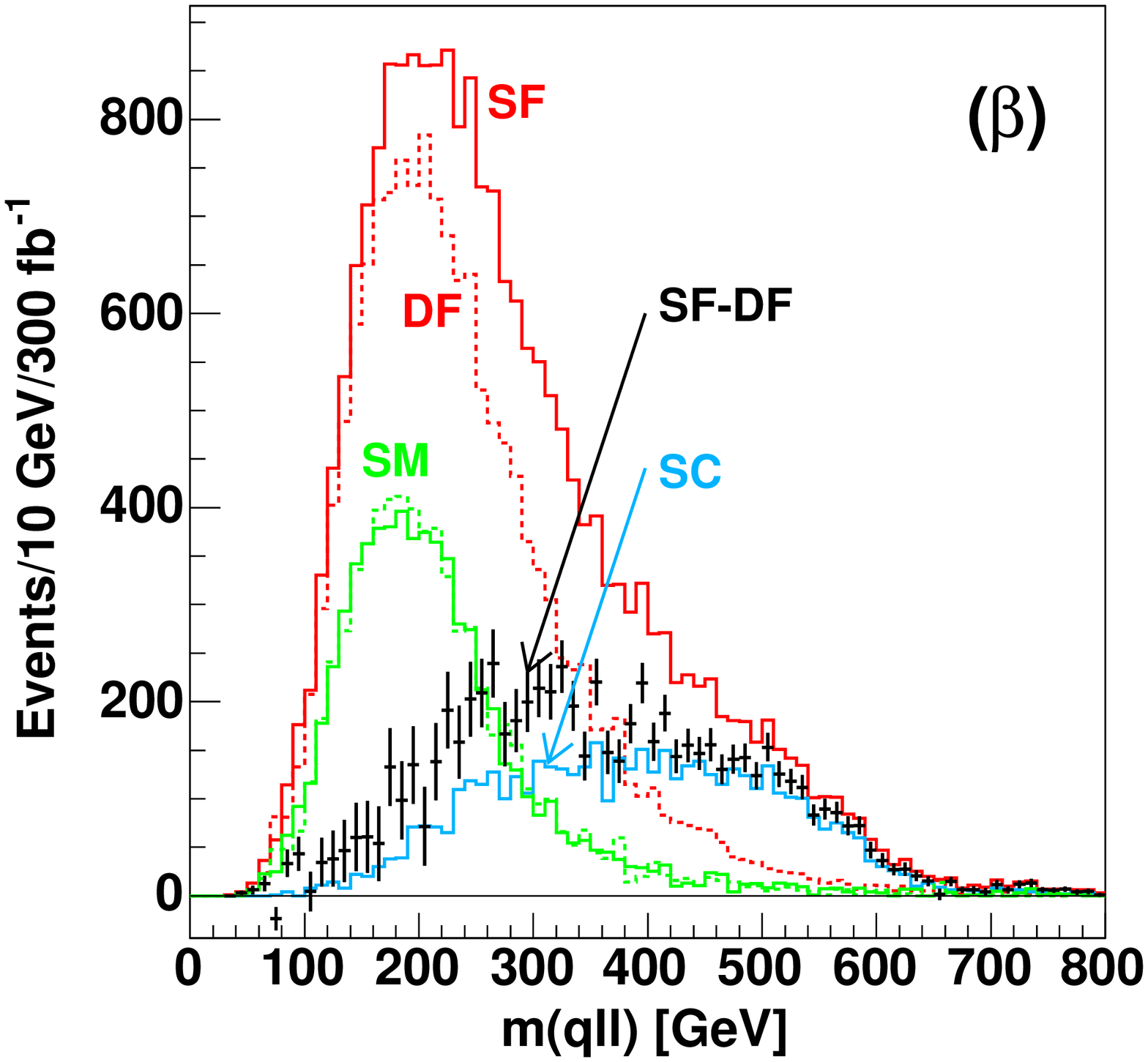}
} } }

}\fi
%%End InstantTeX Picture
\vspace{-6mm}
\caption{Different-flavour subtraction for $\mll$ and $\mqll$ 
at ($\alpha$) (left) and ($\beta$) (right).
The solid/dashed red curves are the same-flavour (`SF')/different-flavour 
(`DF') distributions. 
In black, their difference, the different-flavour-subtracted distribution 
(`SF-DF'), is shown with error bars. 
The blue curve shows the part of the subtracted distribution which contains 
a signal chain (`SC'). 
The solid/dashed green curves (`SM') give 
the Standard Model part (completely dominated by $t\bar t$) 
of the same-flavour/different-flavour distributions. 
They are statistically equal and will cancel each other. 
\label{Fig:OSOF-subtr}}}
%\end{figure}
%%%%%%%%%%%%%%%%%%%%%%%%%%%%%%%%%%%%%%%%%%%%%%%%%%%%%%%%%%%%%%%%%%%%%%

Fig.~\ref{Fig:OSOF-subtr} shows the different-flavour subtraction for the 
$\mll$ (top) and $\mqll$ (bottom) distributions at points
\palpha\ (left) and \pbeta\ (right).
(The jet used in the $\mqll$ distribution is the one of the two 
$p_T$-hardest jets which gives the smaller value of $\mqll$.)
All the plots have the same colour code. 
In red the same-flavour (`SF') distribution is shown by solid curve, 
the different-flavour (`DF') one by dashed curve. 
Their difference, the different-flavour-subtracted distribution (`SF-DF'), 
is shown in black with errors.  
The curve in blue shows the part of the different-flavour-subtracted 
distribution which contains a signal chain (`SC'). 
The reason why the blue distribution of $\mqll$ does not have the form 
of the theory
distribution in Fig.~\ref{Fig:th-curves}, is that the jet is only correctly
selected in roughly half of the cases.  In solid/dashed green the Standard
Model contributions to the same-flavour/different-flavour distributions are
shown (`SM'). They are statistically identical and will cancel each other
through the different-flavour subtraction.

If the samples contained only background events with uncorrelated leptons
(and the different-flavour-subtraction procedure removed all of these),
the different-flavour-sub\-trac\-ted distribution should fall exactly 
on top of the blue line.
When this does not happen, it implies that the sample also contains 
background events with correlated leptons. 
The $Z$-peaks in the $\mll$ distribution at both \palpha\ and \pbeta\ 
are obviously of this type.
The $Z$'s stem predominantly from the decay of heavier gauginos into lighter
ones.  At \palpha\ $\sim80$\% of the peak comes from the heavy gauginos
$\NThree$, $\NQ$ and $\CT$.  At \pbeta\ $\NT$ is sufficiently heavy to decay
into $\NO Z$ and is responsible for $\sim40$\% of the peak. The rest involves
the heavy gauginos.

For $\mll\lesssim20$~GeV a bump is visible in the different-flavour-subtracted
distribution. This excess turns out to come mainly from lepton-uncorrelated
events, predominantly of the type $\NT\to\tauO\tau$ which is abundant in the
two scenarios, and where the taus decay leptonically.  Since a pair of taus
produce same-flavour and different-flavour leptons in identical amounts, the
different-flavour subtraction should take care of this background. When it
does not, it is because of an asymmetry in the reconstruction algorithm of
{ATLFAST} which accepts close same-flavour leptons at a higher rate than close
different-flavour leptons.  Since small $\Delta R$ 
between the leptons means small $\mll$
values, the bump appears at low $\mll$.  Such a reconstruction asymmetry may
also have noticeable effects in other distributions, e.g.\ $\mqll$, for which
the maximum values in our two mass scenarios appear for parallel leptons.  A
dedicated study of the impact detector effects of this type may have on the
endpoint determinations may be worthwhile.

In the $\mqll$ distributions there is an excess of events (compared with the
blue line) at lower masses.  Some of this excess comes from the
lepton-uncorrelated events in the $\mll$ bump.  Typically these land at low
$\mqll$ values.  The main contribution to the excess is however from $\NT$'s
which decay sleptonically, but which originate from $\qR$, $\tO$, $\gl$ or the
heavier gauginos.

At \pbeta\ two effects complicate the endpoint measurements.  One is a
reduction of the SUSY cross-section by one order of magnitude relative to
\palpha. This allows for the Standard Model background to have a larger
impact, see the green curves in Fig.~\ref{Fig:OSOF-subtr}.  Since this
background (practically only $t\bar t$) consists of uncorrelated leptons, it
is dealt with by the different-flavour subtraction. It leaves however an
increased statistical uncertainty in the resulting distribution.
The other effect at \pbeta\ is a reduction in the signal branching ratio, 
mainly from a reduced  BR($\NT\to\lR l$), see Fig.~\ref{fig:squarkchain}.
As a result also the SUSY background becomes larger relative to the signal.

While the Standard Model background is practically negligible at \palpha, 
it is of similar size as the SUSY background at \pbeta.
For the $\mqll$ distribution both background types are important only 
at low values, a fair distance away from the kinematical maximum value. 
The same is true for the $\mql$ distributions.
The endpoint measurements are therefore only minimally affected by the 
background.
An exception is the $\mqllThres$ distribution, where a minimum is measured.

\subsection{Selection cuts}  \label{subsect:selection-cuts}

A fair fraction of the events which pass the precuts, $\sim30\%$ for \palpha\
and $\sim20\%$ for \pbeta, have more than two leptons.  Each opposite-sign
lepton pair which satisfies $p_T>20,10$~GeV is used, either in the
same-flavour or the different-flavour distribution.  As usual, the
different-flavour subtraction takes care of the incorrect combinations.

The selection of the jet to go with the two leptons is more difficult.  In the
precuts at least three jets are required.  If the squarks are considerably
heavier than $\NT$, and the gluinos are not very much heavier than the
squarks, then the two $p_T$-hardest jets are expected to come from the decay
of the two squarks present in nearly all events.  (More specifically we need
$\ssq-\sNT$ somewhat larger than $\sgl-\ssq$.)  Along the entire SPS~1a line
this is the case.  The third jet is then expected to come from a gluino
decay. Specifically, both for \palpha\ and \pbeta\ the correct quark is one of
the two $p_T$ hardest in 94\% of the cases at the parton level. 
It is therefore reasonable to consider only these two jets.

In a realistic setting 
all possible combinations will be investigated, along with all kinds
of precuts.  However, whatever the mass situation, one will not be able to
select the correct jet by some simple cut. Combinatorial background from the
jet selection procedure is something one will have to live with.

%%%%%%%%%%%%%%%%%%%%%%%%%%%%%%%%%%%%%%%%%%%%%%%%%%%%%%%%%%%%%%%%%%%%%%
\FIGURE[ht]{
%\begin{figure}
%%Begin InstantTeX Picture
\let\picnaturalsize=N
\def\picsize{7.1cm}
%If you do not have the picture file add:
%\let\nopictures=Y
%to the beginning of the file.
\ifx\nopictures Y\else{
\let\epsfloaded=Y
\centerline{{\ifx\picnaturalsize N\epsfxsize \picsize\fi
\epsfbox{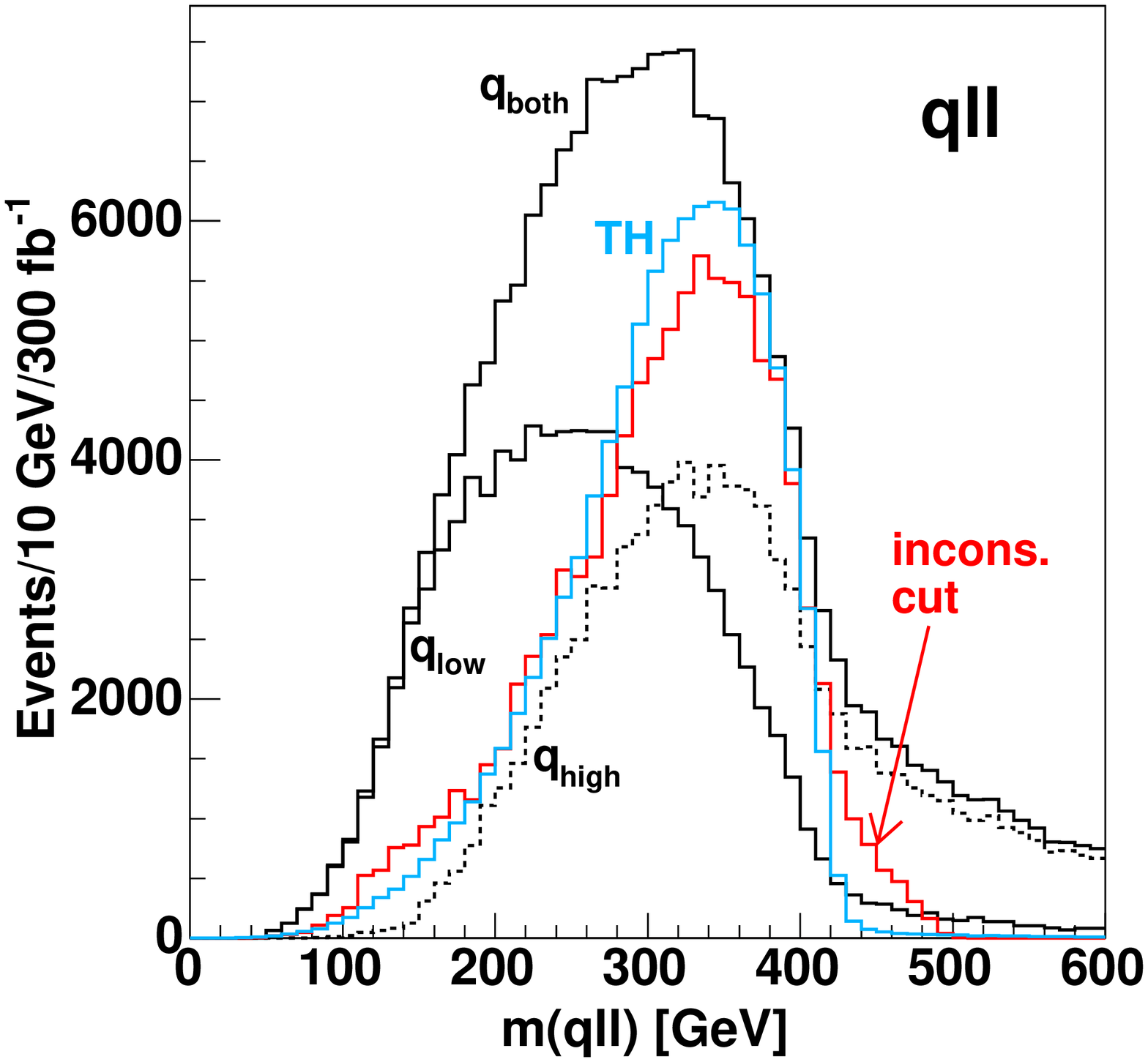}
            {\ifx\picnaturalsize N\epsfxsize \picsize\fi
\epsfbox{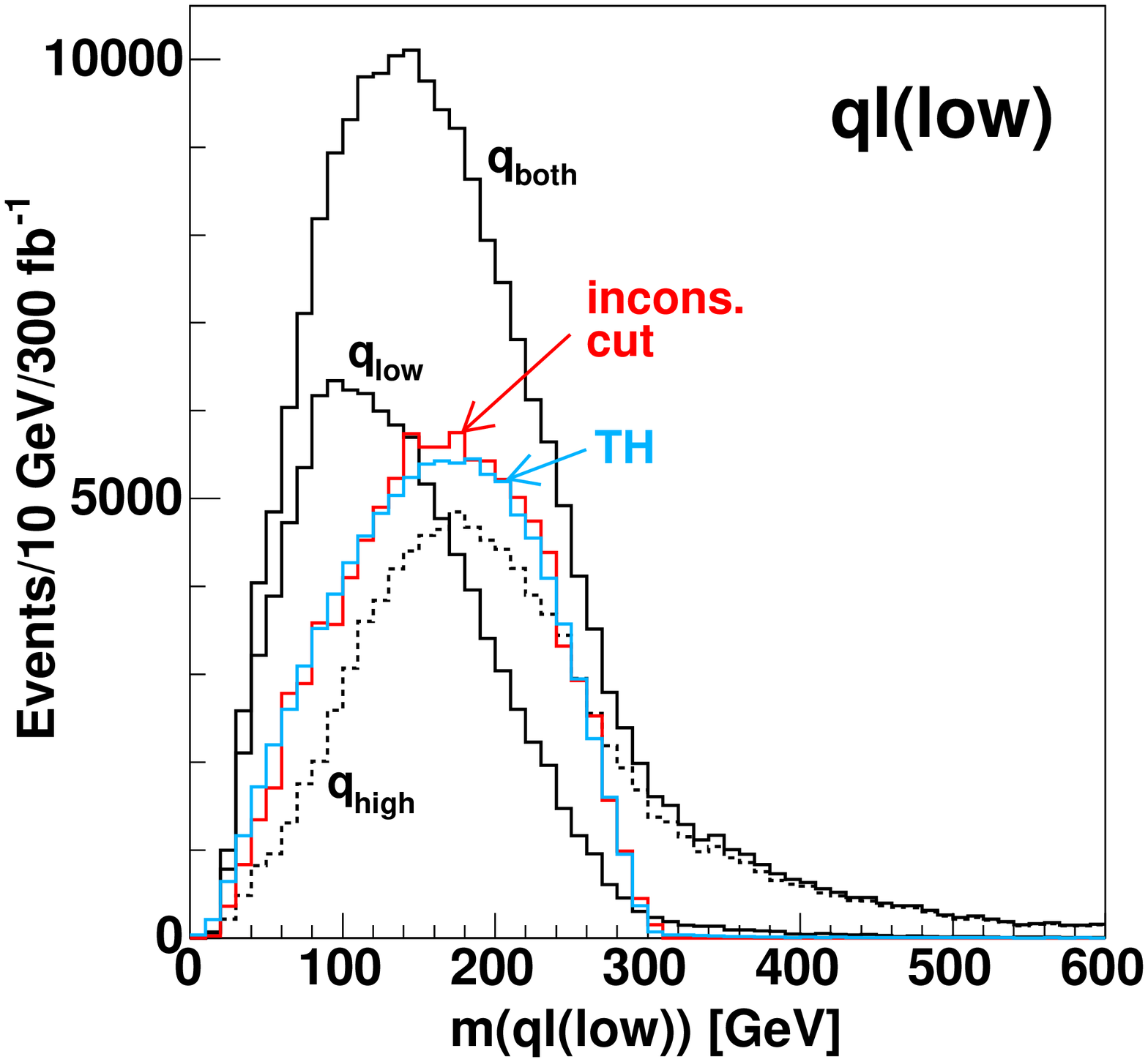}
} } }
}\fi
\vspace{-6mm}
%%End InstantTeX Picture
\caption{Quark selection cuts for $\mqll$ (left) and $\mqlLow$
(right).  Left: The two lower black curves in solid and dashed
show the $\mqLowll$ (`$q_{\low}$') and $\mqHighll$ (`$q_{\high}$')
distributions.  The highest black curve shows the sum of the two,
$\mqBothll$ (`$q_{\both}$'). Notation is defined in the text.  Blue:
theory distribution (`TH').  Red: inconsistency cuts (`incons.~cut')
have been applied on $\mqlLow$ and $\mqlHigh$ (the resulting
distribution is scaled up).  Right: An inconsistency cut is applied on
$\mqll$. The jet chosen for $\mqLowll$ is the one used in $\mqLowlLow$
etc. The inconsistency cut is seen to work very well for the $\mqlLow$
distribution.
\label{Fig:SelCuts}}}
%\end{figure}
%%%%%%%%%%%%%%%%%%%%%%%%%%%%%%%%%%%%%%%%%%%%%%%%%%%%%%%%%%%%%%%%%%%%%%

In Fig.~\ref{Fig:SelCuts} (left) the different-flavour-subtracted
$\mqll$ distribution at \palpha\ is shown for various jet selections.
The two lower black curves are obtained by selecting from the two
$p_T$-hardest jets the one which gives the smaller (solid) and
larger (dashed) value of $\mqll$, for simplicity referred to as
$\mqLowll$ (`$q_{\low}$') and $\mqHighll$ (`$q_{\high}$'),
respectively.  The highest black curve shows the sum of the two,
$\mqBothll$ (`$q_{\both}$').  In blue and marked `TH', with the same
normalisation as $\mqLowll$, is the parton level theory distribution,
as generated by PYTHIA so that it contains widths, production biases
etc.

It is possible to get a reasonable first estimate of the endpoint both
from $\mqLowll$ and $\mqBothll$.  The shape of the theory distribution
is however not reproduced by any of the three experimental
distributions (in black), which is quite natural since they all
contain both combinatorial and lepton-correlated background, and two of them
are also biased by the jet selection procedure.  For \palpha\ the
lepton-correlated background comes mainly from sleptonically decaying
$\NT$'s that do not have a $\qL$ or $\sb$ as a parent.  As can be
seen from Table~\ref{table:sparticlerate} together with 
Fig.~\ref{fig:squarkchain}, $\NT$ comes from $\qL$ or
$\sb$ in 78\% of the times.  If, for a signal event, we assume 50\%
chance of picking the correct jet, a different-flavour-subtracted distribution
will consist of 39\% signal, 39\% combinatorics and 22\% lepton-correlated
background. For \pbeta\ the fractions are 35\%, 35\% and 30\%.
For these numbers $\NT$ is assumed to be the only source of 
lepton-correlated background. 

We have investigated methods to remove or model this combinatorial
background.  For example, the sample may be purified
\cite{Bachacou:1999zb} by using events where only one of the jets can
reasonably be assigned to the signal quark. Consider the case where a
measurement of $\maxmqll$ has been made, and one wants to now measure
$\maxmqlLow$. In constructing the $\mqlLow$ distribution we must
choose one of the two highest $p_T$ jets, and with no other input will
choose the wrong jet 50\% of the time, giving a combinatorial
background. However, in some events, one of the jets will be such that
its intepretation as the signal quark would contradict the previous
measurement of $\maxmqll$, i.e.\ using it to construct $\mqll$ would
give $\mqll>\maxmqll$. In these events, we can be reassured that in
choosing the {\it other} jet we have chosen correctly. If neither of
the jets in the event gives $\mqll$ in contradiction with the previous
measurement, then we cannot be certain that we have the correct jet,
and therefore discard the event. In other words, we only use events
where only one choice of signal quark is consistent with other
endpoint measurements.  This allows one to build a purified sample
where for signal events the wrong jet is only chosen if the correct
jet is not one of the two $p_T$-hardest jets. There will
however still be background events in the sample.

The red curve of Fig.~\ref{Fig:SelCuts} (left) is constructed from the
$\mqll$ distribution by means of `inconsistency cuts' on $\mqlLow$
and $\mqlHigh$ where we insist that only one of the two highest $p_T$
jets has $\mqlLow<310$~GeV and $\mqlHigh<390$~GeV.  Approximately 1/4
of the events pass these cuts; the resulting curve is normalised to
have the same number of events as the $\mqLowll$ distribution.  The
similarity with the blue theory curve has improved, but is still not
excellent.  In particular there is a large tail towards higher values,
which in practice makes it less useful for accurate endpoint
determination.  

In this respect a better result is obtained for the $\mqlLow$
distribution, shown in Fig.~\ref{Fig:SelCuts} (right) with an
inconsistancy cut requiring only one jet with $\mqll<440$~GeV, and
normalisation as $\mqLowlLow$.  The red curve matches quite well with
the theory curve, and there is no disturbing tail.  In fact the red
curve is the one we later use to estimate the endpoint for this
distribution.

While it is certainly reassuring to regain the theoretical distribution,
it is still not clear what is the best way to obtain an accurate measurement 
of the endpoints.
One would probably combine many different methods, looking for convergence as
well as inconsistencies.
For instance, even though we should choose not to fit the distribution from 
the inconsistency cut, since it has less statistics, it teaches us 
that a linear fit to $\mqLowlLow$ from 150 to 300~GeV, which might otherwise
seem reasonable, would not at all be a fit to the signal since it is 
close to linear only in the region 240--300~GeV.

Another approach to obtaining a distribution closer to the original one relies 
not on purification of the actual sample, but on finding an estimate for the 
background, combinatorial and/or lepton-correlated, and then subtracting it 
from the original distribution.
Statistics may be better preserved in this way, but systematics are introduced.
Some attempts have been made with `mixed events', i.e.\ the combination of 
the lepton pair from one event with jets from other events. 
The idea is that since the lepton and the jet sectors of mixed events 
are necessarily uncorrelated, they may mimic both types of background, 
lepton-correlated and combinatorial, 
where also the lepton pair is only weakly correlated with the jet.

A first complication of this method is encountered when the 
four-vectors of different events are to be combined into mixed events. 
What rest frame is the appropriate one? 
This already points to the inexact nature of
such a method.  Its performance is however promising, although somewhat 
variable, as will be demonstrated later in this section.

The mixed event sample is constructed from the events which make up the 
distributions, i.e.\ those which pass the precuts listed
in Sect.~\ref{subsect:eventgen}. 
An additional requirement of having exactly two opposite-sign leptons is 
imposed to avoid ambiguities in the lepton sector. No major differences 
are however expected from leaving out this constraint. 
Each event then provides two leptons, uniquely defined because of the 
additional cut, and its two $p_T$-hardest jets. 
The four-vectors of the selected leptons and jets are here taken in the 
rest frame of the lepton pair. Due to the prominent role of the lepton 
pair, this choice is not unreasonable. On the other hand it is not unique. 
Both the laboratory frame, the rest frame of the entire event or the 
rest frame of the leptons together with the two $p_T$-hardest jets 
are viable candidates. A dedicated study of the characteristics of these 
different possibilities would be worthwhile. 

The mixed sample is then constructed by consecutively combining the 
lepton sector of one event with the jet sector of another. 
In our study each lepton sector was combined with the jet sector of 
five other events. By increasing this number, very high statistics can 
be obtained for the mixed sample. 
For each combination of events all relevant masses were constructed, 
$\mqLowll$, $\mqHighll$, $\mqBothll$, $\mqLowlLow$ etc.

Mixed samples for same-flavour and different-flavour events are constructed
separately.  In the end their difference is taken, in line with the previously
described different-flavour subtraction.  The resulting mixed distributions
are the ones used later in this section.

\subsection{Multiple squark masses}

The theory distributions shown in Fig.~\ref{Fig:th-curves} are only for one
squark mass per scenario. In reality four different squark masses,
$\mdL/\msL$, $\muL/\mcL$, $\mbO$ and $\mbT$ will contribute to the 
distribution.
The blue theory distribution of Fig.~\ref{Fig:SelCuts} is therefore the sum 
of four separate distributions, each with a different endpoint, different 
normalisation and similar though not identical shapes.

While jets from the three lightest quarks, and for most purposes also from
the $c$-quark, will be indistinguishable in ATLAS, jets from $b$-quarks 
will be identified with a certain probability.
The expected rejection factors for incorrectly identifying jets from lighter
quarks and gluons as $b$-jets, are for high luminosity operation given 
in the ATLAS TDR \cite{AtlasTDRvol1}, Fig.~10-41.
Low/high $b$-tagging efficiencies come with high/low rejection factors 
and allow for high-purity $b$/non-$b$-samples, respectively. 
The higher the purity, the smaller the sample. 
In this analysis we have used the following simplistic $b$-tagging 
prescription for both purposes:
For a $b$-tagging efficiency of 50\% the rejection factors against
jets from gluons/three lightest jets and from $c$-jets are set to 
100 and 10, respectively.

With $b$-tagging one can to a certain extent separate the 
$\qL$ and the $\sb$ distributions, thus opening for a disentanglement
of the squark masses.
Nevertheless, even though a high purity separation has been accomplished,
each of the two distributions will still contain contributions from two 
squark masses.
Typically a kink can be observed at the position of the lowest endpoint.
For mSUGRA scenarios $\dL/\sL$ and $\uL/\cL$ only differ by a few GeV,
so the kink will appear very near the end of the distribution. 
Since the proton contains more $u$ than $d$-quarks, $\uL$ 
will be produced at a higher rate than the heavier $\dL$. 
This reduces further the visibility of the kink. 
Then with the general smearing due to physics and detector effects 
in addition to background near the endpoint, it may be very difficult 
to identify such a kink.
In case of the two $b$-squarks the separation will be larger. 
Whether it is possible to identify it or not depends on the rate of 
$\sb$ production as well as the level of impurity from $\qL$-events in
the $b$-tagged distribution.

\subsection{Invariant mass distributions}

SUSY processes for \palpha\ and \pbeta\ as well as the 
Standard Model background have been produced for 300 fb$^{-1}$. 
This corresponds to 3 years at design (high) luminosity.
The mass distributions of the available edges for SPS~1a \palpha\ and \pbeta\
are shown in Figs.~\ref{Fig:mcfit-250}--\ref{Fig:mcfit-400}.

For all plots the black points with errors show the total different-flavour-subtracted 
distribution (`SF-DF'). Solid green marks the SUSY background
(`SUSY'), and in green with error bars the Standard Model background (`SM')
is shown.  The solid blue curve then shows the original theory distribution
(`TH'), normalised to the different-flavour-subtracted distribution.  The
fitted function appears in red.  
In cases where mixed events are used to model the background, 
the smooth function fitted to the high-statistics mixed-event 
sample is shown in dashed red. 
When additional distributions are plotted,
they will be described in the accompanying discussion.

For each distribution the endpoint estimation will be discussed. 
In most cases the edges are fitted to a straight line in combination
with a simple background hypothesis, and in some cases convoluted
with a Gaussian distribution.
At \palpha\ this procedure gives numbers in reasonably good agreement 
with the nominal values.
At \pbeta, where the SUSY cross-section is much smaller and also the 
branching fraction of the signal is reduced, the estimated endpoint
values depend more strongly on the fitting method chosen.
To control and reduce this systematic effect, a better understanding 
of the whole chain is required; physics effects, detector effects,
multiple masses at different rates, background, precuts. 
After some years of LHC operation one can expect these issues 
to be understood sufficiently that the systematics of endpoint estimation
is controlled and corrected for in the fitting procedure, 
up to some small uncertainty.

If this is achieved, it is the statistical error of 
the endpoint values, in combination with the uncertainty on the 
absolute energy scale, expected to be 1\% for jets and 0.1\% for leptons, 
which determine the precision at which masses can be obtained 
for the different scenarios.

In the discussion of the distributions we therefore have the main focus on the 
statistical uncertainty of the endpoint determination, but are 
also concerned with the present magnitude of the systematic uncertainty.
For some of the distributions it may seem bold to state that the 
systematic fit uncertainty will be reduced far below the statistical one. 
One shall however remember that once such distributions become 
available, a great effort will go into investigating them.

%%%%%%%%%%%%%%%%%%%%%%%%%%%%%%%%%%%%%%%%%%%%%%%%%%%%%%%%%%%%%%%%%%%%%%
\FIGURE[ht]{
%\begin{figure}
%%Begin InstantTeX Picture
\let\picnaturalsize=N
\def\picsize{7.1cm}
%If you do not have the picture file add:
%\let\nopictures=Y
%to the beginning of the file.
\ifx\nopictures Y\else{
\let\epsfloaded=Y
\vspace*{-.5mm} %may need more when caption grows
\centerline{{\ifx\picnaturalsize N\epsfxsize \picsize\fi
\epsfbox{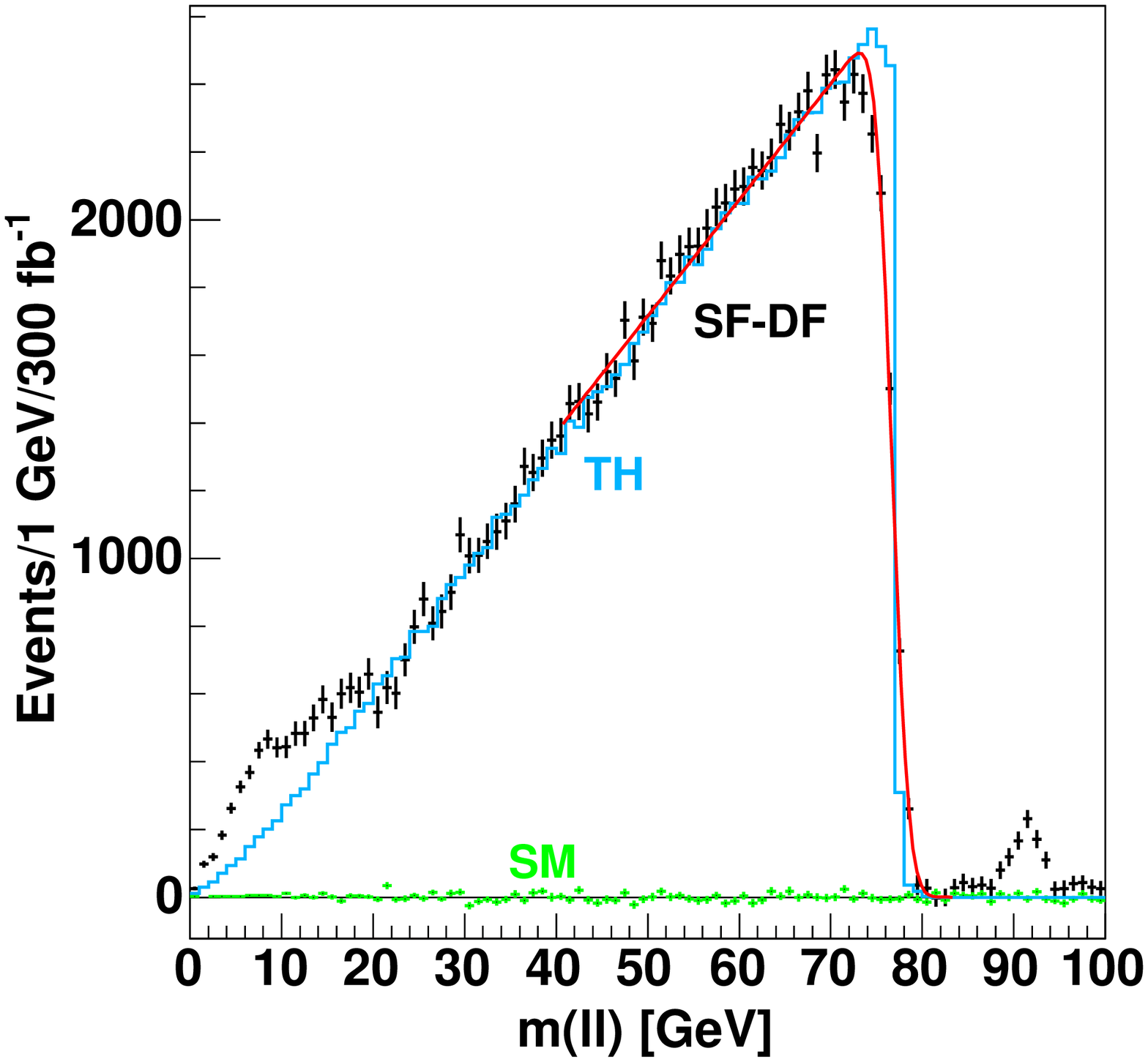}
%\epsfbox{p250_fit_mqll_mixedlow.eps}
            {\ifx\picnaturalsize N\epsfxsize \picsize\fi
\epsfbox{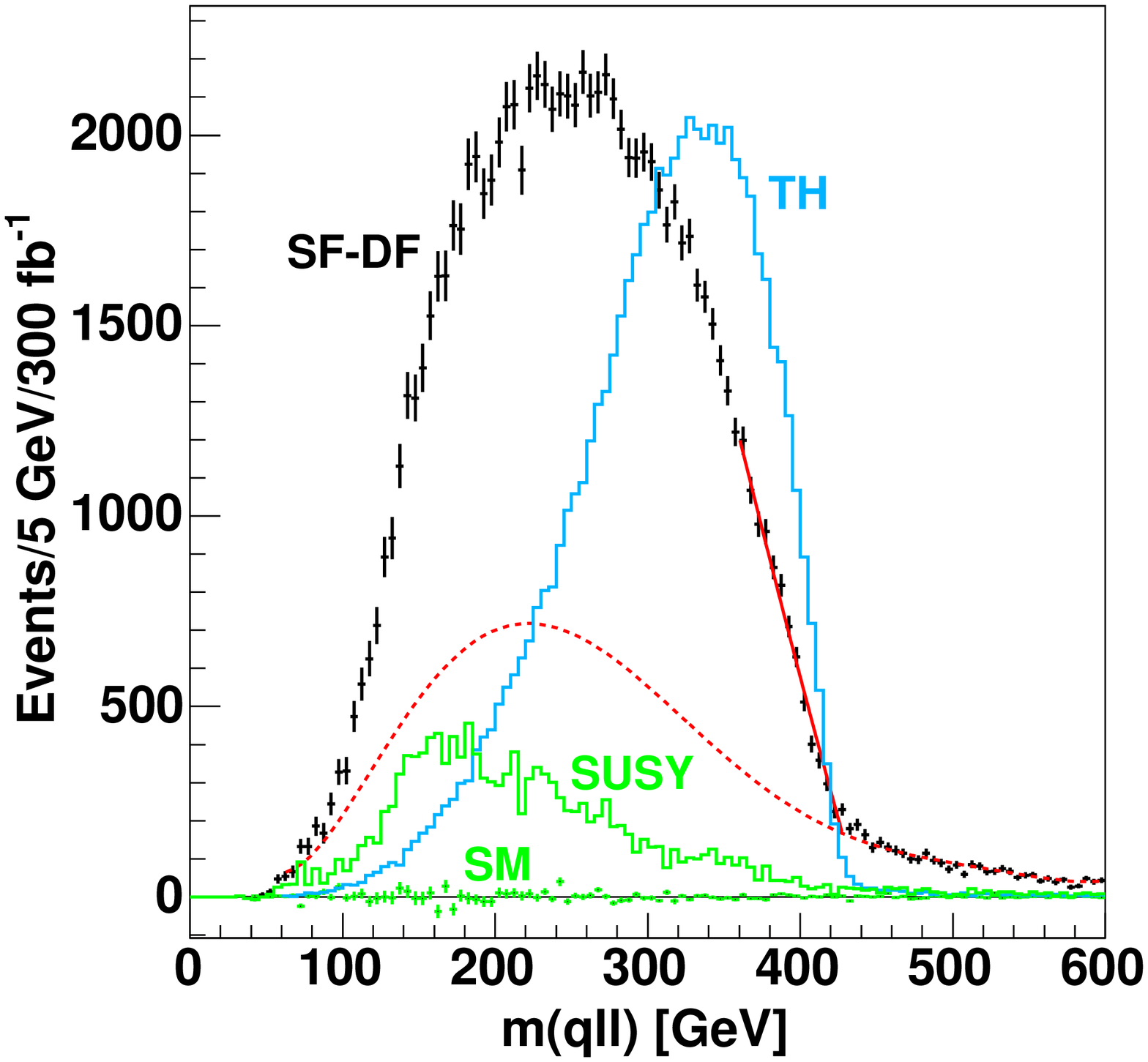}
} } }
%\vspace*{-4mm}
\centerline{{\ifx\picnaturalsize N\epsfxsize \picsize\fi
\epsfbox{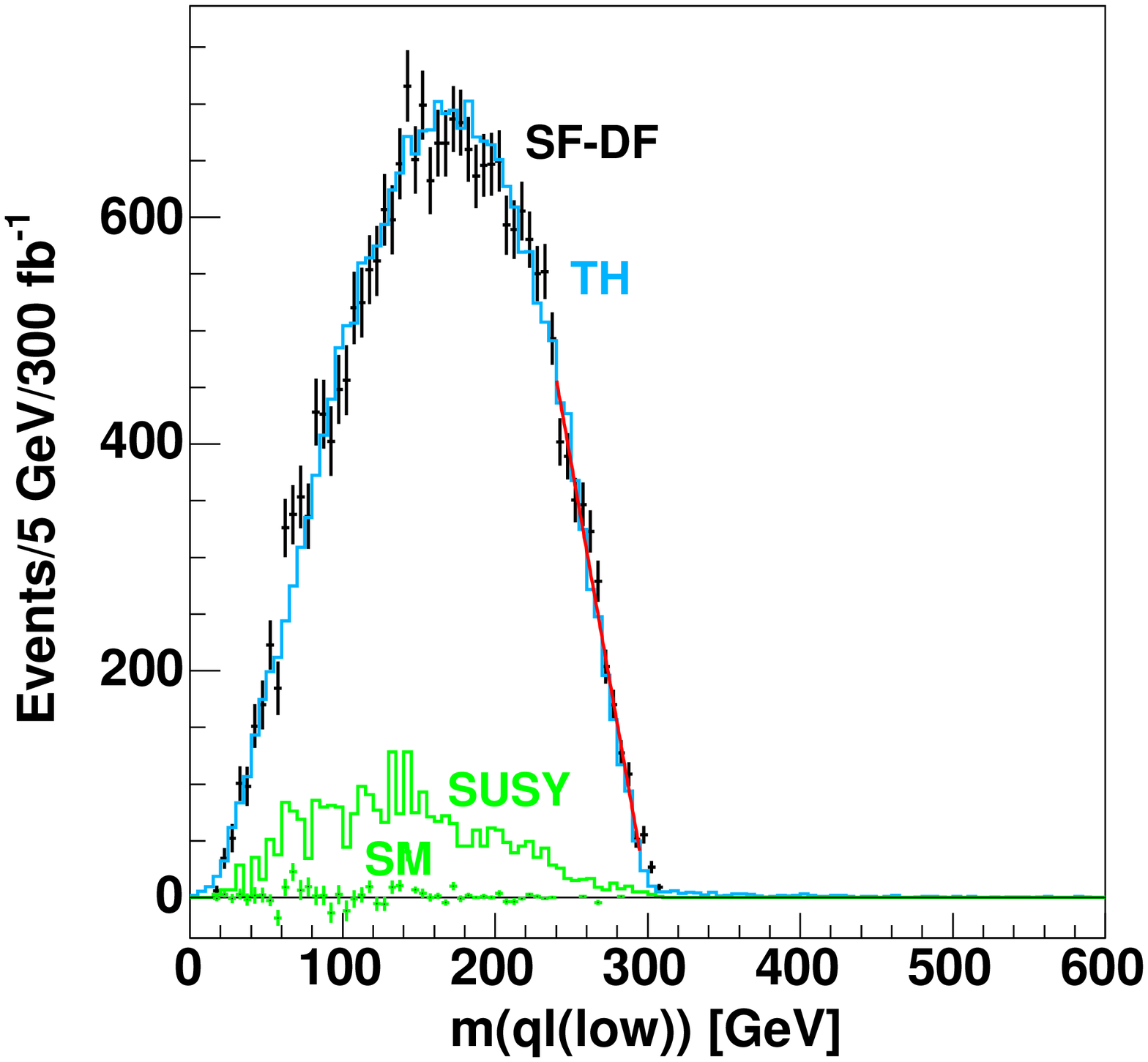}
            {\ifx\picnaturalsize N\epsfxsize \picsize\fi
\epsfbox{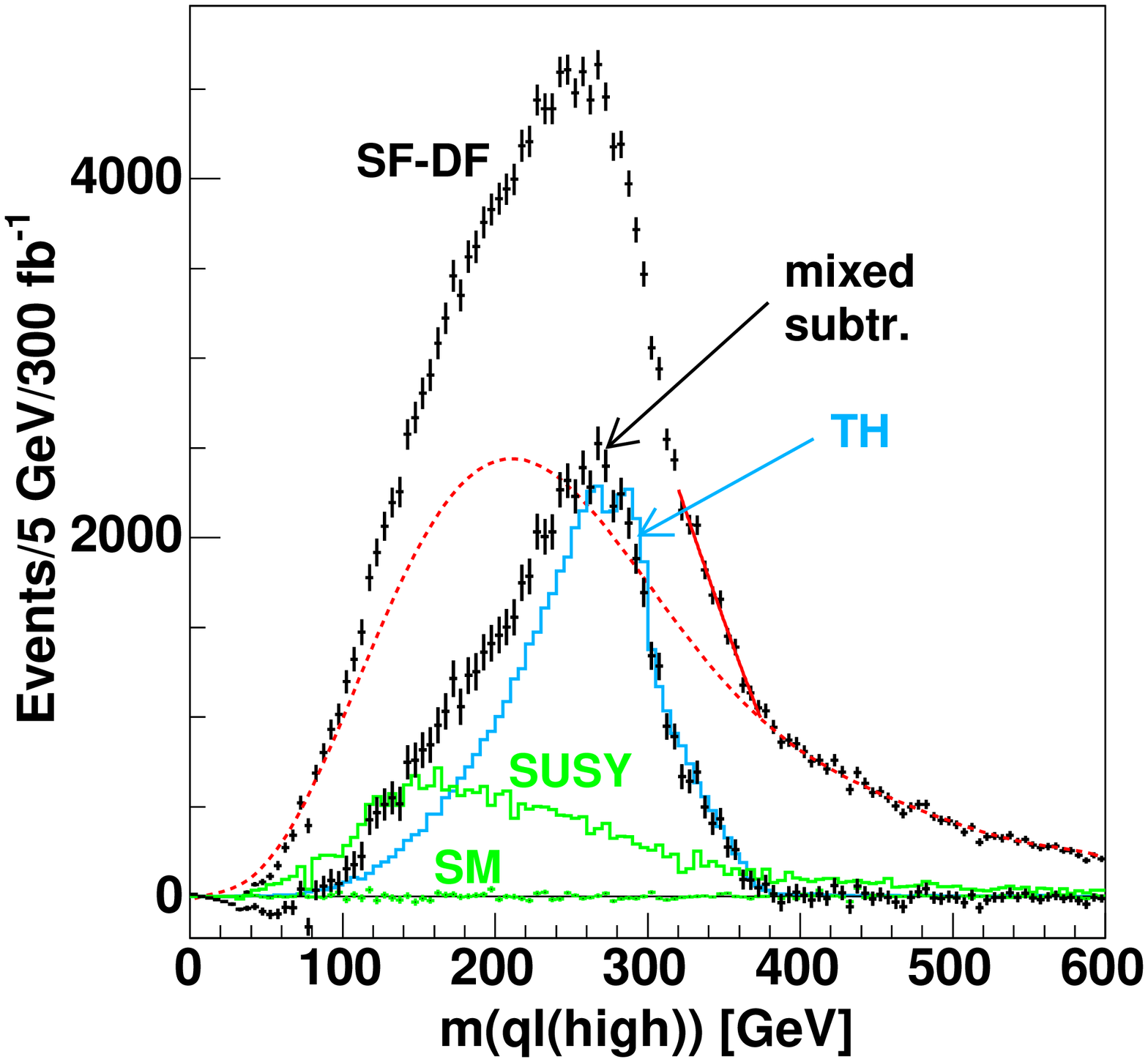}
} } }
\centerline{{\ifx\picnaturalsize N\epsfxsize \picsize\fi
\epsfbox{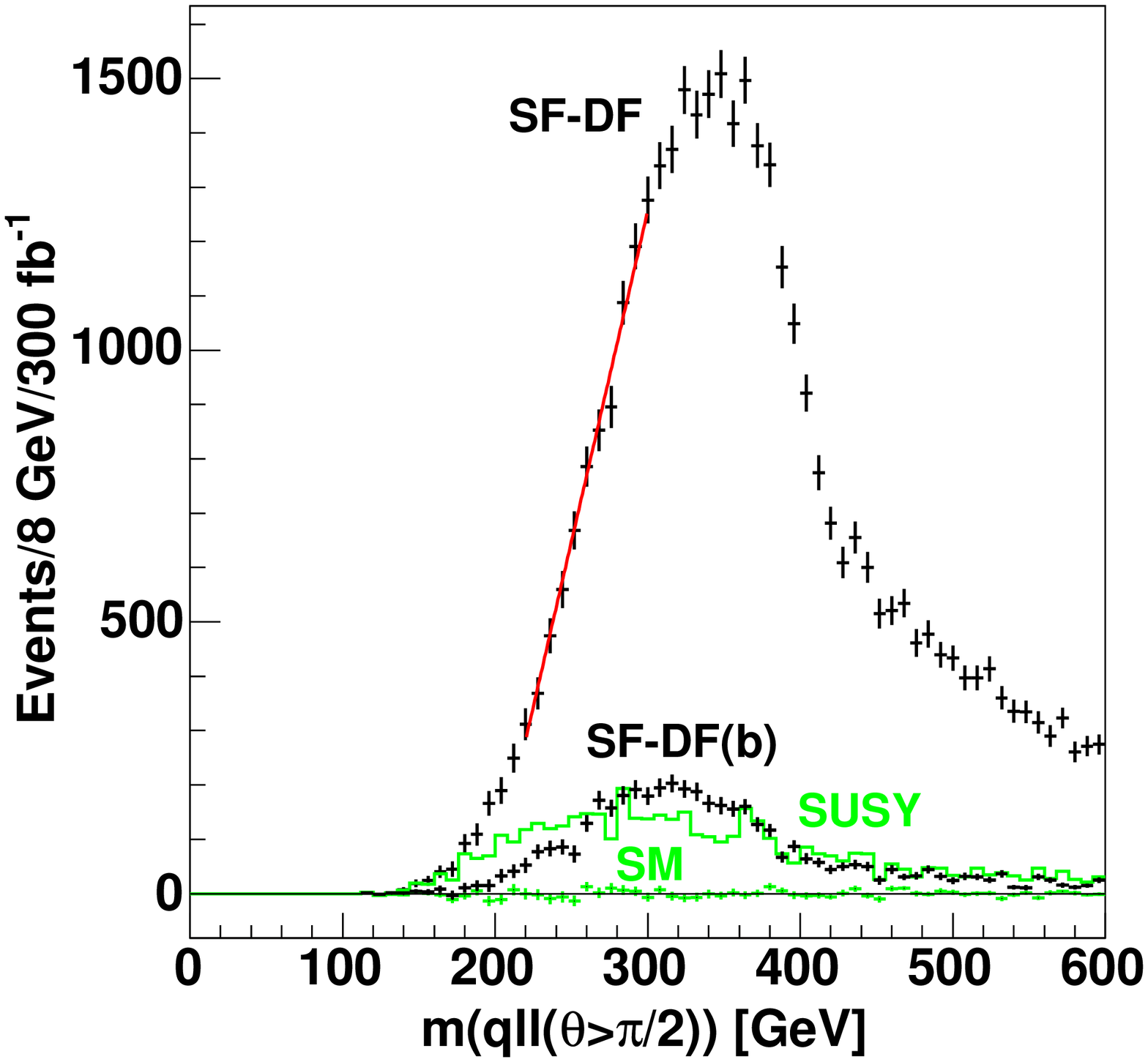}
            {\ifx\picnaturalsize N\epsfxsize \picsize\fi
\epsfbox{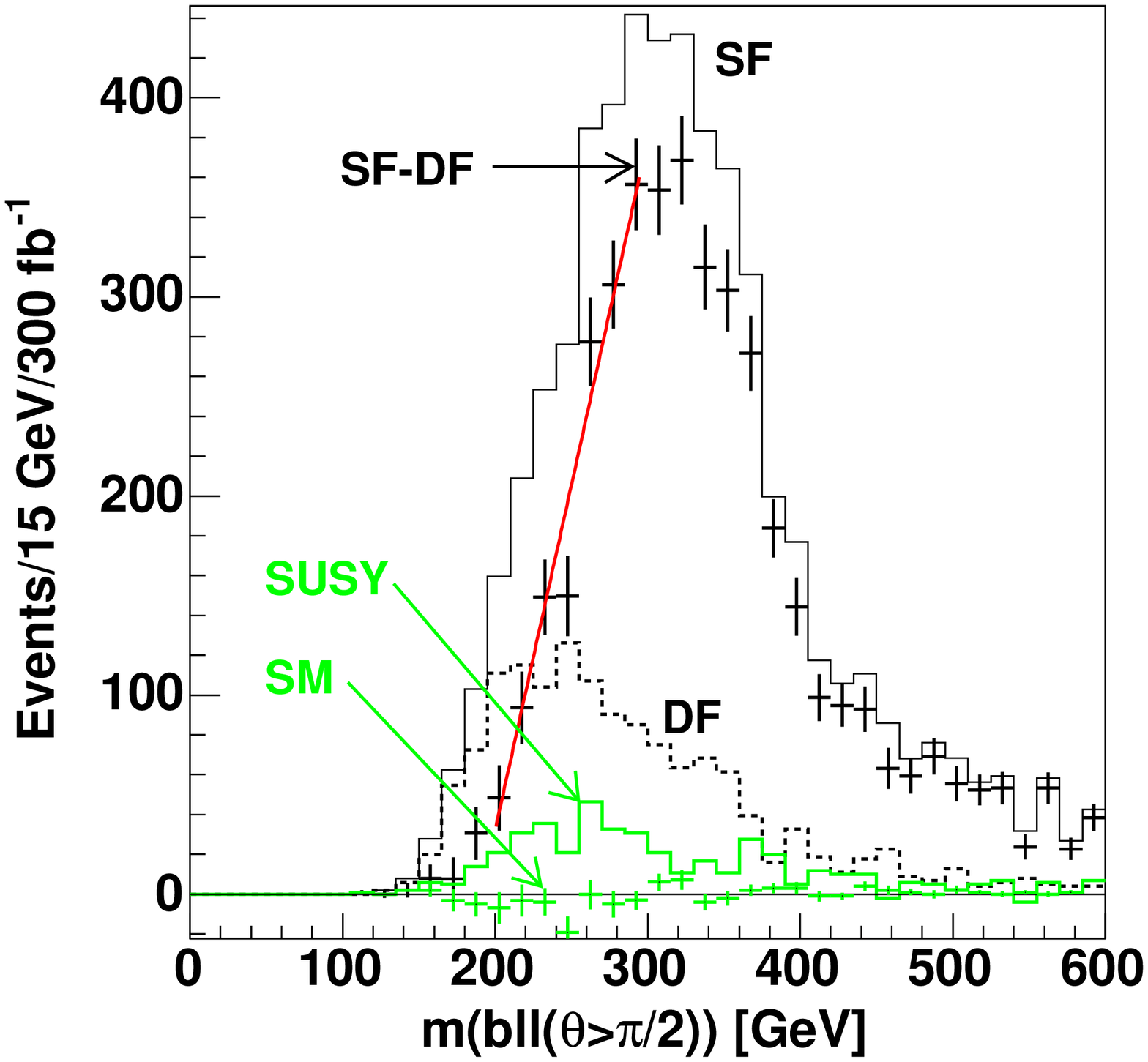}
} } }
}\fi
%%End InstantTeX Picture
\vspace*{-6mm}
\caption{Invariant mass distributions for \SPSOaa. See the 
text for details. \label{Fig:mcfit-250}}} 
%\end{figure}
%%%%%%%%%%%%%%%%%%%%%%%%%%%%%%%%%%%%%%%%%%%%%%%%%%%%%%%%%%%%%%%%%%%%%%

\subsubsection*{\SPSOaa}

\paragraph{$\boldsymbol{\mll}$:}
The $\mll$ distribution at \palpha\ is shown in Fig.~\ref{Fig:mcfit-250} 
(top left). 
It is very close to the expected triangular shape. 
To the right of the endpoint the $Z$ peak is clearly visible. 
These $Z$'s come from SUSY decays. 
The bump at low $\mll$ is discussed in Sect.~\ref{subsect:DFsubtraction}. 

The distribution is fitted with a straight line convoluted with a Gaussian,
\begin{eqnarray}
g(\mll)=\begin{cases}
A\,\mll,\ \mll<\maxmll\\
0,\qquad\mll>\maxmll
\end{cases},\quad
f(\mll)=\frac{1}{\sqrt{2\pi}\sigma}\int_0^\infty d\mll'\, g(\mll')
e^{-\frac{(\mll-\mll')^2}{2\sigma^2}}
\label{Eq:mll-fitfun}
\end{eqnarray}
Both the endpoint, the width of the Gaussian and the normalisation 
are kept free in the fit, 
resulting in an endpoint value of 76.72~GeV at a statistical error 
of 0.04~GeV.
Compared to the nominal value this is 0.35~GeV too low. 
When a more appropriate model for loss of lepton
energy is implemented in the fit function, such a
systematic shift is expected to be reduced to the level
of the statistical error or below.
The fit results are quite stable with respect to variations 
in bin size and range of fit. 
We thus arrive at the well-known result that the $\mll$ endpoint 
is expected to be estimated to high precision.

\paragraph{$\boldsymbol{\mqll}$:}
As for all cases where jets are involved, the $\mqll$ distribution
comes in many versions, depending on the jet selection procedure as well
as the use of $b$-tagging information.
Plotted in Fig.~\ref{Fig:mcfit-250} (top right) is $\mqLowll$ 
(see Sect.~\ref{subsect:selection-cuts}) with no $b$-veto.
Mixed events have been used in an attempt to model the background. 
A $6^\textrm{th}$ degree polynomial 
was first fitted to the high statistics mixed event sample. 
Then, with the normalisation kept free, the polynomial 
was combined with a straight line for the signal part. 
This procedure returns \mbox{$\maxmqll = 427.7(0.9)$~GeV},  
statistical error in parenthesis. 
The nominal value is 425.9/431.1~GeV for $\uL/\dL$. 
Alterations to the bin size and fit range has a 1--2~GeV effect on the 
fit value, but the statistical error remains the same. 

The dashed red line shows the shape of the mixed event sample. 
It matches the $\mqLowll$ distribution very well at values beyond 
the endpoint, justifying its use, 
and is sufficient to obtain a reasonable endpoint fit. 
Similar results can be found by modeling the background tail with an 
exponential or a polynomial. 
If, at some later stage, the entire distribution is to be compared with 
the theory distribution, the size of the background must be known for any 
$\mqll$ value. 
An exponential or a polynomial based on the upper tail can never 
be a good background model for values somewhat below the edge. 
It is here the mixed events may have a role to play. 
In the particular case of the $\mqLowll$ distribution, the mixed sample 
actually appears to be a good background estimate at all values. 
For $\mqBothll$ and some other distributions, it does not do so well. 
Nevertheless, there are many ways to construct the mixed sample. 
Further study of the method seems worthwhile. 

It is tempting to suggest that it is mainly the $\uL/\cL$ endpoint we measure, 
and that the extra events just above the endpoint (see the figure), 
make up the hard-to-detect kink from the $\dL/\sL$ edge. 
At the present level of detail it is impossible to say if such an 
effect could be isolated to give an additional measurement for 
the heaviest mass. 
It does seem difficult, though, as the kink will be washed out by 
other effects, e.g.\ sparticle widths. 
At \palpha\ the intrinsic width of $\uL$ is 5.3~GeV, 
see Fig.~\ref{fig:masses-line} (right), to be compared with 
$\mdL-\muL=5.8$~GeV.
Also detector effects will result in a general smoothing of the distributions.

\paragraph{$\boldsymbol{\mqlLow}$:}
For $\mqlLow$ the mixed sample is fine above the endpoint, but 
overestimates for lower masses. 
While the edge can be fitted and the endpoint measured by a 
mixed sample subtraction, due to its good behaviour in the edge region,
we have here instead used the inconsistency cut $\mqHighll>440$~GeV 
to purify the sample, 
Fig.~\ref{Fig:mcfit-250} (middle left). 
The ability of the inconsistency cut to bring the distribution 
very close to the original one, was already discussed in relation to 
Fig.~\ref{Fig:SelCuts}. 

For a zero background hypothesis 
a straight line fit gives $\maxmqlLow = 300.7 (0.9)$~GeV,
to compare with the nominal 298.5/302.1~GeV for $\uL/\dL$. 
If the few bins around the endpoint are also included, the value
increases by 1--2~GeV. Whether these high-mass events are signal 
or background is not easy to tell from the given distribution, 
since there is virtually no background structure to extrapolate from.
This is because a `consistency cut' has been imposed which requires 
$\mqLowll<440$~GeV, in accordance with the already measured $\mqll$ endpoint. 
From Eq.~(\ref{Eq:sqll=sql+sql+ll}) this implies 
$\mqlLow<440/\sqrt{2}~\GeV=311.1~\GeV$.
The consistency cut takes away a large part of the background, but also 
has the effect that it becomes difficult to see what structure 
the $\mqlLow$ background has. 
If the consistency cut is dropped, the usual background tail appears 
and can be modeled, although at the cost of a slight increase in 
the statistical error.

\paragraph{$\boldsymbol{\mqlHigh}$:}
Following the same procedure as for $\mqll$ the background of the 
$\mqlHigh$ distribution was modeled by the mixed event sample. 
In Fig.~\ref{Fig:mcfit-250} (middle right) the relevant distributions are
shown.  Also the result from subtracting the mixed sample is shown (`mixed
subtr.'); the lower black points with error bars.  This subtracted
distribution follows the original theory distribution (blue) closely in the
edge region, but overestimates at lower values.

In the range $\mqlHigh\in(320,550)$~GeV 
the endpoint was estimated 
to $374.0 (2.0)$~GeV. 
The nominal value is 375.8/380.3~GeV for $\uL/\dL$. 
At \palpha\ the $\mqlHigh$ edge consists of two parts. This is clear 
from the theory distribution (blue), but also in the reconstructed
distributions (black). In the fit only the lowest near-linear stretch
was used.
It is clearly incorrect to apply a straight line for the whole edge, but
if done, the statistical error would be reduced to $\sim1$~GeV. When a good signal 
function is at hand, the whole edge will be described. 
It may therefore be reasonable to expect a statistical 
error of 1~GeV rather than 2~GeV.

\paragraph{$\boldsymbol{\mqllThres}$:}
The $\mqllThres$ distribution differs from the previous ones in that a minimum
is to be measured. This has two important consequences. One can be seen from
Fig.~\ref{Fig:OSOF-subtr} (bottom left), even though this plot is for the
unconstrained $\mqll$ distribution.  The lepton-uncorrelated background,
modeled by the different-flavour distribution (dashed red), sits mostly at low
values, far away from any maximum edge, but quite near the $\mqllThres$
threshold.  While this background type is removed (statistically) by the
different-flavour subtraction, the price is an increased statistical
uncertainty which directly will affect the statistical precision of the
$\minmqllThres$ estimation.

The other consequence of measuring a minimum is related to 
the multiple squark masses. 
As noted earlier, the distributions can be purified 
using $b$-tagging information. For the measurements discussed so far we have 
considered the total sample. This is because the $\qL$-squarks are so 
much heavier than the $\sb$-squarks that the upper edge regions are practically free 
of $\sb$-events anyway. A $b$-veto has therefore no effect, 
at least not at our level of precision.
For the threshold measurement it is the other way around. 
Here the threshold related to the lightest squark, $\bO$, will be the 
threshold for the total distribution. 
A $b$-veto may therefore be important in order to obtain the threshold for 
the non-$\sb$-events.

A third complication is that the edge is very non-linear, as can be 
seen from the theory distributions in Fig.~\ref{Fig:th-curves}. 
Also, the fact that this non-linearity is of the concave type, 
could make it very difficult
to see where the edge ends and where smearing from various sources takes over.
This is all the more relevant since the $\sb$-events come
at a significantly smaller rate than $\qL$-events (ratio $\sim25\%$), 
and may therefore appear as just another contribution to the lower tail,
not easily distinguishable from other effects.

The $\mqllThres$ distribution is shown in Fig.~\ref{Fig:mcfit-250} (bottom
left).  The upper distribution (`SF-DF') shows the non-$b$-tagged $\mqllThres$
sample.  The lower black points (`SF-DF($b$)') mark the $b$-tagged sample.
Only approximately 50\% of the events which contain a $b$ are actually
$b$-tagged.  This means that the non-$b$-tagged distribution contains a
`hidden' $b$-sample, statistically similar to the `visible' $b$-sample both in
size and shape.  (There would be a shift to the left, though, since in the
reconstruction $b$-tagged jets have larger recalibration factors. This is not
accounted for here.)  Analogous to the different-flavour subtraction it is
therefore possible to get the non-$b$-tagged distribution closer to the
original $q$-distribution by subtracting the visible $b$-sample, possibly
scaled up or down according to the $b$-tagging efficiency applied.
A slightly different approach would be to use a much harder $b$-veto and 
no subtraction. The price would be a significant reduction in sample size. 

However, the main problem in the threshold region is not the  
$b$-contamination, but the lepton-correlated SUSY background shown in 
solid green. 
It comes primarily from events where $\NT$ decays sleptonically 
but does not descend from $\qL$ or $\sb$. 
It is this background which needs appropriate modeling. 
The difficulty with having a concave theory distribution together with an 
unknown background is evident from the figure.
It is really difficult to separate the background from the signal.

As a simple estimate of the statistical uncertainty of the edge position, 
a straight line fit somewhat away from the threshold 
was performed on the non-subtracted sample (`SF-DF'), 
returning a statistical error on the endpoint around $1.8$~GeV. 
If the subtracted sample (subtracting `SF-DF($b$)') is used, 
the error increases. 
The actual endpoint obtained from such a simple zero-background hypothesis 
is of course incorrect. 
At present it seems a bit optimistic to expect the systematics of the fit 
to be dominated by the statistical error.

\paragraph{$\boldsymbol{\mbllThres}$:}
The $\mbllThres$ distributions are shown in Fig.~\ref{Fig:mcfit-250} 
(bottom right). The different-flavour-subtracted distribution (`SF-DF') 
is shown in black with error bars. 
The same-flavour (`SF') and different-flavour (`DF') curves are also plotted.  
In the threshold region the same-flavour and different-flavour distributions 
are considerably larger than their 
difference, which necessarily gives large statistical uncertainties
in the different-flavour-subtracted samples.

As for $\mqllThres$ more studies are required to control and reduce 
the systematic error induced by the background and the nonlinear theory curve.
A linear fit was performed in the near-linear region 
to measure the precision with which the position of the edge can be found. 
Variations in the bin size and the fit range yield systematic errors within
1~GeV and statistical error between 4 and 5~GeV.

\paragraph{$\boldsymbol{\mbll,\ \mblLow,\ \mblHigh}$:}
Given that the background can be controlled and a good signal function can 
be found, $\mbllThres$ is a good distribution since the lowest 
threshold is given by $\bO$, which also has a considerable rate.
For the edges $\mbll$, $\mblLow$ and $\mblHigh$ one problem is that it 
is $\bT$ which defines the outer endpoint for the $b$-distributions. 
Since this squark only contributes $\sim 22\%$ of the $\sb$-events, 
the edge above the $\bO$ endpoint may look like background.
(However, since we expect the heavier $\sb$-squark to come at a considerably 
smaller rate as it has less phase space for production, and also
a smaller branching ratio to the wino-like $\NT$, due to its 
larger right-handed component, we can explicitly look for such a 
small tail-like edge.)

A more serious problem is the background from other SUSY events. 
Typically these have the decay Eq.~(\ref{eq:squarkchainq}) in one 
chain, providing two leptons, and $\gl\to\sb b$ or $\gl\to\tO t$ 
in the other, providing $b$-jets. 
The mass distributions of this background stretch sufficiently 
beyond the $\bO$ endpoints that the resulting edge structures 
become difficult to analyse. 
The positions of these three $b$-edges can
typically be measured with a statistical precision of 3--4~GeV, but it
remains unclear whether the systematics of the edge can be
sorted out.
Another approach could be to require a lower $b$-tagging efficiency 
for these distributions in order to get higher rejection factors 
and purer, although smaller, samples.

%%%%%%%%%%%%%%%%%%%%%%%%%%%%%%%%%%%%%%%%%%%%%%%%%%%%%%%%%%%%%%%%%%%%%%
\FIGURE[ht]{
%\begin{figure}
%%Begin InstantTeX Picture
\let\picnaturalsize=N
\def\picsize{7.1cm}
%If you do not have the picture file add:
%\let\nopictures=Y
%to the beginning of the file.
\ifx\nopictures Y\else{
\let\epsfloaded=Y
\vspace*{-1mm}
\centerline{{\ifx\picnaturalsize N\epsfxsize \picsize\fi
\epsfbox{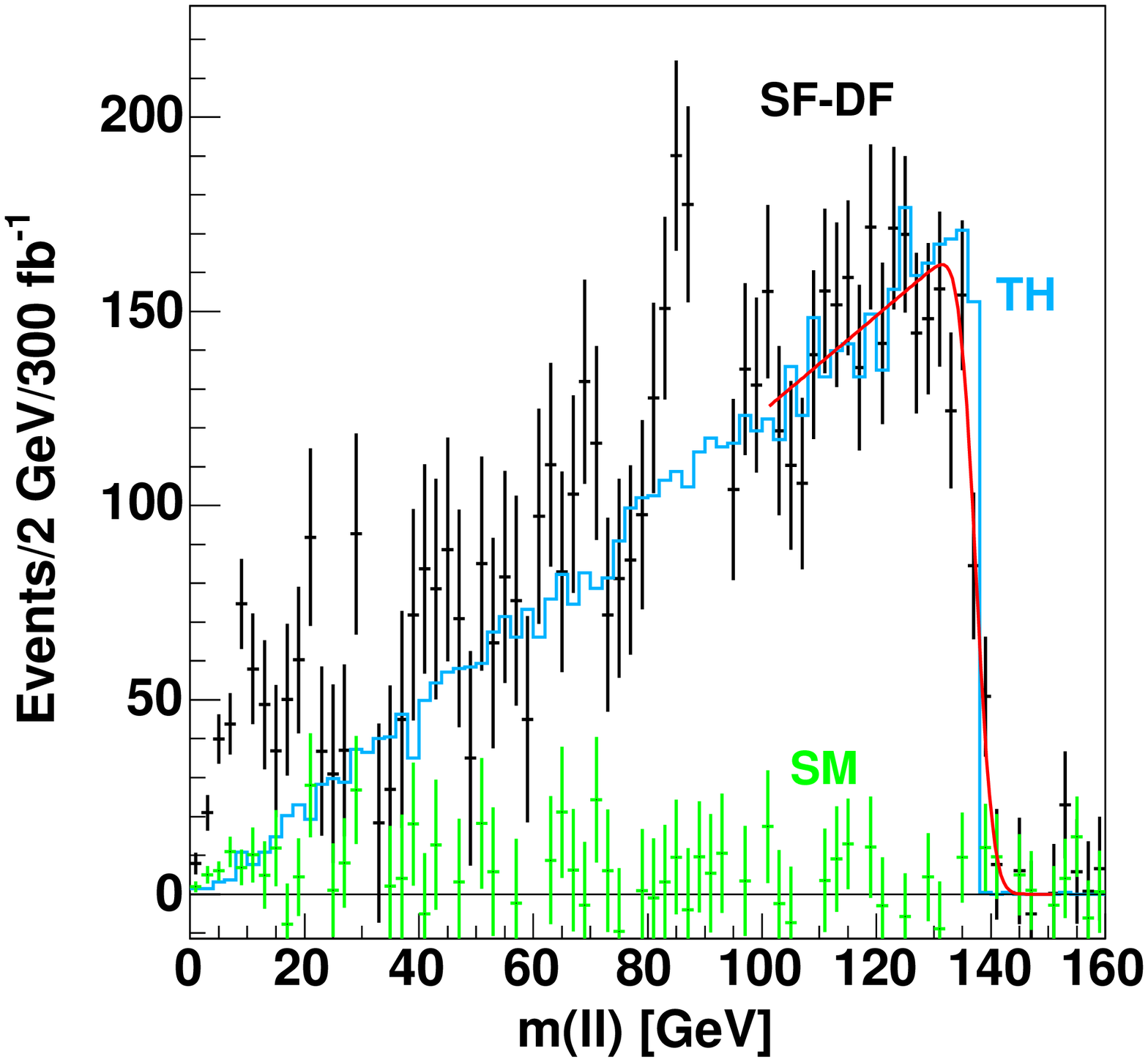}
            {\ifx\picnaturalsize N\epsfxsize \picsize\fi
\epsfbox{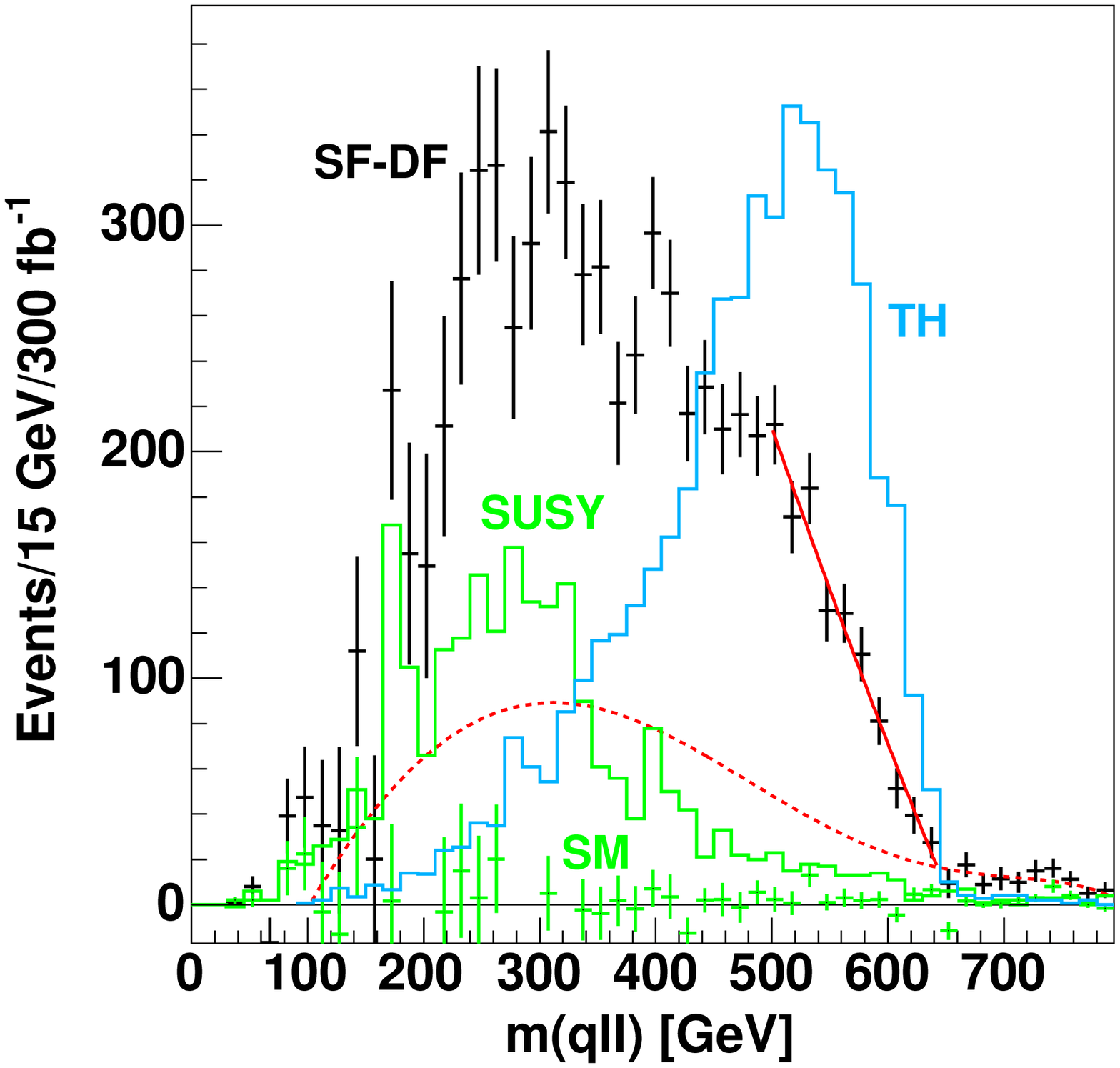}
} } }
%\vspace*{-4mm}
\centerline{{\ifx\picnaturalsize N\epsfxsize \picsize\fi
\epsfbox{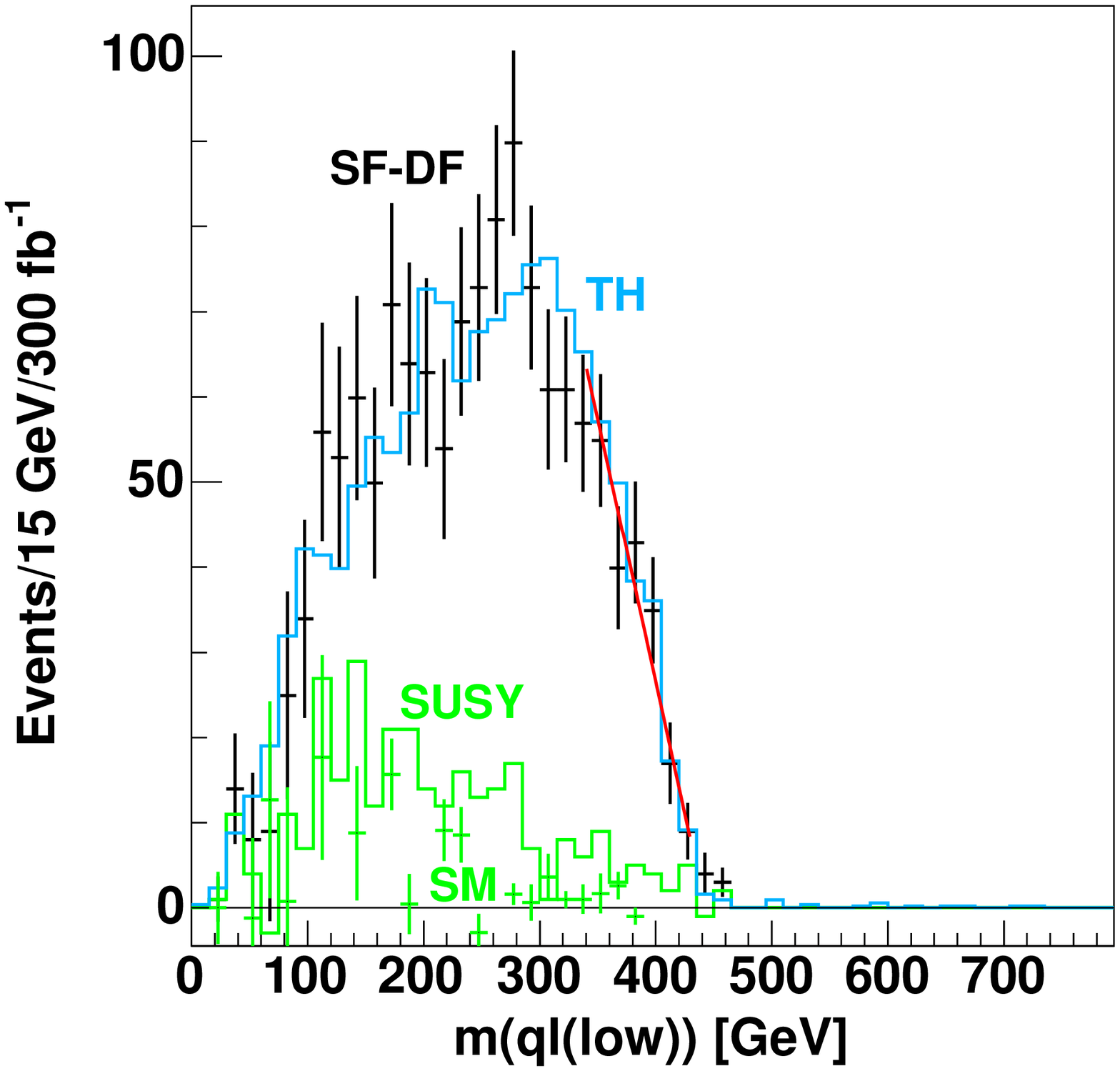}
            {\ifx\picnaturalsize N\epsfxsize \picsize\fi
\epsfbox{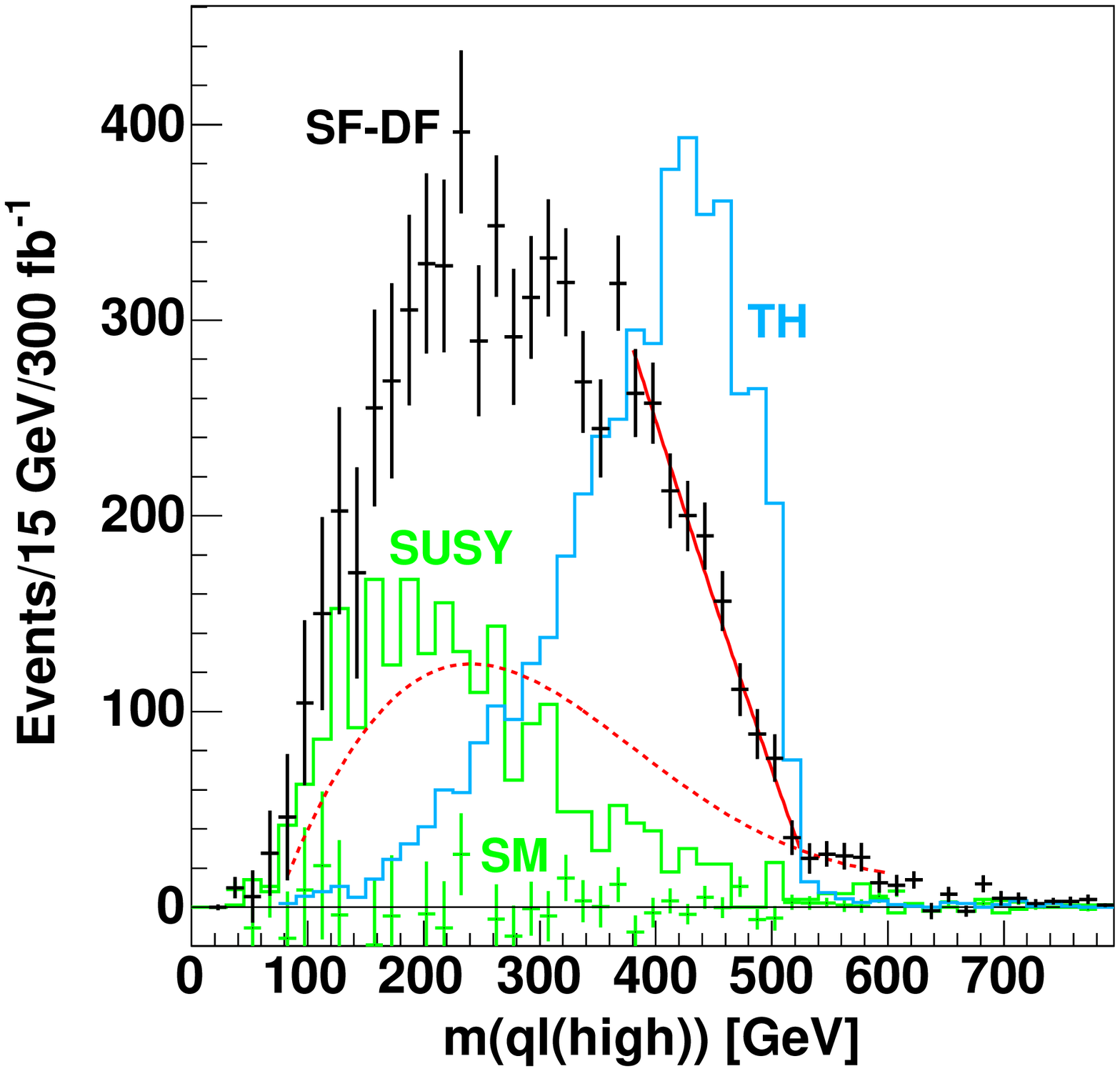}
} } }
\centerline{{\ifx\picnaturalsize N\epsfxsize \picsize\fi
\epsfbox{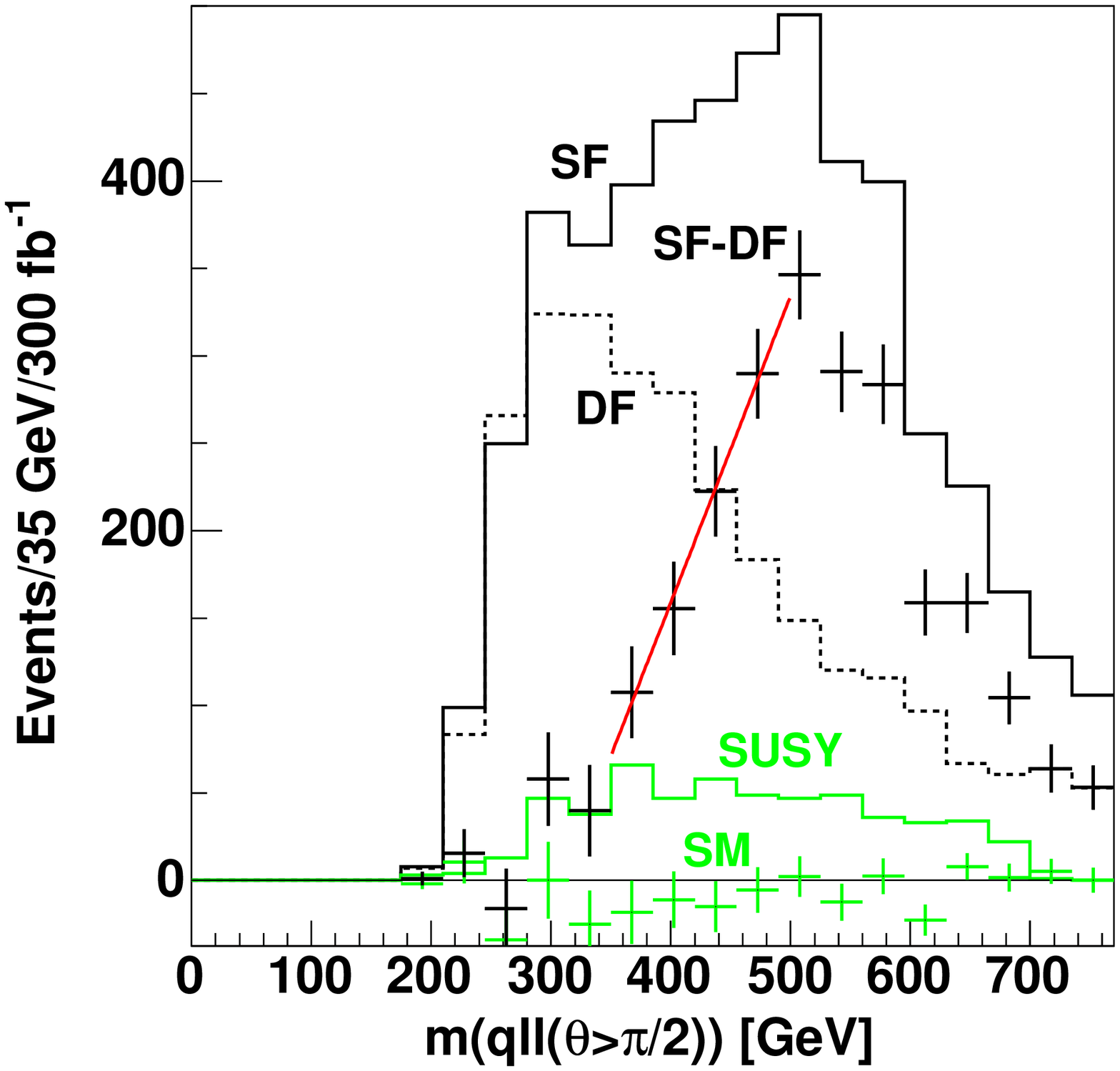}
} }
}\fi
%%End InstantTeX Picture
\vspace*{-6mm}
\caption{Invariant mass distributions for \SPSOab. See the 
text for details. \label{Fig:mcfit-400}}}
%\end{figure}
%%%%%%%%%%%%%%%%%%%%%%%%%%%%%%%%%%%%%%%%%%%%%%%%%%%%%%%%%%%%%%%%%%%%%%

\subsubsection*{\SPSOab}
The mass distributions of \pbeta\ are shown in 
Fig.~\ref{Fig:mcfit-400}.
Throughout, the total samples have been used. 
No $b$-veto has been applied.

\paragraph{$\boldsymbol{\mll}$:}
The $\mll$ distribution is shown in Fig.~\ref{Fig:mcfit-400} (top left).
It has the triangular shape with a $Z$-peak on top.
Some datapoints in the $Z$-peak lie above the range of this plot, 
and can be viewed in Fig.~\ref{Fig:OSOF-subtr} (top right).

As was also the case for \palpha\ there is a bump at lower values. 
The distribution above the $Z$-peak was fitted with the same 
Gaussian-convoluted triangular shape as earlier, Eq.~(\ref{Eq:mll-fitfun}),
giving $\maxmll=137.4(0.5)$~GeV which is 0.5~GeV lower than the 
nominal value.

\paragraph{$\boldsymbol{\mqll}$:} 
The $\mqLowll$ distribution is shown in Fig.~\ref{Fig:mcfit-400} (top
right).  With the mixed sample as a rough background estimate a
straight-line fit gives an endpoint at 640~GeV with a systematic shift
of 1~GeV from varying the bin size and fit range. The statistical
error is 5~GeV.  If, however, the same fitting procedure is
applied to the $\mqBothll$ distribution, the endpoint value increases
to $655\pm2$~GeV and the statistical error is 9~GeV.

Surely more study would bring these values closer, and optimally 
have them converge near the nominal value of 649.1/652.5~GeV for $\uL/\dL$. 
For later use we take an optimistic statistical error of 5~GeV, 
but also include a systematic fit error of 3~GeV for a more 
conservative estimate.

\paragraph{$\boldsymbol{\mqlLow}$:}
The $\mqLowlLow$ distribution is shown in Fig.~\ref{Fig:mcfit-400} 
(middle left). Both a consistency cut $\mqLowll<670$~GeV and 
an inconsistency cut $\mqHighll>670$~GeV are used.
A straight-line fit with no background hypothesis gives a statistical 
error around 6.3~GeV. The actual fit value is 443~GeV, which 
overshoots by a few GeV since the background plateau has not 
been included. 
The nominal value is 436.6/438.9~GeV for $\uL/\dL$.

\paragraph{$\boldsymbol{\mqlHigh}$:}
The $\mqLowlHigh$ distribution with the consistency and the 
inconsistency cut is shown in Fig.~\ref{Fig:mcfit-400} 
(middle right). The mixed sample was again used to roughly 
model the background under the edge. 
A straight-line fit gives an average value of 520.5~GeV with 
systematics from binning and fit range of 3~GeV.
The statistical error is at 5.5~GeV.

The nominal value is 529.9/532.7~GeV for $\uL/\dL$, 
some 10~GeV above our estimate.
One could argue that the current endpoint measurement has 
considerable uncertainties and that this discrepancy is not 
dramatic at the present level of detail. 
However, such an underestimation is actually to be expected.
In Fig.~\ref{Fig:th-curves} the theoretical $\mqlHigh$ distribution 
is shown for $\uL$ at \pbeta. 
There is a long vertical fall towards 517~GeV 
(for $\dL$ it is at 519~GeV), then just before the bottom is reached, 
a small foot appears, as anticipated in Sect.~\ref{subsect:theorycurves}, 
and takes us up by 11--13~GeV. 
To detect such a small foot would require more statistics than is 
available at \pbeta.
Experimentally it is therefore expected to get an endpoint near 517--519~GeV. 
The incorrect endpoint measurement will have important consequences 
for the determination of masses from the endpoints. 
This situation is further discussed in Section~\ref{sect:masses-from-edges}.

\paragraph{$\boldsymbol{\mqllThres}$:}
The $\mqllThres$ distribution is shown in Fig.~\ref{Fig:mcfit-400} (bottom).
The same-flavour (`SF') and different-flavour (`DF') distributions are shown
in solid and dashed black. Clearly, there are large uncertainties from the
different-flavour subtraction.  The statistical uncertainty of the
different-flavour-subtracted sample was estimated with a straight-line fit,
giving 13~GeV.  In addition there will be a systematic error, here
conservatively set to 10~GeV.

\subsubsection*{Endpoint measurement values}

The results of the endpoint estimation for \palpha\ and \pbeta\ 
are summarized in Table~\ref{TABedges}. 
The last column contains an estimate of the systematic error 
from different fitting techniques, ranges and bin widths.
These values are not used in the following, but are included 
for completeness. 

\begin{TABLE}{
\caption{Endpoint values found from fitting the edges in 
Figs.~\ref{Fig:mcfit-250}--\ref{Fig:mcfit-400}, for 300~$\text{fb}^{-1}$. 
The nominal values correspond to the mass of $\uL$, which 
due to the proton content
is produced at higher rates than the heavier $\dL$.
For the thresholds no fit values are shown, 
only the errors. This reflects the fact that a reasonable fit 
function is lacking for this edge.}
\label{TABedges}
\begin{tabular}{|lccccc|}
\hline
  & Nominal & Fit & Energy Scale & Statistical & Syst.\ Fit\\ 
Edge & Value  & Value &Error ($\sigma^\scale$) & Error ($\sigma^\stat$) & Error \\
  &  [GeV] & [GeV] & [GeV] & [GeV] & [GeV] \\
\hline
\palpha\        &&&&&\\
$\maxmll$       &  77.07  &  76.72  &  0.08 &  0.04 &  0.1 \\
$\maxmqll$      & 425.9   & 427.7   &  2.1  &  0.9  &  0.5 \\
$\maxmqlLow$    & 298.5   & 300.7   &  1.5  &  0.9  &  0.5 \\
$\maxmqlHigh$   & 375.8   & 374.0   &  1.9  &  1.0  &  0.5 \\
$\minmqllThres$ & 200.7   &  -      &  1.0  &  2.2  &  2.0 \\
$\minmbllThres$ & 183.1   &  -      &  0.9  &  4.5  &  4.0 \\[0.13cm]
\hline
\pbeta\         &&&&&\\
$\maxmll$       & 137.9   & 137.4   &  0.14 &  0.5  &  0.1 \\
$\maxmqll$      & 649.1   & 647.0   &  3.2  &  5.0  &  3.0 \\
$\maxmqlLow$    & 436.6   & 443.0   &  2.2  &  6.3  &  4.0 \\
$\maxmqlHigh$   & 529.9   & 520.5   &  2.6  &  5.5  &  3.0 \\
$\minmqllThres$ & 325.7   &  -      &  1.6  & 13.0  & 10.0 \\[0.13cm]
\hline
\end{tabular} }
\end{TABLE}

The column with the heading `Energy Scale Error' shows the expected 
error on each endpoint estimation from the uncertainty on the absolute energy 
scale for jets and leptons. 
This effect has not been taken into account in the simulation.
The uncertainties of the energy scales are here set to 1\% for jets and 
0.1\% for electrons and muons, see Ch.~12 of \cite{AtlasTDRvol1}. 
For an invariant mass which consists of only jets or only leptons, 
this will give the same uncertainties, 1\% and 0.1\%, respectively.
If the invariant mass is constructed from one jet and one lepton, the 
endpoint uncertainty is 
\begin{equation}
\frac{\sigma(\mql)}{\mql} = \frac{\sigma(\mql^2)}{2\mql^2}
= \frac{1}{2}\sqrt{\left(\frac{\sigma(E_j)}{E_j}\right)^2 
+ \left(\frac{\sigma(E_l)}{E_l}\right)^2}
= 0.50\%
\end{equation}
where $E_j$ and $E_l$ are the jet and the lepton energies, respectively.
For an invariant mass involving a higher number of jets and leptons, 
the error on the endpoint value from the energy scale uncertainty 
will be different for each event.
The error of $\mqll$ will depend on the relative size of the three terms on 
the right-hand side of Eq.~(\ref{Eq:sqll=sql+sql+ll}).
Since at \palpha\ and \pbeta\ we are in the region where the mass 
ratio $\mqL/\mNT$ dominates the two other mass ratios, 
see Eq.~(\ref{Eq:edge-qll-max})-{\itB(1)}, the quark will usually 
be very energetic, leaving one or both $\mql$ terms to dominate. 
This is particularly true at large values, so near the edge of $\mqll$ 
one can show that the energy scale error will result in an endpoint error 
between 0.35\% and 0.5\% for each event. 
For $\mqllThres$ the average energy scale error will be slightly lower 
in our two scenarios. 
For simplicity we have set the energy scale error to 0.5\%  
for all endpoints involving jets.

%%%%%%%%%%%%%%%%%%%%%%%%%%%%%%%%%%%%%%%%%%%%%%%%%%%%%%%%%%%%%%%%%
\section{Extraction of masses from edges}\label{sect:masses-from-edges}
%%%%%%%%%%%%%%%%%%%%%%%%%%%%%%%%%%%%%%%%%%%%%%%%%%%%%%%%%%%%%%%%%

\subsection{10,000 ATLAS experiments}

In the simulation study described in Sect.~\ref{sect:data-gen-recon} values
for the endpoints and their statistical uncertainties were found, together
with a `systematic fit uncertainty'.  Although not so far from the nominal
endpoint values, the fit values in Table~\ref{TABedges} are somewhat uncertain
due to the as yet not-understood systematics of the fitting procedures.  Also,
the systematic error on the energy scale has not been addressed.

Assuming that one will eventually be able to control the systematics of the
fitting, only the statistical errors together with the systematic error from
the energy scale uncertainty will be what determine the LHC potential to
measure the SUSY masses.  To estimate this potential, consider an ensemble of
typical LHC experiments, i.e.\ where the deviation of each endpoint
measurement from the nominal value is based on a Gaussian distribution of
width equal to the statistical error estimated for that endpoint, as well as a
jet/lepton energy scale error picked from a Gaussian distribution for each
experiment, in line with what is done in \cite{Allanach:2000kt},
\begin{equation}
E_i^\exp = E_i^\nom + A_i \sigma_i^\stat + B\sigma_i^\scale
\label{Eq:EndpointGen}
\end{equation}
Here $E_i$ denotes the position of the $i^{\rm th}$ endpoint.  
The coefficients $A$ and $B$ are picked from a
Gaussian distribution of mean 0 and variance 1.  Each experiment will pick as
many $A$'s as there are endpoint measurements as well as one $B$ for the
$\mll$ endpoint and one for the endpoints involving jets, thus neglecting the
effect of the lepton energy scale error on the latter.

When a set of edges $\vvec E^\exp$ has been found, the task is to find the
masses $\hat{\vvec m}$ which best correspond to the measurements.  If only
four endpoints are measured, the inversion formulae straight away return the
possible mass combinations.  If more endpoints are available, no mass
combination will in general reproduce the edge measurements, and a numerical
approach is required, where the measurements are weighted according to their
uncertainties. Note that in this procedure we do not make use of the fit values
given in Table~\ref{TABedges}.

It should also be emphasised that the systematics 
of the endpoint measurements are here assumed to be under control, 
i.e.\ the `Syst.~Fit Error' of Table~\ref{TABedges} is neglected. 
The precision we will find in this section and the next for 
the determination of masses and mass differences at the LHC 
must be understood in this context. 
If the endpoint systematics turn out to be comparable to the 
combined statistical and energy scale errors, then the precision 
will be worse.

\subsection{Mass estimation via $\LSfun$}

In our case, where the jet energy scale error produces a correlation between
the endpoint measurements, the method of least squares is appropriate. The
best mass estimate $\hat{\vvec m}$ is then the one which minimises the
function
\begin{equation}
\LSfun = 
[\vvec{E}^\exp-\vvec{E}^\theory(\vvec{m})]^T 
\mat{W}
[\vvec{E}^\exp-\vvec{E}^\theory(\vvec{m})]
\label{Eq:LSfun}
\end{equation}
where $\vvec{E}^\theory(\vvec{m})$ contains the theoretical edge values for a
set of masses $\vvec{m}$.  The weight matrix $\mat{W}$ is the inverse of the
correlation matrix or error matrix of the observations, which is given by the
variances and covariances of the endpoint measurements,
\begin{eqnarray}
\nonumber
(\mat{W}^{-1})_{ii} &=& 
\sigma_{ii}^\stat + \sigma_{ii}^\scale 
= (\sigma_i^\stat)^2 + (\sigma_i^\scale)^2 \\
(\mat{W}^{-1})_{ij} &=& \sigma_{ij}^\scale 
= \av{E_i^\exp E_j^\exp}-\av{E_i^\exp}\av{E_j^\exp} 
= \sigma_i^\scale\sigma_j^\scale ,\quad i\neq j \\
\nonumber
(\mat{W}^{-1})_{i1} &=& 0 ,\quad i\neq 1
\end{eqnarray}
where $j=1$ refers to $\maxmll$, which to a good approximation 
is uncorrelated with the other measurements. 
The covariances are similar to the variances in size, and so cannot be
neglected.  If the endpoint measurements were uncorrelated, $\mat{W}$ would
become diagonal, and the least-squares method would reduce to the normal
$\chi^2$ minimum method.

The ensemble distributions obtained by such a procedure can be interpreted as
probability density functions.  From these the `inverse probability problem'
can be addressed, which is that of stating something about the true masses on
the basis of the ones obtained in one experiment.  We will be interested in
the mean values of the ensemble distribution, their standard deviations,
skewness, as well as the correlation between masses.

\subsection{Minima of $\LSfun$\label{subsect:MinimaofLSfun}}

Because many endpoints are given by different expressions for 
different mass regions, 
see Eqs~(\ref{Eq:edge-qll-max})--(\ref{Eq:edge-ql-lowhigh}), 
the minimisation function
$\LSfun$ is a composite function. For the endpoint measurements used in this
paper, $\LSfun$ is made up of nine individual functions, $\LSfun_{\itB(i,j)}$,
one for each of the nine regions {\itB(i,j)}.  Considered separately each
function $\LSfun_{\itB(i,j)}$ has one or more minima.  For these to also be
minima of the composite function (`physical minima'), they need to be situated
in the region of validity (`home region') {\itB(i,j)} of the corresponding
function.  Physical minima can also occur on the borders between regions, in
which case they will be referred to as `border minima'.

If the threshold endpoints are left out, there are four measurements for four
masses.  The clear failure of the endpoint measurements of \SPSOaa\ and
$(\beta)$ to comply with Eq.~(\ref{Eq:sqll=sql+sql}) already discards the
three regions where these four measurements are not sufficient to determine
the masses.  In each of the other six regions the minima can be sought by use
of the inversion formulae. Such solutions correspond to $\LSfun=0$.  In cases
where no physical solutions are found in this way, border minima exist at
$\LSfun>0$, and will have to be found by a least square minimisation.

When the threshold measurement is included, the system of equations becomes
overconstrained.  This will give a non-uniform increase in the value of
$\LSfun$, which may destroy or create minima.  Another effect will be to move
the minima of $\LSfun$ around in mass space, possibly moving them into or out
of their home regions.  One way to picture the effect is to `tune in' the new
measurement by letting its uncertainty go from infinity, in which case the
measurement has no effect, to the value specified in Table~\ref{TABedges}.
The masses and height of each $\LSfun$ minimum will then move continuously
from the old to the new position.

Even though composite, $\LSfun$ is continuous, so its realisations in two
neighbouring regions attain the same value at their common border.  Assume
that the endpoint measurements are such that one of the realisations has a
minimum at the border (not a so-called border minimum).  Consider first the
case of no threshold measurement.  Since $\LSfun=0$, also the other
realisation must have a minimum at the border.  If now one of the endpoint
measurements is shifted up or down, the two minima will be driven off the
border and also separated in mass space.  To which side of the border they
move, and whether they go to the same side or not, will depend on the actual
parameters.  If the minima are on the same side, as is the case at \pbeta,
only one of them will be a physical minimum. If they are on opposite sides, as
is the case at \palpha, either both will be physical minima, or neither, in
which case there will be a border minimum.

This means that if a mass scenario is situated close to one of the borders,
the endpoints it produces may also have been produced by a set of masses from
the neighbouring region, provided the minimum there is a physical minimum.

If the threshold measurement is added, this picture is no longer exact.  Since
$\LSfun$ does not vanish at the minimum, the two regions will in general no
longer have a common minimum at the border.  However, the threshold
measurement often has less weight than the other measurements, so the above
picture still has some validity: Near a border two minima will be
lurking. Both, one or none may be physical minima.

\subsection{\SPSOaa}

If we neglect the threshold measurements to start with, and thereby also wait
with the $\bO$ mass, there are for each `experiment' two solutions.  One
solution is in region {\itB(1,1)}, which is the home region of the nominal
masses, and one is in region {\itB(1,2)}.  This is an example of the situation
described above: \SPSOaa\ has $2\slR/(\sNO+\sNT)=1.01$.  If this ratio becomes
less than one, the region changes, as is seen from
Eq.~(\ref{Eq:edge-ql-lowhigh}). A reduction of $\mlR$ by 0.7~GeV would put the
mass set on the border to region {\itB(1,2)}.  In both regions the mass
distributions are close to Gaussian.  The ensemble means of the home region
solution essentially equal the nominal values.  The {\itB(1,2)} solution has
central values some 15--20~GeV below the nominal ones.  Without additional
information both solutions are equally good and it would not be possible to
determine which one to choose. One would have to state that the SUSY masses
are summarized by either the {\itB(1,1)} set or by the {\itB(1,2)} set.

If the non-$b$ threshold endpoint is included as a fifth measurement in a
least square minimisation, the situation changes.  While a {\itB(1,1)} minimum
exists for practically all the `experiments', the occurence of a {\itB(1,2)}
minimum is now slightly reduced to 85--90\%, but it is in the $\LSfun$ value
of the minima the effect is most apparent.  The overconstraining fifth
measurement lifts the two minima asymmetrically from zero.  As could be
expected, the {\itB(1,1)} minimum is more often in accordance with the data,
but there is always a non-negligible {\itB(1,2)} contribution.  The difference
in $\LSfun$ value between a given minimum and the global minimum,
$\Delta\LSfun$, is a measure of the relevance of the minimum.
Table~\ref{table:p250Nsol} shows how many minima are available, on average,
for various $\Delta\LSfun$ cuts, and how these are shared between regions
{\itB(1,1)} and {\itB(1,2)}.

%%%%%%%%%%%%%%% iW=4
\begin{TABLE} { %with b1
\label{table:p250Nsol}
\begin{tabular}{|l|c|cc|}
\hline
  & $\#$ Minima & {\itB(1,1)} & {\itB(1,2)}  \\
\hline
$\Delta\LSfun\leq 0$ & 1.00 & 90\%  & 10\% \\
$\Delta\LSfun\leq 1$ & 1.12 & 94\%  & 17\% \\
$\Delta\LSfun\leq 3$ & 1.30 & 97\%  & 33\% \\
$\Delta\LSfun\leq99$ & 1.88 & 99\%  & 88\% \\
\hline
\end{tabular}
\caption{Number of minima for various $\Delta\LSfun$ cuts
and their whereabouts.} }
\end{TABLE}

If only the lowest minimum is chosen, the wrong solution is returned in 
10\% of the experiments. However, if two minima exist and are close
in $\LSfun$ value, one 
would have to consider both. In a certain fraction of experiments, depending 
on the $\Delta\LSfun$ cut, there would thus be two solution sets, 
e.g.\ for $\Delta\LSfun\leq1$ we will have two solutions in 12\% 
of the experiments. 
Whether or not it would be possible to select one of the solutions, 
and preferably the correct one, hinges on other measurements.
In this case, where the masses of the two sets are quite close, 
they might be very difficult to distinguish, by e.g.\ cross-section considerations.

%%%%%%%%%%%%%%% iW=4, iChi=2, iClC={1,2}
\begin{TABLE} {
\label{table:p250masses}
\begin{tabular}{|c|r|rrr|rrr|}
\hline
  &  &  & {\itB(1,1)} & &  &  {\itB(1,2)} &  \\
  & Nom & $\av{m}$ & $\sigma\hspace{1.5ex}$ & $\gamma_1$ & $\av{m}$ & $\sigma\hspace{1.5ex}$  & $\gamma_1$ \\
\hline
$\mNO    $ &   96.1\spcA  &   96.3\spcA &  3.8\spcA &  0.2  &   85.3 &  3.4\spcA &  0.1  \\
$\mlR    $ &  143.0\spcA  &  143.2\spcA &  3.8\spcA &  0.2  &  130.4 &  3.7\spcA &  0.1  \\
$\mNT    $ &  176.8\spcA  &  177.0\spcA &  3.7\spcA &  0.2  &  165.5 &  3.4\spcA &  0.1  \\
$\mqL    $ &  537.2\spcA  &  537.5\spcA &  6.1\spcA &  0.1  &  523.2 &  5.1\spcA &  0.1  \\
$\mbO    $ &  491.9\spcA  &  492.4\spcA & 13.4\spcA &  0.0  &  469.6 & 13.3\spcA &  0.1  \\
\hline
$\mlR-\mNO$ &  46.92  &  46.93 & 0.28 &  0.0 &  45.08 & 0.72 & $-$0.2  \\
$\mNT-\mNO$ &  80.77  &  80.77 & 0.18 &  0.0 &  80.18 & 0.29 & $-$0.1  \\
$\mqL-\mNO$ &  441.2\spcA  &  441.3\spcA &  3.1\spcA &  0.0  &  438.0 &  2.7\spcA &  0.0  \\
$\mbO-\mNO$ &  395.9\spcA  &  396.2\spcA & 12.0\spcA &  0.0  &  384.4 & 12.0\spcA &  0.1  \\
\hline
\end{tabular}
\caption{\SPSOaa: Minima for $\Delta\LSfun\leq1$ in regions {\itB(1,1)} 
and {\itB(1,2)}. 
Ensemble means, $\av{m}$, and root-mean-square distances 
from the mean, $\sigma$, are in GeV. 
The three lightest masses are very correlated. 
The mass of $\qL$ is fairly correlated to the lighter masses, 
but $\mbO$ is essentially uncorrelated. 
The distributions are very close to symmetric.} }
\end{TABLE}

The upper part of Table~\ref{table:p250masses} shows the ensemble means 
of the masses, 
$\av{m}$, the root-mean-square distances from the mean, $\sigma$,  
and skewness $\gamma_1$\footnote
{Skewness is defined by $\gamma_1 %=\av{(x-\bar x)^3}/\av{(x-\bar x)^2}^{3/2} 
=\mu_3/(\mu_2)^{3/2}=\mu_3/\sigma^3$, where $\mu_i 
=(x-\bar x)^i$ is the $i^{\rm th}$ moment about the mean.}  
of the two solutions for $\Delta\LSfun\leq1$.  
The values are relatively stable with respect to the $\Delta\LSfun$ cut: 
The same table for $\Delta\LSfun\leq99$
would for {\itB(1,1)} show a decrease in the masses by 0.1--0.2~GeV, and for
{\itB(1,2)} an increase by 1--1.3~GeV.

The inclusion of the threshold measurement has very little effect on the 
ensemble values of the {\itB(1,1)} solution. 
For the {\itB(1,2)} solution, to better comply with the additional measurement, 
the masses have increased, and are now 10--15~GeV below the nominal ones. 
Also the $\mbll$ threshold was included in the fit which returned the values 
of Table~\ref{table:p250masses}.
It is measured with much less precision than the other endpoints, so
its inclusion has practically no effect on the other masses, only on $\mbO$, 
for which it is the only measurement here.

The fact that the ensemble means of {\itB(1,1)} reproduce the nominal values,
relates to the good average performance of the ensemble of experiments.  The
probability of doing well with only one experiment relies in addition on the
spread of the ensemble values, given by~$\sigma$.  For \SPSOaa\ the high
precision of the endpoint measurements translates into rather small $\sigma$ 
values.  From the table we see e.g.\ that in $\sim68\%$ of the experiments the
mass of $\NO$ from the {\itB(1,1)} solution will lie within 3.8~GeV of the
nominal value.  The root-mean-square distances from the ensemble means 
are in principle unknown, as seen from
one experiment.  They can however be approximated by the procedure of
simulating $10^4$ experiments, where the measured values play the role as
`nominal'. This will engender a systematic shift, but $\sigma$ and any skewness
should be fairly well approximated. The root-mean-square distances from the 
mean values also have their
counterparts in the 1$\sigma$ errors returned by the fit of each `experiment'.
To within a few percent they are found to be identical.  This means that this
information is available for the experiment actually performed.  One can then
make the inverse statement: For a given experiment one can with $\sim68\%$
confidence state that the nominal value of $\mNO$ lies within 3.8~GeV of the
mass returned.

%%%%%%%%%%%%%%%%%%%%%%%%%%%%%%%%%%%%%%%%%%%%%%%%%%%%%%%%%%%%%%%%%%%%%%
\FIGURE[ht]{
%\begin{figure}
%%Begin InstantTeX Picture
\let\picnaturalsize=N
\def\picsize{16cm}
%If you do not have the picture file add:
%\let\nopictures=Y
%to the beginning of the file.
\ifx\nopictures Y\else{
\let\epsfloaded=Y
%\vspace*{-4mm}
\centerline{{\ifx\picnaturalsize N\epsfxsize \picsize\fi
\epsfbox{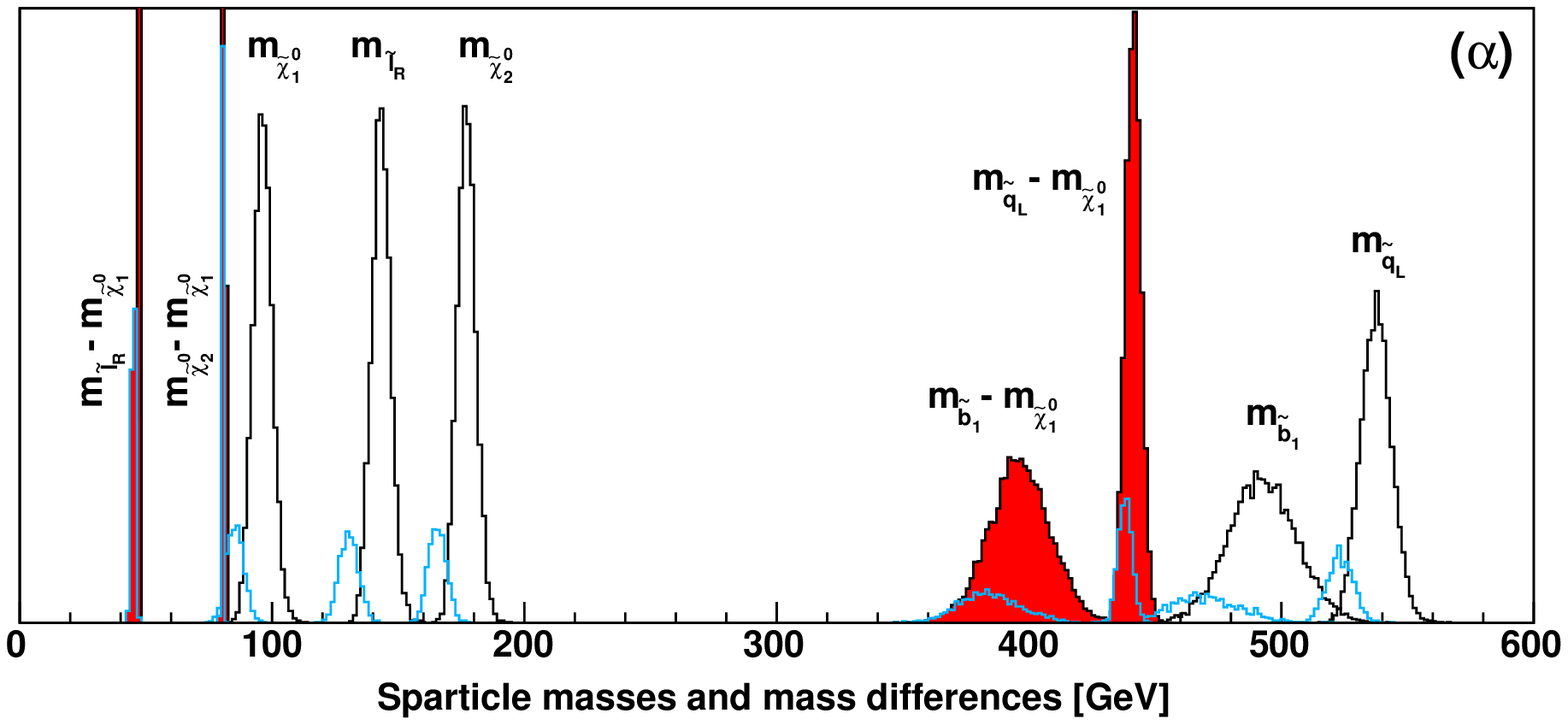}}}
}\fi
%%End InstantTeX Picture
\vspace*{-7mm}
\caption{Sparticle masses and mass differences at \SPSOaa\ for solutions with
$\Delta\Sigma\leq1$.  The unfilled distributions in black show from left to
right $\mNO$, $\mlR$, $\mNT$, $\mbO$ and $\mqL$ for solutions in the nominal
region {\itB(1,1)}. We will have such a solution in 94\% of the experiments, 
see Table~\ref{table:p250Nsol}.
The unfilled distributions in blue show the same masses
for solutions in region {\itB(1,2)}. 
Such a solution occurs in 17\% of the experiments,  
and the masses returned are lower. 
The smaller rate of the {\itB(1,2)} solutions is reflected in the smaller 
area under the blue curves.  
The ratio of probabilities between {\itB(1,2)} and 
{\itB(1,1)} solutions is $17\%/94\%=18\%$. The area under one of the blue 
curves is 18\% of the area under the corresponding black curve.  
The filled distributions show from left to right
$\mlR\!-\!\mNO$, $\mNT\!-\!\mNO$, $\mbO\!-\!\mNO$ and $\mqL\!-\!\mNO$.  
Again, the most populated distributions (black curves) are for solutions 
in region {\itB(1,1)}, the least populated (blue curves) for {\itB(1,2)} 
solutions.
For mass differences there is more overlap between the {\itB(1,1)} and 
{\itB(1,2)} solutions, in particular for $\mlR\!-\!\mNO$ and $\mNT\!-\!\mNO$, 
of which only the lower parts of the distributions are visible. 
Mass differences are better determined than the masses themselves, 
reflected here by the narrower distributions of the former. 
The exception is $\mbO$ which largely decouples from the other masses.
\label{fig:p250masses}}}
%\end{figure}
%%%%%%%%%%%%%%%%%%%%%%%%%%%%%%%%%%%%%%%%%%%%%%%%%%%%%%%%%%%%%%%%%%%%%%

Due to the way masses enter in the endpoint expressions, the fit returns
masses which have a strong positive correlation.  If one mass is low at the
minimum of the $\LSfun$ function, so the others tend to be and by a similar
amount.  In the lower part of Table~\ref{table:p250masses} ensemble mean 
and root-mean-square values
of mass differences are shown. It is clear that the three lightest
sparticles are very correlated. Fix one and the others are given very
accurately.  The squark masses are less correlated.  Also the results in
region {\itB(1,2)} are closer to {\itB(1,1)} and the nominal ones when
considering mass differences.
Fig.~\ref{fig:p250masses} shows the ensemble distributions corresponding to 
Table~\ref{table:p250masses}.

Because of this strong correlation between the masses, not only the mean
values and their 1$\sigma$ uncertainties, but the entire error matrix should
be considered if one wants to use the result obtained by this method as input
for other analyses.

A less involved solution would be to use less dependent variables, e.g.\ 
$\mNO$ to set the scale, then differences for the remaining masses, 
$\mlR-\mNO$, $\mNT-\mNO$, $\mqL-\mNO$ and $\mbO-\mNO$.

Due to the high cross-section, most of the endpoints are determined with high precision, 
which in turn gives narrow and approximately symmetric ensemble distributions. 
The masses are thus determined with quite high precision.
As a result of the strong correlations between in particular the lighter masses, 
even better estimates can be obtained for other combinations of the variables, 
e.g.\ mass differences.
At \SPSOaa\ there is however a fair chance that two sets of masses do equally well 
in the minimisation procedure. Other considerations must in that case be 
made in order to choose between them, or both must be kept.

\subsection{\SPSOab}
In combination with the theory plots of Section~\ref{sect:edge-meas} 
we found in Section~\ref{sect:data-gen-recon} that the $\maxmqlHigh$ value 
of \pbeta\ would most probably on the average be underestimated by 11--13~GeV.
We will find that this has a dramatic effect on the masses returned. 
However, for easier comparison with nominal values we first consider the situation 
without such a systematic effect. Then, afterwards, the impact of the mismeasurement 
will be shown, together with a way to mend the situation. 

As in the previous case we start out without the threshold measurement. 
Also here two solutions are available for all `experiments', 
one in {\itB(1,1)}, the 
other in either {\itB(1,2)} or {\itB(1,3)}. 
The nominal region for \pbeta\ is {\itB(1,2)}, but 
it is quite close to {\itB(1,3)}. This becomes clear from an inspection
of the `border parameter', 
\begin{equation}
\label{eq:border}
b=\frac{\slR}{\mNO\,\mNT}, 
\end{equation}
which is 1 on the border between these two regions.
For \pbeta\ we have $b=1.02$. If $\mlR$ is reduced by 2.5~GeV, 
this ratio becomes unity and the the mass set sits on the border, see 
Eq.~(\ref{Eq:edge-ql-lowhigh}).
The other point, \palpha, was near the border between {\itB(1,1)} and {\itB(1,2)}. 
There, both solutions (or none) were available. 
Here, the derivatives of $\LSfun$ are such that only one of the two solutions 
is available. In 71\% of the cases we get {\itB(1,2)}, in 29\% we get {\itB(1,3)}. 
While the `low-mass' solution, {\itB(1,2)} or {\itB(1,3)}, 
is in the vicinity of the nominal masses,
the {\itB(1,1)} solution, which is always present, usually sits at much higher masses, 
$\av{\mNO}=514$~GeV.
Because the two solutions are so separated one may hope that the incorrect one 
will be sufficiently disfavoured by other measurements, e.g.\ cross-sections, 
that it can be discarded. 

For \palpha\ the solution in the nominal region on the average reproduced 
the nominal values to within 0.2--0.5~GeV. 
Here the {\itB(1,2)}/{\itB(1,3)} solution has a mean of $\mNO$ at 183~GeV, 
some 22~GeV above the nominal value. The most probable value of the 
ensemble distribution is much closer to the nominal value. 
The distributions are infested with considerable skewness. 
On the way from endpoint measurement to mass determination a systematic effect 
which favours higher masses has been introduced. 
In statistical language our {\em estimators} of the true masses are not {\em consistent}: 
they do not converge to the nominal values. 
This of course has implications for the interpretation of the masses we obtain. 
What can be said about the true masses on the basis of the measured ones?
We will return to the reasons for the skewness later. 

When the threshold measurement is included, the {\itB(1,1)} minimum usually yields a 
large $\LSfun$ value. 
Only in a small fraction of the experiments does it challenge the other minima. 
In the other sector there is either one minimum, 
positioned in {\itB(1,2)}, {\itB(1,3)} or on the border (B), 
or there are two minima, in {\itB(1,2)} and {\itB(1,3)}. 
These minima are usually in good agreement 
with the threshold measurement and have low $\LSfun$ values. 

%%%%%%%%%%%%%%% iW=4
\begin{TABLE} {
\label{table:p400Nsol}
\begin{tabular}{|l|c|c|ccc|c|}
\hline
  &           &       &       & 1 sol &   & 2 sol        \\
  & $\#$ Min & {\itB(1,1)}  & {\itB(1,2)}  & {\itB(1,3)}  & B & {\itB(1,2)}\&{\itB(1,3)}  \\
\hline
$\Delta\LSfun\leq 0$  &  1.0   &    3\%   &   60\%   &   25\% &   12\% &    0\% \\
$\Delta\LSfun\leq 1$  &  1.2   &    5\%   &   52\%   &   18\% &   12\% &   16\% \\
$\Delta\LSfun\leq 3$  &  1.4   &   13\%   &   46\%   &   14\% &   12\% &   28\% \\
$\Delta\LSfun\leq99$  &  2.3   &   99\%   &   41\%   &   13\% &   12\% &   34\% \\
\hline
\end{tabular}
\caption{\SPSOab: 
Average number of minima and the fraction of experiments with the 
specified solution types, for different $\Delta\LSfun$ cuts.}
}
\end{TABLE}

Table~\ref{table:p400Nsol} shows the average number of minima for different
$\LSfun$ cuts. The three rightmost sections show the fraction of experiments
which have the specified solution type. The two rightmost sections exclude one
another.  Either there is one solution in the low-mass sector, or there are
two.  For the one-solution case the whereabouts of the minimum is also shown.
The home region of the nominal masses, {\itB(1,2)}, is seen to dominate.  As
it may well be possible to discard the {\itB(1,1)} minimum on the basis of
other observations, it is logically separated from the low-mass minima. E.g.\
in 13\% of the cases, regardless of the low-mass solution type, there is a
{\itB(1,1)} minimum at $\Delta\LSfun\leq3$.  To get the average number of
minima shown in column 2, sum horizontally, adding twice the two-solution
percentage.  For small $\Delta\LSfun$ cuts the two rightmost sectors do not
add up to 1.  This simply means that in some cases the global minimum lies in
{\itB(1,1)}, and no low-mass minimum is available in the given $\Delta\LSfun$
range.  For the current set of endpoint measurements the {\itB(1,1)}
contamination is seen to be very moderate.  However, the systematic fit error
(column 6 of Table~\ref{TABedges}) is here assumed to be zero.  If it should
become impossible to obtain the threshold value with such optimistic
precision, the fraction of {\itB(1,1)} solutions at low $\LSfun$ will grow
rapidly.

%%%%%%%%%%%%%%% iW=4, dChiM=1.000000
\begin{TABLE} {
\label{table:p400masses}
\begin{tabular}{|c|r|rrr|rrr|rrr|rrr|}
\hline
\multicolumn{2}{|c|}{}  & \multicolumn{3}{|c|}{}  & \multicolumn{3}{|c|}{1 solution} 
&   \multicolumn{6}{|c|}{2 solutions}     \\
\cline{3-14}
\multicolumn{2}{|c|}{}  & \multicolumn{3}{|c|}{{\itB(1,1)}}  
& \multicolumn{3}{|c|}{{\itB(1,2)/(1,3)}/B}  
& \multicolumn{3}{|c|}{{\itB(1,2)}} & \multicolumn{3}{|c|}{{\itB(1,3)}} \\
\hline
& Nom &$\av{m}$&$\sigma$&$\gamma_1$&$\av{m}$&$\sigma$&$\gamma_1$
&$\av{m}$&$\sigma$&$\gamma_1$&$\av{m}$&$\sigma$&$\gamma_1$ \\
\hline
$\NO    $ &161 & 438&  88&  0.9 & 175&  35&  1.0 & 161&  22&  0.3 & 166&  27&  0.6  \\
$\lR    $ &222 & 518&  85&  0.7 & 236&  37&  0.8 & 221&  24&  0.3 & 223&  28&  0.5  \\
$\NT    $ &299 & 579&  85&  0.7 & 313&  35&  1.0 & 299&  22&  0.3 & 304&  27&  0.6  \\
$\qL    $ &826 &1146& 104&  0.8 & 843&  44&  0.9 & 826&  30&  0.3 & 835&  36&  0.5  \\
\hline
$\lR-\NO$ & 61 &  81& 1.8& \hspace{-2mm}$-$0.3 &  61& 4.4&  0.4 &  61& 1.9& \hspace{-2mm}$-$0.2 &  57& 1.3& \hspace{-2mm}$-$0.2  \\
$\NT-\NO$ &138 & 141& 0.9&  0.1 & 138& 0.6&  0.2 & 138& 0.5&  0.0 & 138& 0.5&  0.0  \\
$\qL-\NO$ &665 & 708&  17&  0.1 & 668&  10&  0.5 & 665&   9&  0.1 & 669&  10&  0.2  \\
\hline
\end{tabular}
\caption{\SPSOab: Nominal masses (`Nom') and
$\Delta\LSfun\leq1$ ensemble distribution values for the three solution 
types. High-mass sector: The {\itB(1,1)} solutions return masses far 
beyond the nominal values. 
Low-mass sector: For the one-solution case the values are based on the 
common distribution of {\itB(1,2)}, {\itB(1,3)} and border (B) solutions. 
In the two-solution case the ensemble variables of both solutions are shown.
Ensemble means, $\av{m}$, and root-mean-square values, $\sigma$, are in GeV.} }
\end{TABLE}

In Table~\ref{table:p400masses} $\av{m}$, $\sigma$ and $\gamma_1$ of the ensemble 
masses and mass differences are shown for the different solution types and the
cut $\Delta\LSfun\leq1$.  The masses of the {\itB(1,1)} solution are much
higher than what the low-mass minima give.  Even though the distributions are
broad, allowing for low values to occur, it is very rare that the masses
stretch down to the low-mass sector.  In section 4 of the table the low-mass
one-solution values are shown.  Since in such a case only one acceptable
solution is available (discarding {\itB(1,1)}), and since \pbeta\ anyway is
situated close to the border, it makes sense to show the combined distribution
of the {\itB(1,2)}, {\itB(1,3)} and border (B) minima.  From
Table~\ref{table:p400Nsol} this situation is seen to occur in
52\%+18\%+12\%=82\% of the experiments.  The mean values of the masses lie
some 15~GeV above the nominal ones.  This is an improvement compared to the
non-threshold situation, but it remains an undesirable feature.  The mass
distributions are skewed and the most probable value is found close to the
nominal value.  The root-mean-square values of the ensemble distributions are large, 
nearly an order of magnitude larger than at \palpha.

The rightmost sections show the values for the two-solution type. 
The ensemble means of the two distributions do not differ too much, 
and they are much closer to the nominal values than is the case for the 
one-solution type.
Since the values are rather close, it will probably be quite difficult to 
find other 
measurements which favours one of the sets. On the other hand, since the 
root-mean-square values are larger than the differences between the two solutions 
(also within one experiment), taking the average value, perhaps somehow 
weighted with the $\LSfun$ value, might be a possible compromise.

%%%%%%%%%%%%%%%%%%%%%%%%%%%%%%%%%%%%%%%%%%%%%%%%%%%%%%%%%%%%%%%%%%%%%%
\FIGURE[ht]{
%\begin{figure}
%%Begin InstantTeX Picture
\let\picnaturalsize=N
\def\picsize{16cm}
%If you do not have the picture file add:
%\let\nopictures=Y
%to the beginning of the file.
\ifx\nopictures Y\else{
\let\epsfloaded=Y
%\vspace*{-4mm}
\centerline{{\ifx\picnaturalsize N\epsfxsize \picsize\fi
\epsfbox{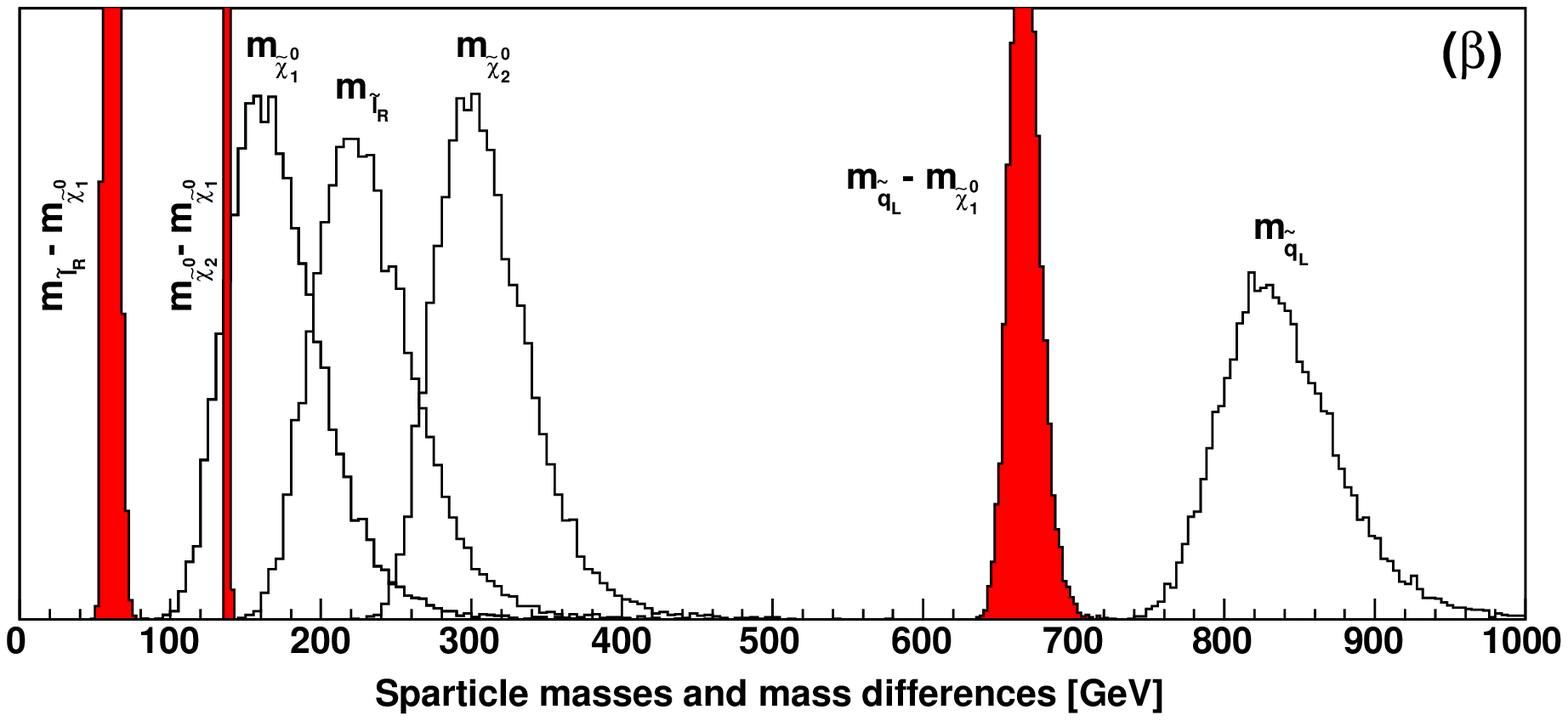}}}
%} }
}\fi
%%End InstantTeX Picture
\vspace*{-7mm}
\caption{Sparticle masses and mass differences at \SPSOab. 
All masses of solutions with $\Delta\Sigma\leq1$ which lie in 
regions {\itB(1,2)}, {\itB(1,3)} and on their common border are shown. 
From left to right the unfilled distributions show $\mNO$, 
$\mlR$, $\mNT$ and $\mqL$. The filled distributions show the narrower mass 
differences $\mlR-\mNO$, $\mNT-\mNO$ and $\mqL-\mNO$. 
Skewness of mass distributions is visible. 
\label{fig:p400masses}}}
%\end{figure}
%%%%%%%%%%%%%%%%%%%%%%%%%%%%%%%%%%%%%%%%%%%%%%%%%%%%%%%%%%%%%%%%%%%%%%

Again, the mass correlation is very strong. 
This is evident from the lower part of Table~\ref{table:p400masses}, 
where the ensemble distributions of mass differences come with much 
smaller root-mean-square distance to the mean values than what the 
masses themselves do. 
They are also very close to the nominal values. 
Even more seems to be gained by using mass differences here than at \palpha. 
Fig.~\ref{fig:p400masses} shows the ensemble distributions for the masses 
of all solutions with $\Delta\Sigma\leq1$ and which lie in the regions 
{\itB(1,2)}, {\itB(1,3)} or on their common border. 
See figure caption for details. 

\subsubsection*{Skewness}
The ensemble distributions are not symmetric.  While the most probable values
are close to the nominal values, the means lie above.  For \palpha\ the
tendency of such an asymmetry is small, but for \pbeta\ the effect is large.
The reason why we naively would expect a symmetric distribution 
around the nominal masses in the first place, is that the endpoint 
measurements are generated symmetrically.
For complex functions like Eqs.~(\ref{eq:inv(11)mN1})--(\ref{eq:inv(43)mqL}) 
symmetric fluctuation of the endpoint arguments will produce 
near-symmetric variation of the function only for small fluctuations. 
As the arguments fluctuate more, the deviation from symmetry in the 
function values grows. 
At \palpha\ the endpoint fluctuations are so small that the effect is 
negligible. 
For \pbeta, where the endpoint fluctuations are larger, 
the effect of the `asymmetric propagation' is 
a noticeable increase of 3--4~GeV for the ensemble means.

%%%%%%%%%%%%%%%%%%%%%%%%%%%%%%%%%%%%%%%%%%%%%%%%%%%%%%%%%%%%%%%%%%%%%%
\FIGURE[ht]{
%\begin{figure}
%%Begin InstantTeX Picture
\let\picnaturalsize=N
\def\picsize{8.5cm}
%If you do not have the picture file add:
%\let\nopictures=Y
%to the beginning of the file.
\ifx\nopictures Y\else{
\let\epsfloaded=Y
%\vspace*{-4mm}
\centerline{{\ifx\picnaturalsize N\epsfxsize \picsize\fi
\epsfbox{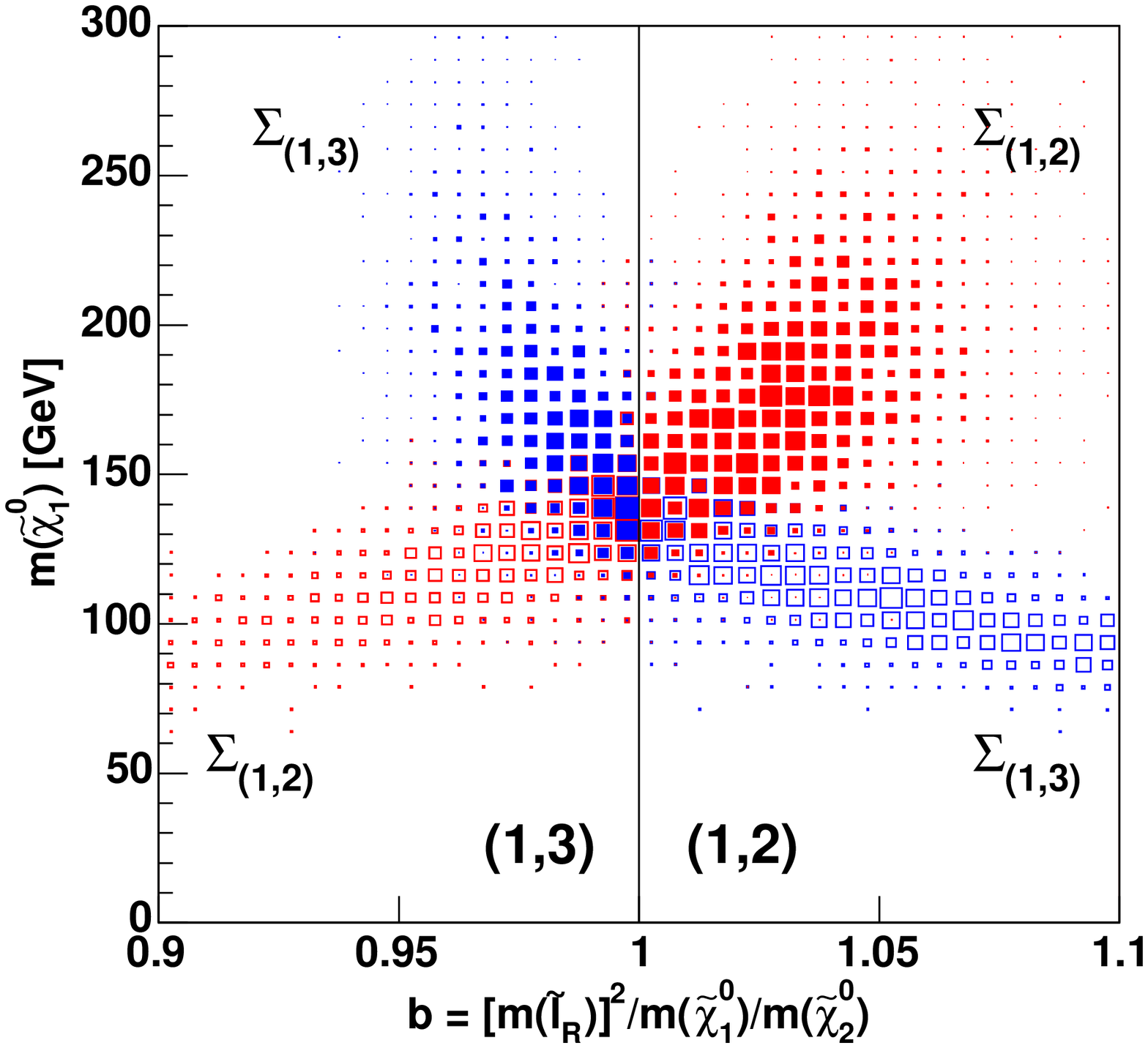}
} }
}\fi
%%End InstantTeX Picture
\vspace{-6mm}
\caption{Border effect between region {\itB(1,2)} and {\itB(1,3)}; 
the mass of $\NO$ as a function of the border parameter $b$, see 
Eq.~(\ref{eq:border}), demonstrating border effects. Filled boxes 
represent physical solutions, empty boxes represent unphysical solutions. 
\label{Fig:p400BorderEffect}}}
%\end{figure}
%%%%%%%%%%%%%%%%%%%%%%%%%%%%%%%%%%%%%%%%%%%%%%%%%%%%%%%%%%%%%%%%%%%%%%

This is however not sufficient to explain the low-mass $\av{\mNO}$ 
of 183 and 173~GeV ($\Delta\LSfun\leq 99$)
without and with the threshold measurement, respectively. 
`Border effects' need to be considered. 
As described earlier, \pbeta\ lies in {\itB(1,2)} but close to the border to {\itB(1,3)}. 
First consider the situation without the threshold measurement. 
There is then always only one low-mass solution. 
If the {\itB(1,2)} solution is physical, i.e.\ lies in {\itB(1,2)}, 
then the true minimum of $\LSfun_{\itB(1,3)}$
also lies in region {\itB(1,2)} and so is unphysical, and vice versa, 
as described in Sect.~\ref{subsect:MinimaofLSfun}. 

In Fig.~\ref{Fig:p400BorderEffect} the mass of $\NO$ is plotted as a function
of the border parameter, $b$, of Eq.~(\ref{eq:border}), for both physical and
unphysical minima of $\LSfun_{\itB(1,2)}$ and $\LSfun_{\itB(1,3)}$.  The
minima of $\LSfun_{\itB(1,2)}$ are shown in red, from upper right to lower
left.  The $\LSfun_{\itB(1,3)}$ solutions are in blue.  Filled boxes are
physical solutions, i.e.\ $\LSfun_{\itB(1,2)}$ (red) for $b>1$ and
$\LSfun_{\itB(1,3)}$ (blue) for $b<1$.  Empty boxes are unphysical solutions.
An asymmetry arises from the accidental fact that for both functions the lower
masses tend to lie in the unphysical region.  The average of the entire
$\LSfun_{\itB(1,2)}$ distribution, both physical and unphysical minima,
returns 164~GeV, the nominal value plus the 3~GeV of the asymmetric
propagation effect.  It is then obvious that when the unphysical {\itB(1,2)}
solutions are replaced by the physical {\itB(1,3)} solutions which lie at
higher masses, the ensemble mean increases.  Here, this effect brings the
average value for $\mNO$ to 183~GeV, an additional increase of nearly 20~GeV.

When the threshold measurement is added, the border effect is reduced, giving
an average value of 173~GeV (for one solution).  This is because a measurement
will, unless there is any bias, on the average be conservative.  It will try
to keep the masses at their nominal values.  Here, as the border effect is
pushing the masses upwards, away from the nominal values, the threshold is
holding back.

If, in a realistic situation, a set of endpoints has been measured, and 
the resulting set of masses is found to lie close to a border, caution 
should be exercised. 
Ad hoc procedures seem necessary for such a case. 
At least one is in a position to be aware of the danger. 
It is probably appropriate to consider unphysical minima as well.

\subsubsection*{Mismeasured $\maxmqlHigh$}
As pointed out at the start of this subsection it is very likely that, for
\pbeta, $\maxmqlHigh$ will be understimated by 11--13~GeV.  Without the
threshold measurement, the effect of this as returned by the inversion
formulae, is an increase of 50~GeV for the three lightest masses and 60~GeV
for the squark!  If the threshold measurement is included, the effect is
reduced to 30--40~GeV for the lighter masses, depending on the solution type,
and 40--55~GeV for the squark.  The increase is still surprisingly large and
represents a serious threat to the applicability of the method in the case
where the nominal masses happen to sit near a border.  (Not to be confused
with the previously mentioned border effect.)

It may however be possible to discover afterwards that such a mismeasurement 
has or may well have been made. If the masses obtained are used to generate 
the corresponding theory distributions, one may spot the problem. 
A quick test showed that in our case the $\mqlHigh$ theory distribution 
as generated with the obtained masses, did have a small foot. 
This is then a sign that the original $\maxmqlHigh$ measurement may have been
wrong, since no such foot was seen and included in the fit.
This test probably has quite wide applicability. The reason is that even
though, as we have seen, the masses may change a lot in the case of a
mismeasured $\maxmqlHigh$, mass {\em differences} are not so far off the
nominal values.  The main change is therefore in the overall scale.  For
theory distributions, the shapes are independent of the overall scale.  Only
mass ratios are relevant.  The theory shapes based on the masses obtained from
the partly mismeasured endpoints, will therefore usually not deviate too much
from the nominal shapes.

If one is certain that a mismeasurement has been made, either from the test
described above or from other considerations, not only does one know that the
$\maxmqlHigh$ measurement is not to be trusted.  There is also a way to pursue
the matter further.  It relies on the fact that the vertical fall we see and
measure, is the endpoint of the $\mqlN$ distribution. We have the explicit
expression for this endpoint, and so we can just replace the $\maxmqlHigh$
expression and redo the least square minimisation.  \\

At \SPSOab\ many new aspects of the endpoint method have emerged.  For a large
fraction of the experiments a total of three minima were competing for a given
set of endpoint values, some returning masses near the nominal values, some
returning much higher masses.  A generic characteristic is the skewness of the
ensemble distributions.  The problems and the causes of the asymmetry have
been discussed, but a way to handle the asymmetry has not been proposed.  The
ensemble distributions are much broader than at \SPSOaa.  Mass differences are
however quite precisely determined.  The problem of mismeasuring
$\maxmqlHigh$, which will be a danger in a large fraction of mass space, has
been discussed and a solution proposed.  Throughout, we have also seen that
the inclusion of a fifth measurement in general improves on the results.

%%%%%%%%%%%%%%%%%%%%%%%%%%%%%%%%%%%%%%%%%%%%%%%%%%%%%%%%%%%%%%%%%%%%%%
\section{Linear Collider inputs} \label{sect:lhc-lc}
%%%%%%%%%%%%%%%%%%%%%%%%%%%%%%%%%%%%%%%%%%%%%%%%%%%%%%%%%%%%%%%%%%%%%%

As compared with the LHC, the Linear Collider
\cite{Aguilar-Saavedra:2001rg,Abe:2001nn} will more directly provide very
precise measurements of the lightest neutralino mass (and possibly also the
second lightest). Thus, the LC input fixes the scale, whereas the cascade
decays primarily provide mass differences.  This kind of input, if it becomes
available during the analysis of LHC data, will have a dramatic influence on
the over-all analysis \cite{lhc-lc}.

To each LHC experiment a corresponding Linear Collider experiment is
considered.  The Linear Collider measurement of the LSP, $m_{\NO}^{\rm LC}$,
was for every experiment picked randomly from a Gaussian distribution of mean
equal to $m_{\NO}^\nom$, and standard deviation set by the expected
uncertainty, $\sigma^{\rm LC}_{\mNO}$.  We used $\sigma^{\rm LC}_{\mNO} =
0.05$~GeV, in accordance with \cite{Aguilar-Saavedra:2001rg}.  Such a small
uncertainty practically means fixing the mass at the nominal value.  Still,
for completeness the measurement was appropriately included in the least
square minimisation by adding the term $[(\mNO-m_{\NO}^{\rm LC})/\sigma^{\rm
LC}_{\mNO}]^2$ to the $\LSfun$ function, Eq.~(\ref{Eq:LSfun}).

%%%%%%%%%%%%%%% iW=4
\begin{TABLE} {
\label{table:p400Nsol_LC}
\begin{tabular}{|l|c|ccc|c|}
\hline
  &           &       & 1 sol &   & 2 sol        \\
  & $\#$ Min & {\itB(1,2)}  & {\itB(1,3)}  & B & {\itB(1,2)}\&{\itB(1,3)}  \\
\hline
 $\Delta\LSfun\leq 0$  &  1.0 &   65\%   &   26\% &    9\% &    0\% \\
 $\Delta\LSfun\leq 1$  &  1.2 &   52\%   &   18\% &    9\% &   21\% \\
 $\Delta\LSfun\leq 3$  &  1.4 &   36\%   &   10\% &    9\% &   45\% \\
 $\Delta\LSfun\leq99$  &  1.6 &   24\%   &    7\% &    9\% &   59\% \\
\hline
\end{tabular}
\caption{Number of minima for \SPSOab\ with LC input.} }
\end{TABLE}

%%%%%%%%%%%%%%% iW=4, iChi=2, iClC={1,2} foc
\begin{TABLE} {
\label{table:p250masses_LC}
\begin{tabular}{|c|r|rr|}
\hline
\multicolumn{1}{|c|}{} & \multicolumn{1}{|c|}{} & \multicolumn{2}{|c|}{\itB(1,1)} \\
  & Nom & $\av{m}\ $ & $\sigma\hspace{1ex} $ \\
\hline
$\NO    $ &   96.05  &   96.05 &  0.05  \\
$\lR    $ &  142.97  &  142.97 &  0.29  \\
$\NT    $ &  176.82  &  176.82 &  0.17  \\
%$\qL    $ &  537.25  &  537.22 &  2.49  \\
$\qL    $ &  537.25  &  537.2\spcA  &  2.5\spcA  \\  %ROUNDED
%$\bO    $ &  491.92  &  492.06 & 11.68  \\ 
$\bO    $ &  491.92  &  492.1\spcA  & 11.7\spcA   \\  %ROUNDED
\hline
\end{tabular}
\caption{The \SPSOaa\ masses with LC input. All values in GeV. 
Region {\itB(1,2)} solutions now occur only in $\sim1$\% of the 
cases and are left out.} }
\end{TABLE}
%%%%%%%%%%%%%%%
%%%%%%%%%%%%%%%% iW=4, iChi=2, iClC={1,2} foc
%\begin{TABLE} {
%\label{table:p250masses_LC}
%\begin{tabular}{|c|r|rr|rr|}
%\hline
%  &  &  {\itB(1,1)} &  &  {\itB(1,2)} & \\
%  & Nom & $\av{m}$ & $\sigma$  & $\av{m}$ &  $\sigma$  \\
%\hline
%$\NO    $ &   96.05  &   96.05 &  0.05 &   96.04 &  0.05  \\
%$\lR    $ &  142.97  &  142.97 &  0.29 &  141.90 &  0.79  \\
%$\NT    $ &  176.82  &  176.82 &  0.17 &  176.34 &  0.36  \\
%$\qL    $ &  537.25  &  537.22 &  2.49 &  537.55 &  2.67  \\
%$\bO    $ &  491.92  &  492.06 & 11.68 &  488.87 & 11.59  \\
%\hline
%\end{tabular}
%\caption{The \SPSOaa\ masses with LC input. All values in GeV.} }
%\end{TABLE}
%%%%%%%%% Nov20: [table above: (1,2) solution removed. With only 1% occurence,
%%%%%%%%%         not sufficient statistics to give good RMS values.]

For \palpha\ the fixing of $\mNO$ reduces the occurrences of multiple minima
to the per mille level for any usable minima, $\Delta\LSfun\leq3$. 
In nearly all cases, 98-99\%, it is the home region
{\itB(1,1)} minimum which survives.  This seems reasonable since without the
LC measurement the {\itB(1,2)} minimum has $\av{\mNO}$ at some 10~GeV below
the nominal value, see Table~\ref{table:p250masses}.

At \pbeta\ the high-mass minimum {\itB(1,1)} is absent, disfavoured as it is
by the Linear Collider measurement.  The number of {\itB(1,3)} minima has
increased, which is reflected in an increase of the two-solution case, compare
Tables~\ref{table:p400Nsol} and ~\ref{table:p400Nsol_LC}.  For small
$\Delta\LSfun$ cuts this increase is not drastic.  Fixing $\mNO$ thus does not
help us to uniquely determine one minimum, contrary to what one might have
expected.

The Tables~\ref{table:p250masses_LC}--\ref{table:p400masses_LC} show the mean
and root-mean-square values of the ensemble distributions 
at \palpha\ and \pbeta.  For both
SUSY scenarios the ensemble means fall very close to the nominal masses, even
for minima not situated in the nominal home region.  The uncertainty on the
scale, which for the LHC alone is the main contribution to the spread of the
ensemble distributions, is set to zero by the LC measurement.  The root-mean-square 
values are therefore strongly reduced.  Without Linear Collider measurements, the
mass differences were more accurately determined in that they were less
dependent on the mass scale.  A comparison of the root-mean-square values 
of the masses in
Tables~\ref{table:p250masses_LC}--\ref{table:p400masses_LC} with the 
root-mean-square values of the mass differences in
Tables~\ref{table:p250masses}--\ref{table:p400masses} shows to which extent
mass differences are scale independent variables.  When a Linear Collider
measurement is available, the mass differences no longer out-perform the
masses themselves in terms of precision.  With the fixing of the scale also
the skewness of the distributions has vanished.
%%%%%%%%%%%%%%% iW=4, dChiM=1.000000
\begin{TABLE} {
\label{table:p400masses_LC}
\begin{tabular}{|c|c|cc|cc|cc|}
\hline
%\multicolumn{1}{|c|}{} & \multicolumn{1}{|c|}{} & \multicolumn{2}{|c|}{1 sol} 
%& \multicolumn{2}{|c}{\itB(1,2)} & \multicolumn{2}{c|}{\itB(1,3)} \\
%\hline
\multicolumn{1}{|c|}{} & \multicolumn{1}{|c|}{} & \multicolumn{2}{|c|}{1 solution} 
& \multicolumn{4}{|c|}{2 solutions} \\
\cline{3-8}
%\hline
\multicolumn{1}{|c|}{} & \multicolumn{1}{|c|}{} 
& \multicolumn{2}{|c|}{{\itB(1,2)/(1,3)}/B}
& \multicolumn{2}{|c|}{\itB(1,2)} & \multicolumn{2}{|c|}{\itB(1,3)} \\
%  &     &       1 sol &     & {\itB(1,2)}      &       & {\itB(1,3)} &       \\
\hline
  & Nom &$\av{m}$&$\sigma$&$\av{m}$&$\sigma$&$\av{m}$&$\sigma$ \\
\hline
$\NO    $ &161.02 & 161.02 &   0.05 & 161.02 &   0.05 & 161.02 &   0.05  \\
$\lR    $ &221.86 & 221.15 &   3.26 & 222.22 &   1.32 & 217.48 &   1.01  \\
$\NT    $ &299.05 & 299.15 &   0.57 & 299.11 &   0.53 & 299.05 &   0.52  \\
%$\qL    $ &826.29 & 826.12 &   6.33 & 825.89 &   5.78 & 828.60 &   5.52  \\ 
$\qL    $ &826.29 & 826.1\spcA  &   6.3\spcA  & 825.9\spcA  &   5.8\spcA  & 828.6\spcA  &   5.5\spcA   \\  %ROUNDED
\hline
\end{tabular}
\caption{The \SPSOab\ masses with LC input. Nominal, ensemble averages,
$\av{m}$, and root-mean-square deviations from the mean, $\sigma$, 
are all in GeV.} }
\end{TABLE}

%%%%%%%%%%%%%%%%%%%%%%%%%%%%%%%%%%%%%%%%%%%%%%%%%%%%%%%%%%%%%%%%%%%%%%
\section{Conclusions} \label{sect:conc}
%%%%%%%%%%%%%%%%%%%%%%%%%%%%%%%%%%%%%%%%%%%%%%%%%%%%%%%%%%%%%%%%%%%%%%

We have investigated the measurement of supersymmetric
masses from the decay chain $\tilde q \to \tilde \chi_2^0 q \to
\lR lq \to \tilde \chi_1^0 llq$, in the Snowmass mSUGRA scenario SPS~1a. 
Since the lightest neutralino $\tilde \chi_1^0$ is the LSP in most
interesting mSUGRA scenarios, it will escape detection and only the
quark and two leptons are available for the construction of invariant
mass distributions. Nevertheless, the kinematic endpoints of these
distributions have a well defined dependence on the masses of the
particles in the decay chain and their measurement allows the
extraction of the masses either by analytic inversion or numerical
fit. The analytic expressions for the endpoints in terms of the masses
were confirmed and presented together with their analytic inversions. 

In order to measure the endpoints of the invariant mass distributions
pertaining to the chosen decay chain, one must have the correct mass
hierarchies for the decay chain and a large enough cross-section. To
ensure that this decay chain could be used over a wide range of
scenarios, we performed a scan over the SUSY parameters. We found that
as long as $m_0$ was not too large in comparison to $m_{1/2}$, a large
proportion of the allowed parameter space would display the correct
mass hierarchy. Furthermore, on examination of the sparticle
production cross-sections and decay branching ratios we found that a
large cross-section for the decay was available over much of this
region. The Snowmass mSUGRA SPS~1a line/point falls into this
region and is a good candidate for study. However, we noted that the
cross-section for the decay chain is particularly high for the SPS~1a
point, and it is instructive to examine a second point on the SPS~1a
line with a less optimistic cross-section. We have denoted this new
point as SPS~1a~($\beta$) while the original point became
SPS~1a~($\alpha$).

The LHC measurements of the endpoints were simulated using PYTHIA and
ATLFAST. Hard kinematic cuts remove practically all Standard Model 
backgrounds, except $t\bar t$. Up to statistical fluctuations the 
powerful different-flavour subtraction then cancels the remaining $t\bar t$ as 
well as any lepton-uncorrelated background from other SUSY channels. The 
resulting distributions are however contaminated by lepton-correlated 
background from $\NT$'s not taking part in the decay chain under 
study and combinatorial background from choosing the wrong jet. 
The inconsistency cut was shown to address the latter part with 
great efficiency, giving distributions much closer to the theoretical 
ones. Also mixed events were studied, revealing their potential to 
describe the background. More study is however needed. 

The endpoints were found by simplistic fitting of the edges, 
usually with a straight line together with a reasonable background 
estimate. A Gaussian smearing was sometimes included as a first 
approximation of the various smearing effects which take place. 
The statistical precision of the edge was sought rather 
than an accurate determination of the endpoint. Still, the 
endpoints were seen to be in reasonably good agreement with the 
nominal values.  
However, the fitting procedure is clearly an area for improvement. 
On one hand, more realistic study of how the detector affects the 
distributions, and in particular the end regions, is called for.
On the other hand, further study of the many realisations of the 
theory curves seems necessary. It is important to find good fit 
functions for the signal. A central part of such a programme 
is the incorporation of multiple squark masses at different rates. 
At a less ambitious scale the theory distributions should be studied 
for sheer acquaintance. The importance of such an awareness was 
demonstrated for the $\maxmqlHigh$ measurement at \pbeta. 

In order to turn the endpoint measurements into particle masses, and
understand the resulting errors of this procedure, we
considered an ensemble of 10,000 `gedanken' experiments. For each 
experiment a numerical fit for the particle masses was performed, 
using the method of least squares, thus appropriately handling 
the correlation between measurements due to the common jet energy 
scale error. Where available the analytic expressions for the masses
in terms of the endpoints were used to provide starting points for the
fits.

The least squares function was found to often have two or even 
three minima of comparable importance, a consequence of the multiple 
realisations for many of the endpoints. 
Without the threshold measurement there are for both scenarios 
usually two equally good minima, one in the correct 
region and one in another region, giving different masses. 
When the threshold endpoint is added, the minimum in the correct 
region is usually preferred. 
Still, in a noticeable fraction of experiments there will be 
more than one solution. Due to less precise measurements this applies 
more to \pbeta\ than to \palpha.

At \palpha\ the minima of the correct region give masses very close to the
nominal ones.
The other (incorrect) region gives masses some 10--15~GeV lower. 
The ensemble distributions are symmetric.  At \pbeta\ there is one
high-mass solution. The more precisely the threshold endpoint is determined,
the less important this false minimum becomes. The low-mass solutions, one or
two, are closer to the nominal values, but the distributions are skewed. This
is a combined effect of the large endpoint uncertainties and the so-called
border effect.

The obtained masses of the three lightest particles are found to be very
strongly related.  
Furthermore it was seen that mass
differences are better variables, in the sense that they are less correlated
than the masses themselves.  Due to the form of the endpoint expressions, the
LHC will measure mass differences at high precision, but leave the overall
scale less certain.  A Linear Collider measurement of the LSP mass effectively
sets the scale, which is why the precision of the masses improve drastically
when the LHC and the Linear Collider measurements are combined.

\acknowledgments
This work has been performed partly within the ATLAS Collaboration,
and we thank collaboration members for helpful discussions.
We have made use of the physics analysis framework and tools
which are the result of collaboration-wide efforts.
It is a great pleasure to thank Giacomo Polesello for
his contributions in the early stages of this work,
and continued interest and advice.
BKG would like to thank Steinar Stapnes for useful discussions. 
DJM would like to thank Ben Allanach for useful discussions.
This research has been supported in part by the Research Council of Norway.

%%%%%%%%%%%%%%%%%%%%%%%%%%%%%%%%%%%%%%%%%%%%%%%%%%%%%%%%%%%%%%%%%%%%%%

\end{document}